\documentclass[]{aa_old}

\usepackage{graphicx}
\usepackage{natbib}
\usepackage{amssymb}

\def\lsim{\mathrel{\rlap{\lower4pt\hbox{\hskip1pt$\sim$}}
    \raise1pt\hbox{$<$}}}                
\def\gsim{\mathrel{\rlap{\lower4pt\hbox{\hskip1pt$\sim$}}
    \raise1pt\hbox{$>$}}}                

\begin{document}

\title{The evolution of the luminosity functions in the FORS Deep
  Field from low to high redshift: II. The red bands \thanks{Based on
    observations collected with the VLT on Cerro Paranal (Chile) and
    the NTT on La Silla (Chile) operated by the European Southern
    Observatory in the course of the observing proposals 63.O-0005,
    64.O-0149, 64.O-0158, 64.O-0229, 64.P-0150, 65.O-0048, 65.O-0049,
    66.A-0547, 68.A-0013, and 69.A-0014.}
}
%

\author{
Armin Gabasch\inst{1,2}
\and
Ulrich Hopp\inst{1,2}
\and
Georg Feulner\inst{1,2}
\and
Ralf Bender\inst{1,2}
\and
Stella Seitz\inst{1}
\and
Roberto~P. Saglia\inst{2}
\and
Jan Snigula\inst{1,2}
\and
Niv Drory\inst{3}
\and
Immo Appenzeller\inst{4}
\and
Jochen Heidt\inst{4} 
\and
D\"orte Mehlert\inst{4}
\and
Stefan Noll\inst{2,4}
\and
Asmus B\"ohm\inst{5}
\and
Klaus J\"ager\inst{5}
\and
Bodo Ziegler\inst{5}
}

\institute{
Universit\"ats--Sternwarte M\"unchen, Scheinerstr. 1, D$-$81679 M\"unchen,
Germany 
\and
Max--Planck--Institut f\"ur Extraterrestrische Physik,
Giessenbachstra\ss e, D-85748 Garching, Germany
\and
McDonald Observatory, University of Texas at Austin, Austin, Texas 78712
\and 
Landessternwarte Heidelberg, K\"onigstuhl,
D$-$69117 Heidelberg, Germany
\and 
Institut f\"ur Astrophysik, Friedrich--Hund--Platz 1,  
37077 G\"ottingen, Germany 
}

\authorrunning{Gabasch et al.}
\titlerunning{The evolution of the luminosity functions in the FDF: II. The red bands}

\offprints{A.~Gabasch}
\mail{gabasch@usm.uni-muenchen.de}
\date{Received --; accepted --}

\abstract{We present the redshift evolution of the restframe galaxy
  luminosity function (LF) in the red r', i', and z' bands as derived
  from the FORS Deep Field (FDF), thus extending the results published
  in \cite{gabasch:1} to longer wavelengths.  Using the deep and
  homogeneous I-band selected dataset of the FDF we are able to follow
  the red LFs over the redshift range \mbox{$0.5 < z < 3.5$}.  The
  results are based on photometric redshifts for 5558 galaxies derived
  from the photometry in 9 filters achieving an accuracy of
  \mbox{$\Delta z / (z_{spec}+1) \le 0.03$} with only $\sim 1$\%
  outliers.  A comparison with results from the literature shows the
  reliability of the derived LFs.  Because of the depth of the FDF we
  can give relatively tight constraints on the faint-end slope
  $\alpha$ of the LF: The faint-end of the red LFs does not show a
  large redshift evolution and is compatible within $1\sigma$ to
  $2\sigma$ with a constant slope over the redshift range \mbox{$0.5
    \lsim z \lsim 2.0$}.  Moreover, the slopes in r', i', and z' are
  very similar with a best fitting value of $\alpha=-1.33 \pm 0.03$
  for the combined bands.  There is a clear trend of $\alpha$ to
  steepen with increasing wavelength: \mbox{$\alpha_{UV \& u'}=-1.07
    \pm 0.04$} $\rightarrow$ \mbox{$\alpha_{g' \& B}=-1.25 \pm 0.03$}
  $\rightarrow$ \mbox{$\alpha_{r' \& i' \& z'}=-1.33 \pm 0.03$}.  We
  subdivide our galaxy sample into four SED types and determine the
  contribution of a typical SED type to the overall LF.  We show that
  the wavelength dependence of the LF slope can be explained by the
  relative contribution of different SED-type LFs to the overall LF,
  as different SED types dominate the LF in the blue and red bands.
  Furthermore we also derive and analyze the luminosity density
  evolution of the different SED types up to $z
  \sim 2$.\\
  We investigate the evolution of M$^\ast$ and $\phi^\ast$ by means of
  the redshift parametrization \mbox{$M^\ast(z)= M^\ast_0 + {a}
    \ln(1+z)$} and \mbox{$\phi^\ast(z)= \phi^\ast_0 (1+z)^{b}$}.
   Based on the FDF data, we find only a mild brightening
    of M$^\ast$ ($a_{r'} \sim -0.8$, and $a_{i',z'} \sim -0.4$) and
    decrease of $\phi^\ast$ ($ b_{r',i',z'} \sim -0.6$) with increasing
    redshift.  Therefore, from \mbox{$\langle z \rangle\sim 0.5$} to
    \mbox{$\langle z \rangle\sim 3$} the characteristic luminosity
    increases by $\sim$~0.8, $\sim$~0.4 and $\sim$~0.4 magnitudes in
    the r', i', and z' bands, respectively.  Simultaneously the
    characteristic density decreases by about 40\% in all analyzed
    wavebands. 
  A comparison of the LFs with semi-analytical galaxy formation models
  by \citet{kauffmann:2} shows a similar result as in the blue bands:
  the semi-analytical models predict LFs which describe the data at
  low redshift very well, but show growing disagreement with
  increasing redshifts.

   }

\maketitle

\section{Introduction}
\label{sec:lfred:intro}

One of the major efforts in extragalactic astronomy is to derive and
analyze the restframe galaxy luminosity function in different
bandpasses and redshift slices in order to follow the time evolution
of the galaxy populations by a statistical approach.  This is of
particular importance because the energy output at different
wavelengths is dominated by stars of different masses. While galaxy
luminosities measured in the ultraviolet are sensitive to the energy
output of hot, short-living O and B type stars and therefore to
ongoing star formation \citep{tinsley:2,madau:96,mad_poz_dick1}, the
optical and NIR luminosities provide constraints on more evolved
stellar populations \citep{hunter:1}.  
This can be used, in principle, to derive the evolution of
  basic galaxy properties as the stellar mass function \citep[see
  e.g.][ and references therein]{drory:5}, the star formation rate
  density \citep[see e.g.][ and references therein]{gonzales:1} or the
  specific star formation rate \citep[see e.g.][ and references
  therein]{feulner:3}. The determination of these quantities, however, is
  based on assumptions, e.g.\ on the shape of the initial mass function or on the
  details in modeling the stellar population like age, chemical
  composition, and star formation history. Hence studying the LF at
  different wavelengths and cosmic epochs offers a more direct
  approach to the problem of galaxy evolution.\\
As the LF is one of the fundamental observational tools, the amount of
work spent by different groups in deriving accurate LFs is
substantial.  Based on either spectroscopic redshifts, drop-out
techniques, or photometric redshifts, it has been possible to derive
luminosity functions at different redshifts in the ultraviolet \& blue
band (\citealt{baldry:1,croton:1,arnouts:3,budavari:1,treyer:1}; see
also \citealt{gabasch:1} and references therein), in the red bands
\citep{lin:2, lin:1,
  brown:1,shapley:1,combo17:1,chen:1,ilbert:2,dahlen:1,trentham:1} as
well as in the near-IR bands \citep{loveday:2, Kochaneketal01,
  Coleetal2001,
  balogh:1, drory:1, huang:2, feulner:1,pozzetti:3,dahlen:1}.\\
The evolution of the characteristic luminosity and density of galaxy
populations can be analyzed by fitting a Schechter function
\citep{schechter:1} to the LF.  The redshift evolution of the three
free parameters of the Schechter function, the characteristic
magnitude M$^\ast$, the density $\phi^\ast$, and the faint-end slope
$\alpha$ can be used to quantitatively describe the change of the LF
as a function of redshift.  
 Unfortunately, the Schechter parametrization of the LF cannot
  account for possible excesses at the bright and faint end or other
  subtle shape deviations.  Furthermore, the Schechter parameters are
  highly correlated making it challenging, but not impossible, to
  clearly separate the evolution of the different parameters (see e.g.
  \citealt{andreon:1} for a discussion).\\
The evolution of the LFs is also very suitable to constrain the free
parameters of theoretical models (e.g. semi-analytical or smoothed
particle hydrodynamics models).  Ideally a comparison between model
predictions and observations should be done simultaneously for
different wavebands (UV, optical, NIR) and for different redshift
slices as different stellar populations are involved in generating the
flux in the different bands. Therefore, the FDF \citep{fdf_data}
provides a unique testing ground for model predictions, as the depth
and the covered area allow relatively precise LF measurements from the
UV to the z'-band up to high redshift in a very
homogeneous way.\\
In this paper we extend the measurements of the blue luminosity
functions presented in \citet[][ hereafter FDFLF~I]{gabasch:1} to the
red r', i', and z' bands.  In Sect.~\ref{sec:lfred:lumfkt} we derive
the LFs and show the best fitting Schechter parameters M$^\ast$,
$\phi^\ast$, and $\alpha$ in the redshift range \mbox{$0.5 < z <
  3.5$}. We also present a detailed analysis of the slope of the LF as
a function of redshift and wavelength. Furthermore, we analyze the
contributions of different SED types to the overall LF and present the
evolution of the type dependent luminosity density up to redshift
$z\sim 2$.  Sect.~\ref{sec:lfred:evol_parameter} shows a parametric
analysis of the redshift evolution of the LF, whereas a comparison
with the LFs of other surveys as well as with model predictions is
given in Sect.~\ref{sec:lfred:lit} and in Sect.~\ref{sec:lfred:model},
respectively. We summarize our work in
Sect.~\ref{sec:lfred:summary_conclusion}.\\
Throughout this paper we use AB magnitudes and adopt a $\Lambda$
cosmology with \mbox{$\Omega_M=0.3$}, \mbox{$\Omega_\Lambda=0.7$}, and
\mbox{$H_0=70 \, \mathrm{km} \, \mathrm{s}^{-1} \,
  \mathrm{Mpc}^{-1}$}.

\section{Luminosity functions in the r', i', and z' bands}
\label{sec:lfred:lumfkt}

The results presented in this paper are all based on the deep part of
the I-band selected catalog of the FDF \citep{fdf_data} as introduced
in FDFLF~I.  Galaxy distances are determined by the photometric
redshift technique \citep{bender:1} with a typical accuracy of
\mbox{$\Delta z / (z_{spec}+1) \approx 0.03 $} if compared to the
spectroscopic sample \citep{noll:1, boehm:1} of more than 350 objects.
To derive the absolute magnitude for a given band (which will be
briefly summarized below) we use the best fitting SED as determined by
the photometric redshift code, thus reducing possible systematic
errors which could be introduced by using k-corrections applied to a
single observed magnitude. To account for the fact that some fainter
galaxies are not visible in the whole survey volume we perform a
$V/V_{max}$ \citep{Schmidt1} correction.  The errors of the LFs are
calculated by means of Monte-Carlo simulations and include the
photometric redshift error of every single galaxy as well as the
statistical error (Poissonian error).  To derive precise Schechter
parameters we limit our analysis of the LF to the magnitude bin where
$V/V_{max}\le 3$.
We also show the uncorrected LF in the various plots as open circles.
We do not assume any evolution of the galaxies within the
  single redshift bins, since the number of galaxies and the distance
  determination based on photometric redshifts would not be able to
  constrain it.
The redshift binning was chosen such that we have good
statistics in the various redshift bins and that the influence of
redshift clustering was minimized. In order to have good statistics at
the bright end (rare objects) of the LF we had to slightly change some
of the redshift bins if compared to FDFLF~I. The new redshift binning
together with the number of galaxies in every bin is shown in
Table~\ref{tab:lfred:binning}.  As can be seen from
Table~\ref{tab:lfred:binning}, the redshift intervals are
  approximately the same size in $\ln(1+z)$ and  most of the results
we are going to discuss in this paper are based on 700 -- 1000
galaxies per redshift bin.

\begin{table}[]
\caption
{\label{tab:lfred:binning}Number of galaxies in the FDF for the
  redshift intervals used for computing the LFs. Note
  that we derive the LF in all redshift bins, but
  exclude the lowest ($z < 0.45$) and highest redshift bin ($z >
  3.81$) from our analysis of the LF evolution, since
  the lowest redshift bin corresponds to too small a volume while the
  $z>3.81$ bin suffers from extrapolation errors.}
\begin{center}
\begin{tabular}{c|cr}
\hline\hline
redshift &  number     & fraction\\
interval & of galaxies & of galaxies\\
\hline
 0.00 -  0.45  &    808 & 14.54 \%    \\  
 0.45 -  0.85  &   1109 & 19.95 \%    \\  
 0.85 -  1.31  &   1029 & 18.51 \%    \\  
 1.31 -  1.91  &    880 & 15.83 \%    \\  
 1.91 -  2.61  &    816 & 14.68 \%    \\  
 2.61 -  3.81  &    718 & 12.92 \%    \\  
 $>3.81$       &    196 &  3.53 \%    \\   
 unknown       &    2   &  0.04 \%    \\   
\hline
\end{tabular}
\end{center}
\end{table}

\subsection{The slope of the LF as a function of redshift}
\label{sec:lfred:lumfkt:slope_z}

\begin{table*}[tbh]
\caption{\label{tab:lfred:slope_single}Slope of the LF
  for all wavelengths and all redshifts as derived from a 3-parameter
  Schechter fit.}
\begin{center}
\begin{tabular}{c||c|c|c|}
$z$ & $\alpha$ (r') & $\alpha$ (i') & $\alpha$ (z') \\ 
\hline
 0.45 -- 0.85  &$-$1.37 (+0.04 $-$0.04)  &  $-$1.37 (+0.04 $-$0.03) &  $-$1.39 (+0.04 $-$0.04) \\
 0.85 -- 1.31  &$-$1.25 (+0.06 $-$0.04)  &  $-$1.27 (+0.06 $-$0.05) &  $-$1.34 (+0.06 $-$0.04) \\
 1.31 -- 1.91  &$-$1.30 (+0.16 $-$0.09)  &  $-$1.50 (+0.13 $-$0.10) &  $-$1.45 (+0.12 $-$0.09) \\
 1.91 -- 2.61  &$-$1.01 (+0.15 $-$0.14)  &  $-$1.03 (+0.17 $-$0.14) &  $-$0.97 (+0.17 $-$0.12) \\
 2.61 -- 3.81  &$-$0.98 (+0.17 $-$0.17)  &  $-$1.03 (+0.15 $-$0.13) &  $-$1.01 (+0.15 $-$0.13) \\
\end{tabular}
\end{center}
\end{table*}

\begin{table}[tbh]
\caption[Slope $\alpha$ of the LFs for the different
wavebands as determined from an error-weighted fit to the data]
{\label{tab:lfred:slope_fixed} 
In the upper part of the Table we show the slope $\alpha$ of the luminosity
  functions for the different wavebands as determined from an
  error-weighted fit to the data under
  the assumption that $\alpha(z)=\mathrm{const.}$ In the
  lower part of the Table we show the best values of $\alpha$ after
  combining the data of all bands.}
\begin{center}
\begin{tabular}{l|c}
filter  & $\alpha(z)=const.$\\
\hline
\rule[+3mm]{-1.4mm}{2mm}
\rule[-3mm]{0mm}{2mm}{r' }        & $-1.30 \pm 0.05$\\
\rule[-3mm]{0mm}{2mm}{i' }        & $-1.33 \pm 0.05$\\
\rule[-3mm]{0mm}{2mm}{z'  }        & $-1.35 \pm 0.05$\\
\hline
\rule[+3mm]{-1.4mm}{2mm}
\rule[-3mm]{0mm}{2mm}{r' \& i' \& z'} & $-1.33 \pm 0.03$\\
\end{tabular}
\end{center}
\end{table}

To investigate the redshift evolution of the faint-end slope of the
LF, we fit a three parameter Schechter function (M$^\ast$,
$\phi^\ast$, and $\alpha$) to the data points. The best fitting slope
$\alpha$ and the corresponding $1\sigma$ errors for the 3 wavebands
are reported in Table~\ref{tab:lfred:slope_single} for the various
redshift bins.\\
It can be inferred from Table~\ref{tab:lfred:slope_single} that there
is only marginal evidence for a change of $\alpha$ with redshift (at
least up to $z\sim 2$ where we are able to sample the LF to a suitable
depth).  Under the assumption that $\alpha$ does not depend on
redshift, Table~\ref{tab:lfred:slope_fixed} (upper part) yields the
slopes' best error-weighted values in the redshift range from
\mbox{$\langle z \rangle\sim 0.65$} to \mbox{$\langle z \rangle \sim
  1.6$} (including also the higher redshift bins changes $\alpha$ only
marginally).  Since the slopes in all bands are very similar we derive
a combined slope of $\alpha_{r' \& i' \& z'}=-1.33 \pm 0.03$
(Table~\ref{tab:lfred:slope_fixed}, lower
part).\\
Almost all of the slopes listed in Table~\ref{tab:lfred:slope_single}
are compatible within \mbox{$1\sigma - 2\sigma$} with $\alpha =-1.33
\pm 0.03$. Therefore, we fixed the slope to this value for the further
analysis. Please note, that this slope is steeper than for the blue
bands ($\alpha_{UV \& u'}=-1.07$ and $\alpha_{g' \& B}=-1.25$), but it
follows the trend observed in FDFLF~I: With increasing wavelength the
slope steepens, i.e. the ratio of faint to bright galaxies increases.
This trend is illustrated best in Fig.~\ref{fig:lfsedtype:slopes},
where we combine the results derived in FDFLF~I with those of this
work and plot the wavelength dependence of the LF slope.  As we will
show in Sect.~\ref{sec:lfred:lumfkt:slope_lambda}, this effect can be
explained by the contribution of different galaxy populations to the
overall LF in the various wavebands.\\

\subsection{I selection versus I+B selection}

We checked the dependence of our results on the selection band by
comparing the I band selected catalog and the I+B selected FDF
catalog. The combined catalog has been described in \citet{fdf_data}
and reaches limiting magnitudes of $ I \sim 26.8$ and $B \sim 27.6$.
In the combined sample $M^\ast$ agrees within its $1\sigma$ errors
with the values derived from the I-band catalog only. The slope
$\alpha$ tends to be slightly steeper in the combined sample, but by
not more than $1\sigma$.  The larger number of objects in the combined
catalog mostly influences the characteristic density $\phi^\ast$ which
is a factor of 1.05 to 1.20 larger (depending on the redshift bin).
Given the errors of $\phi^\ast$, this is in the order of $1\sigma$ to
$2\sigma$.

\subsection{The slope of the LF as a function of wavelength}
\label{sec:lfred:lumfkt:slope_lambda}

\begin{figure}[tbp]
  \centering
  \includegraphics[width=0.5\textwidth]{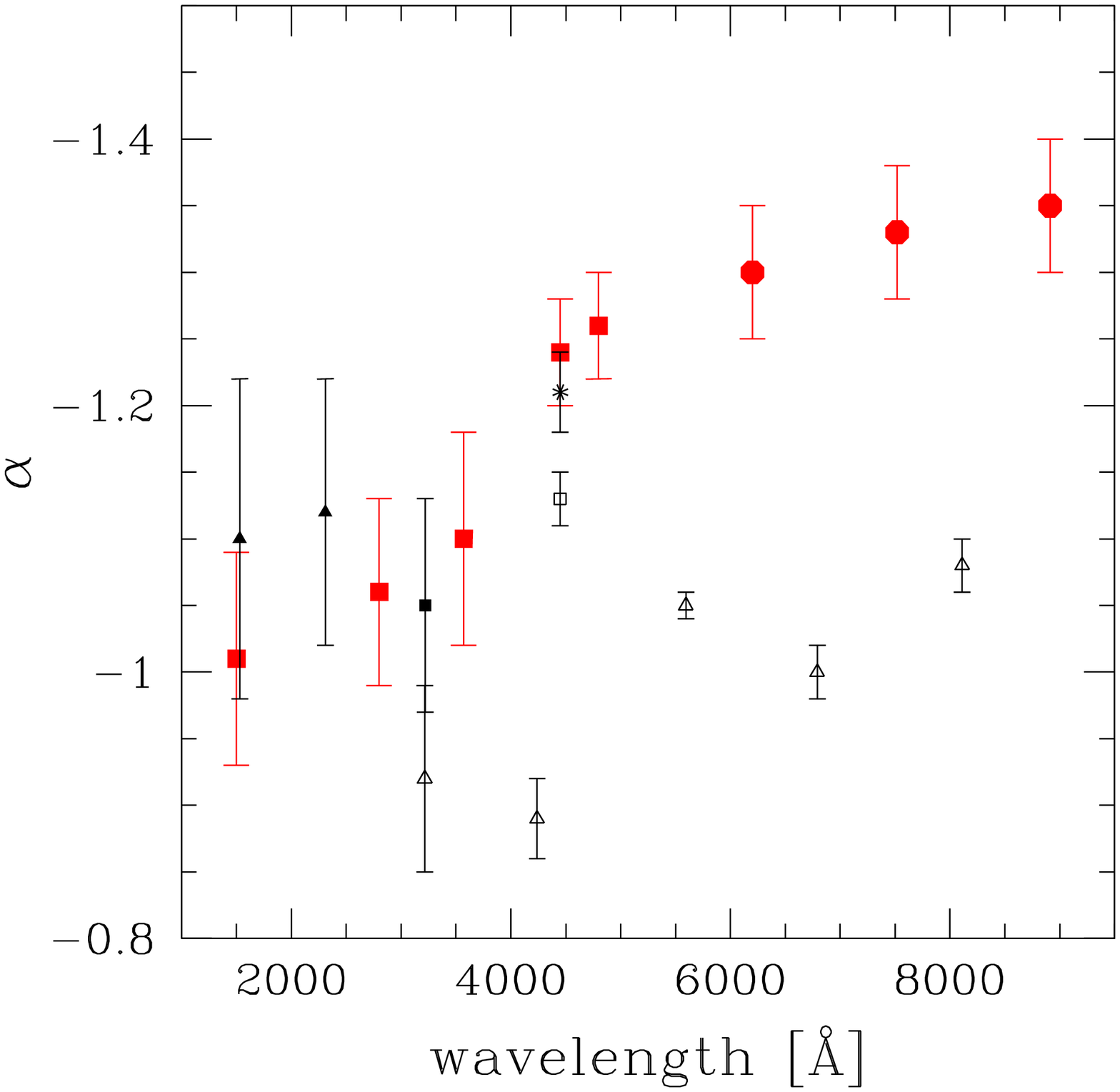}
\caption{\label{fig:lfsedtype:slopes} 
    Slope of the LF as a function of wavelength. The filled red
    squares denote the values derived in FDFLF~I whereas the filled
    red dots are taken from this work (Table
    \ref{tab:lfred:slope_fixed}, upper part). Local slope values (black) are
    shown as filled squares \citep{baldry:1}, open squares
    \citep{driver:1}, filled triangles \citep{budavari:1}, open
    triangles \citep{blanton:2}, and as an asterisk
    \citep{norberg:1}.
}
\end{figure}
\begin{figure}[tbp]
\centering
\includegraphics[width=0.5\textwidth]{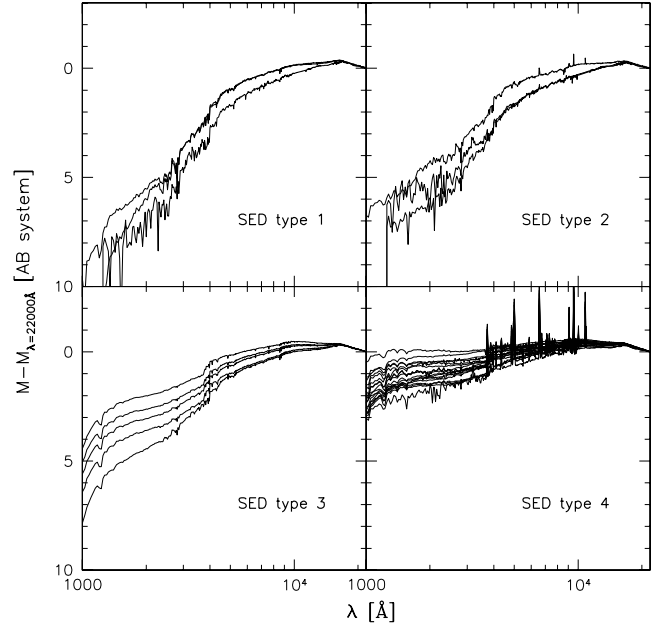}
\caption{\label{fig:lfsedtype:sed} SEDs grouped according to their spectral
  type. See text for details.  }
\end{figure}

\begin{figure*}[tbp]
\includegraphics[width=\textwidth]{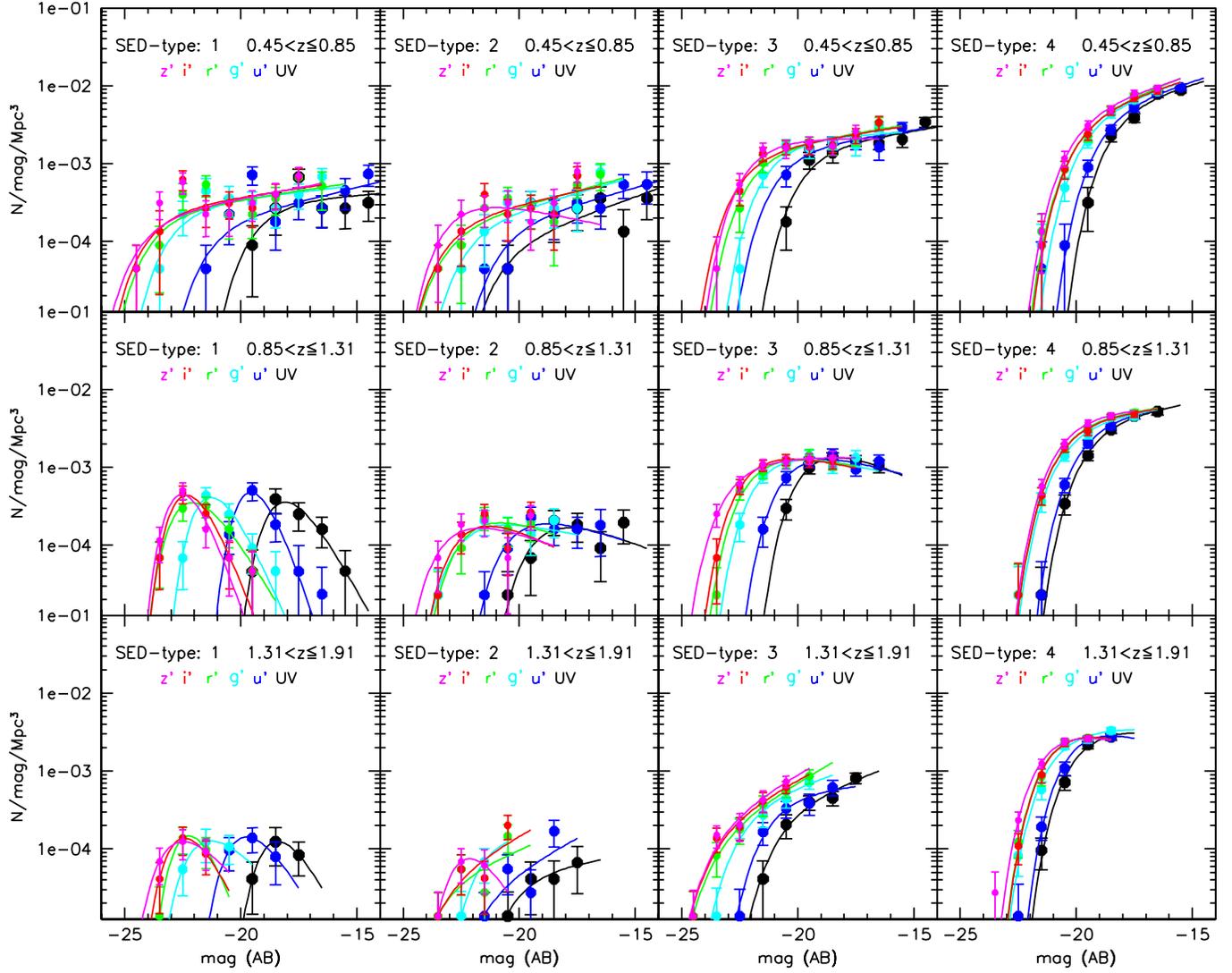} 
\caption{\label{fig:lfsedtype:16panel_lf} 
  LFs for the four SED types in the redshift intervals \mbox{$ 0.45 < z
    \le 0.85 $}~(upper panels), \mbox{$ 0.85 < z \le 1.31 $}~(middle
  panels), and \mbox{$ 1.31 < z \le 1.91 $}~(lower panels): SED type
  increases from left to right. The filters are color coded and
  denoted in the upper part of the various panels. For clarity, the
  three parameter Schechter function fit to the data is shown as
  solid line with the same color coding as the LF.}
\end{figure*}

\begin{figure*}[tbp]
\centering
\includegraphics[width=0.35\textwidth]{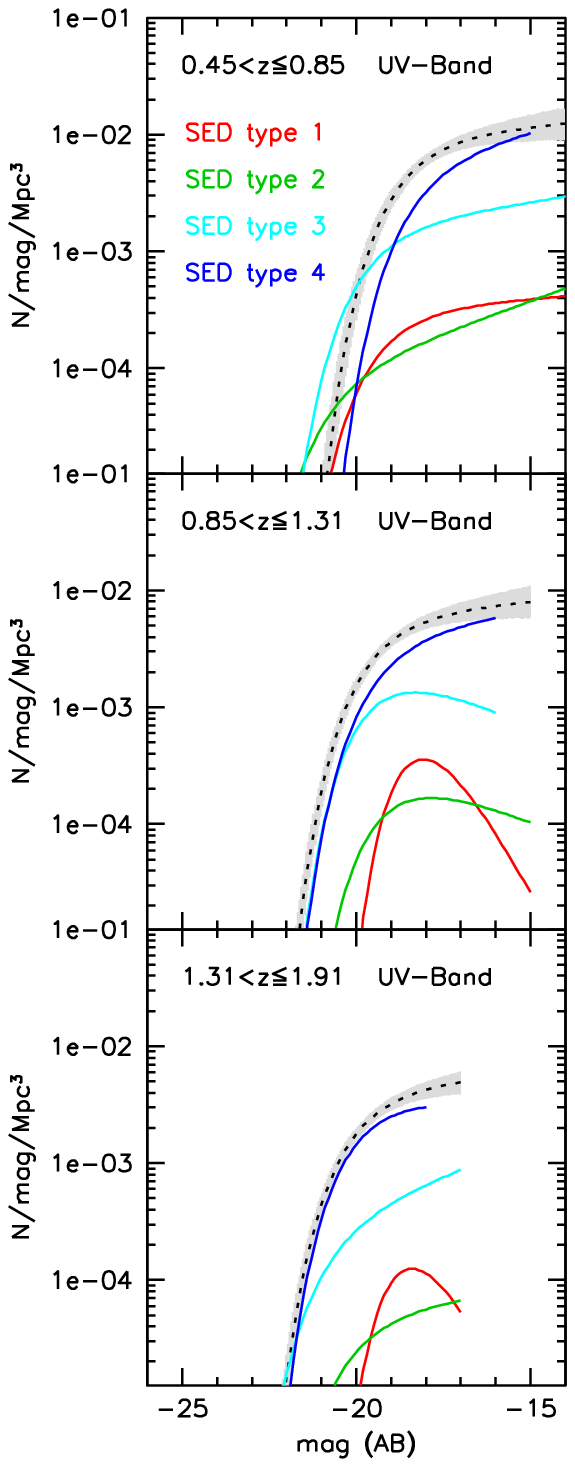} 
\includegraphics[width=0.35\textwidth]{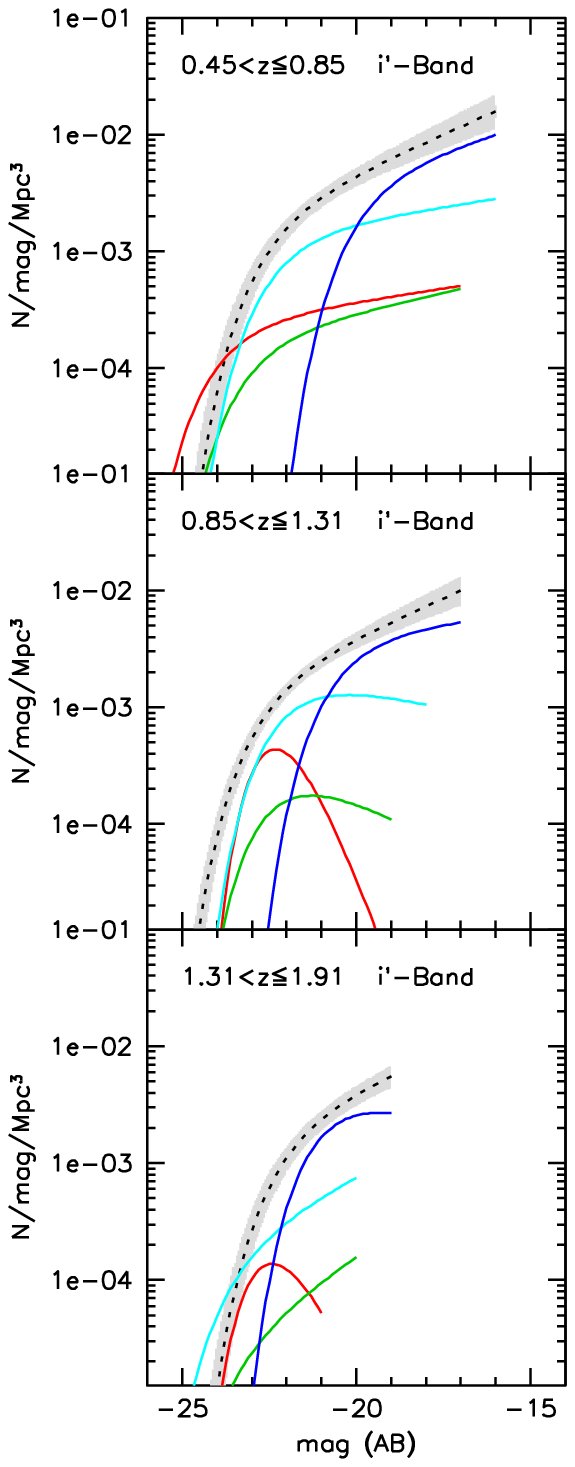} 
\caption{\label{fig:lfsedtype:uv_i_schechter} Schechter
  functions fitted to the LFs in the UV (left panels) as well as in
  the i'-band (right panels) for the redshift intervals \mbox{$ 0.45 <
    z \le 0.85 $}~(upper panels), \mbox{$ 0.85 < z \le 1.31 $}~(middle
  panels), and \mbox{$ 1.31 < z \le 1.91 $}~(lower panels).  The solid
  lines show the best fitting Schechter functions for the four SED
  types. The SED type is color coded and denoted in the upper left
  panel.  The dotted black line shows the total LF (as fitted to the
  data) whereas the shaded region represent the corresponding
  1$\sigma$ error of the latter.  For the slope values of the
  different SED types see Table~\ref{tab:lfred:slope_single_sed}.
  Please note that we show the SED type LFs to the limiting magnitude
  where the $V/V_{max}$ begins to contribute by at most a factor of
  1.5, being more conservative than for the overall LF for which we
  allow a correction factor of 3.}
\end{figure*}

\begin{figure*}[tbp]
\centering
\includegraphics[width=0.8\textwidth]{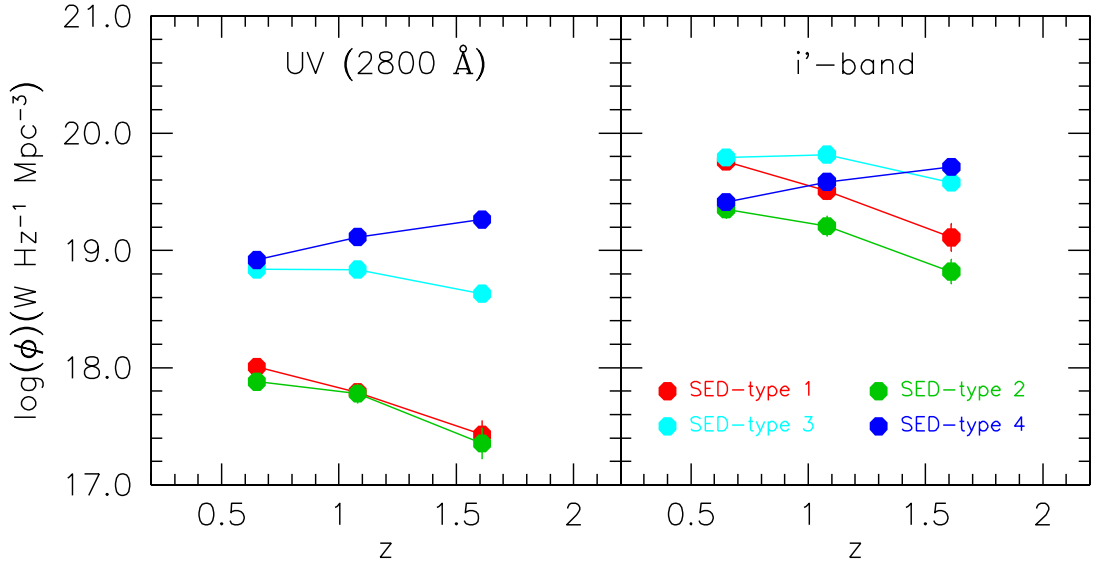} 
\caption{\label{fig:lfsedtype:lumdens} 
  Luminosity densities in the UV (left panel) and i' (right panel)
  bands for the four SED types in the redshift intervals \mbox{$ 0.45 < z
    \le 0.85 $}, \mbox{$ 0.85 < z \le 1.31 $}, and \mbox{$ 1.31 < z
    \le 1.91 $}. The luminosity densities are completeness corrected
  to zero luminosity (ZGL, see text for details). The values are
  listed in Table~\ref{tab:lfred:lumdens}}
\end{figure*}

\begin{table*}[tbp]
\caption{\label{tab:lfred:lumdens} 
Luminosity densities in the UV and i'
  bands for the four SED types. See also Fig.~\ref{fig:lfsedtype:lumdens}}
\begin{center}
\begin{tabular}{c|c|c|c|c|c}
           filter & SED type & redshift & luminosity density & error & completeness correction (ZGL)\\ 
 & & & W Hz$^{-1}$ Mpc$^{-3}$ & W Hz$^{-1}$ Mpc$^{-3}$ & \%\\
\hline\hline
  UV (2800\AA)&   1  &  0.45 -- 0.85  & 1.01e+18 & 1.5e+17 & 0.1\\  
              &      &  0.85 -- 1.31  & 6.16e+17 & 1.0e+17   & 0.7\\  
              &      &  1.31 -- 1.91  & 2.68e+17 & 7.6e+16 & 8.9\\  
&&&&&\\                                       
              &   2  &  0.45 -- 0.85  & 7.58e+17 & 1.3e+17 & 0.5\\  
              &      &  0.85 -- 1.31  & 6.00e+17 & 1.2e+17 & 2.0\\  
              &      &  1.31 -- 1.91  & 2.26e+17  & 7.0e+16   & 9.0\\  
&&&&&\\                                      
              &   3  &  0.45 -- 0.85  & 6.88e+18 & 7.9e+17 & 0.6\\  
              &      &  0.85 -- 1.31  & 6.86e+18 & 5.4e+17 & 2.0\\  
              &      &  1.31 -- 1.91  & 4.27e+18 & 5.6e+17 & 4.7\\  
&&&&&\\                                      
              &   4  &  0.45 -- 0.85  & 8.29e+18 & 5.1e+17 & 8.4\\  
              &      &  0.85 -- 1.31  & 1.30e+19 & 8.4e+17 & 6.9\\  
              &      &  1.31 -- 1.91  & 1.83e+19 & 1.9e+18 & 21.4\\ 
&&&&&\\                                      
\hline                                       
 &&&&&\\                                       
  i'          &   1  &  0.45 -- 0.85  &  5.72e+19 & 8.7e+18 & 0.3 \\ 
              &      &  0.85 -- 1.31  &  3.22e+19 & 5.4e+18 & 1.5 \\ 
              &      &  1.31 -- 1.91  &  1.29e+19 & 3.5e+18 & 10.2\\ 
 &&&&&\\                                       
              &   2  &  0.45 -- 0.85  &  2.24e+19 & 3.9e+18 & 0.8 \\ 
              &      &  0.85 -- 1.31  &  1.61e+19 & 3.2e+18 & 3.6 \\ 
              &      &  1.31 -- 1.91  &  6.60e+18 & 1.6e+18 & 11.6\\ 
 &&&&&\\                                      
              &   3  &  0.45 -- 0.85  &  6.21e+19 & 6.9e+18 & 0.4 \\ 
              &      &  0.85 -- 1.31  &  6.56e+19 & 5.1e+18 & 1.6 \\ 
              &      &  1.31 -- 1.91  &  3.79e+19 & 5.4e+18 & 7.4 \\ 
 &&&&&\\                                       
              &   4  &  0.45 -- 0.85  &  2.59e+19 & 1.5e+18 & 5.8 \\ 
              &      &  0.85 -- 1.31  &  3.84e+19 & 2.4e+18 & 5.7 \\ 
              &      &  1.31 -- 1.91  &  5.17e+19 & 4.5e+18 & 18.9\\ 
\end{tabular}
\end{center}
\end{table*}

\begin{table*}[tbp]
\caption{\label{tab:lfred:slope_single_sed} 
Slope of the UV (2800 \AA\,) and i' band LF
  for the different SED types from a 3-parameter Schechter fit. The Schechter
  functions are shown in Fig.~\ref{fig:lfsedtype:uv_i_schechter}}
\begin{center}
\begin{tabular}{c|c||c|c|c|c}
$z$ & filter & $\alpha$ for SED type 1  & $\alpha$ for SED type 2 & $\alpha$ for SED type 3 & $\alpha$ for SED type 4 \\ 
\hline
&&&&&\\
 0.45 -- 0.85  & UV & $-$1.06 (+0.16 $-$0.10)  & $-$1.27 (+0.08 $-$0.04)  & $-$1.12 (+0.11 $-$0.07)& $-$1.19 (+0.13 $-$0.11)\\
 0.45 -- 0.85  & i' & $-$1.11 (+0.15 $-$0.02)  & $-$1.17 (+0.22 $-$0.03)  & $-$1.12 (+0.12 $-$0.11)& $-$1.23 (+0.07 $-$0.10)\\
&&&&&\\
 0.85 -- 1.31  & UV & +0.38   (+0.60 $-$0.37)  & $-$0.71 (+0.62 $-$0.27)  & $-$0.68 (+0.17 $-$0.15)& $-$1.14 (+0.12 $-$0.08)\\
 0.85 -- 1.31  & i' & +1.04   (+0.65 $-$0.68)  & $-$0.62 (+0.79 $-$0.32)  & $-$0.84 (+0.15 $-$0.13)& $-$1.09 (+0.11 $-$0.06)\\
\end{tabular}
\end{center}
\end{table*}

To better understand the filter-dependence of the LF slope shown in
Fig.~\ref{fig:lfsedtype:slopes}, we analyze the contribution of
different galaxy types to the overall LF.
Thus, we subdivide our galaxy sample into four SED types and analyze
the type-dependent LF, i.e. we determine the contribution of a typical
SED type to the overall LF.
The SEDs are mainly grouped according to the UV-K color (see
Fig.~\ref{fig:lfsedtype:sed}): for increasing spectral type (\mbox{SED
  type 1 $\rightarrow$ SED type 4}) the SEDs become bluer, i.e. the UV
flux (and thus the recent star formation rate) increases if compared
to the K-band flux.  \citet{pannella:1} analyzed the morphology of
about 1400 galaxies in the FDF down to $I \sim 25$ mag on HST (ACS)
data, and find a good correlation between the four main SED types and
the
morphology of the galaxies (at least up to redshift $z \sim 1.5$).\\
The four SED types also show a sequence in the restframe U-V color
often used to discriminate between blue and red galaxies \citep[see
e.g.][ and references therein]{giallongo:1}. As the restframe U-V
color includes the 4000 \AA\ break it is quite sensitive to galaxy
properties as age and star formation.  The U-V color lies in the range
between 2.3 -- 1.9, 2.0 -- 1.6, 1.6 -- 0.9, and 0.9 -- 0 for SED type
1, 2, 3, and 4, respectively. Therefore, in a rough classification one
can refer to SED types 1 and 2 (SED type 3 and 4) as red (blue)
galaxies.  
We use the same SED cuts at all redshifts (see below), i.e. we do
  not use the time evolution of the galaxy color bimodality \citep[see
  e.g.][]{giallongo:1} to redefine the main SED type of a galaxy as a
  function of redshift.

We show in Fig.~\ref{fig:lfsedtype:16panel_lf} the LFs for the four SED
types in three redshift intervals: \mbox{$ 0.45 < z \le 0.85 $},
\mbox{$ 0.85 < z \le 1.31 $}, and \mbox{$ 1.31 < z \le 1.91 $}. The
SED type increases from the left panel to the right panel, i.e. the
extremely star-forming galaxies are shown in the rightmost panel. The
LFs for the different filters are color coded and denoted in the upper
part of the various panels.  We show every LF to the limiting
magnitude where the $V/V_{max}$ begins to contribute by at most a
factor of 1.5, being more conservative as for the overall LF
($V/V_{max} \le 3$ for every bin).  For clarity, a Schechter function
fit to the data is shown as solid line using the same color coding as
for the LF.

First of all it is clear from Fig.~\ref{fig:lfsedtype:16panel_lf},
that the faint-end of the LF is always dominated by SED type 4
galaxies. This is true for \textit{all} analyzed bands.  If we focus
on the bright end of the SED type 4 LFs, we only see a relatively
small variation between the different filters. On the other hand, the
difference between the filters for SED type 1 (for the bright end) is
very large.  Although (because of the low number density) SED type 1
does not contribute at all to the faint-end of the LFs, the picture
changes for the bright end.  While for the bright end of the LF in the
\textit{UV} (black line), SED type 1 and 4 galaxies have about the
same number density, in the
\textit {red bands} SED type 1 galaxies dominate the LF. \\
This trend applies for all three redshift bins, although it is more
pronounced at lower redshift.  It explains naturally the change of the
LF slope as a function of waveband. This can be best seen in
Fig.~\ref{fig:lfsedtype:uv_i_schechter} where we concentrate on only
two filters.  There we show the Schechter functions fitted to the LFs
in the UV as well as in the i'-band for the redshift intervals \mbox{$
  0.45 < z \le 0.85 $}, \mbox{$ 0.85 < z \le 1.31 $}, and \mbox{$ 1.31
  < z \le 1.91 $}. We plot the single Schechter functions for all four
SED types as well as for the overall LF. In the UV the overall LF
(dotted line) is completely dominated by the SED type 4 galaxies.  On
the other hand the overall LF in the i'-band is mainly dominated by
SED type 1 to type 3 at the bright end, and SED type 4 at the
faint-end. This results in a steeper slope for the overall LF.\\
Please note that in Fig.~\ref{fig:lfsedtype:16panel_lf} and
Fig.~\ref{fig:lfsedtype:uv_i_schechter} we show the SED type LFs and
Schechter functions to the limiting magnitude where the $V/V_{max}$
begins to contribute by at most a factor of 1.5, being more
conservative than for the overall LF for which we allow a correction
factor of 3. Furthermore, all Schechter functions in
Fig.~\ref{fig:lfsedtype:uv_i_schechter} are fits to the data points.
This is also true for the overall Schechter function which is
\textit{not} the sum of the individual SED type Schechter functions
and explains why, at the bright end, the overall Schechter function is
in some plots below individual SED type
Schechter functions.\\
Another interesting aspect which can be inferred from
Fig.~\ref{fig:lfsedtype:uv_i_schechter} is the fast decrease in number
density of bright SED type 1 to 3 galaxies if compared to SED type 4
galaxies (for increasing redshift).  Therefore at high redshift ($z
\sim 2$) SED type 4 galaxies start to dominate also the overall
i'-band LF.
This can be seen best, if one follows the redshift evolution of the
type dependent \textit{luminosity density} (LD), i.e. the integrated
light emitted by the different SED types. The results (for the UV and
i' bands) are shown in Fig.~\ref{fig:lfsedtype:lumdens}. We calculated
the LD as described in \citet{gabasch:sfr}: First, we
derive the LD at a given redshift by summing the
completeness corrected (using a $V/V_{max}$ correction) luminosity of
every single galaxy up to the absolute magnitude limits.  Second, we
apply a further correction (to zero galaxy luminosity) ZGL, to take
into account the missing contribution to the LD of the fainter
galaxies.  To this end we use the best-fitting Schechter function for
a slope $\alpha$ constant with redshift. For every SED type we derive
$\alpha(z)=const.$ by an error-weighted averaging of the slopes given
in Table~\ref{tab:lfred:slope_single_sed}. This results in slopes
between $\alpha=-0.98$ and $\alpha=-1.25$.  For the FDF the ZGL
corrections are at most 22\% in size (see last column in
Table~\ref{tab:lfred:lumdens}).  The small ZGL correction employed
here stems from the faint magnitude limits probed by our deep FDF data
set and the relatively flat slopes of the Schechter function.  Errors
are computed from Monte Carlo simulations that take into account the
probability distributions of the
photometric redshifts and the Poissonian error.\\
As shown in the left panel of Fig.~\ref{fig:lfsedtype:lumdens} the
contribution of type 1 and 2 galaxies to the UV LD is negligible at
all analyzed redshifts. SED type 3 and 4 completely dominate the UV
output and although the number density of these galaxies decreases
with increasing redshift the luminosity
density (and thus the SFR) increases.\\
If we analyze the i'-band LD, in the lowest redshift bin SED type 1
and 3 dominate (by a factor of about three if compared to type 2 and
4) and have about the same LD. At higher redshifts the relative
contribution of the different SED types changes because the LD of type
1 and 2 galaxies decreases with increasing redshift and SED type 3 and
4 take over.\\
A detailed analysis of the type dependent LF will be presented in a
future paper (Gabasch et. al., in preparation) where we combine the
I-band selected MUNICS catalog (MUNICS IX, Feulner et. al., in
preparation, $\sim$ 900 arcmin$^2$) with the FDF ($\sim$ 40
arcmin$^2$) catalog. This overcomes the small volume of the FDF at
lower redshift making it possible to include also rare bright objects
in the analysis of the LF. First results in the MUNICS fields will be
presented in MUNICS IX.

\subsection{The redshift evolution of the LFs}
\label{sec:lfred:lumfkt:restframe}

\begin{figure*}[tbp]

  \includegraphics[width=0.33\textwidth]{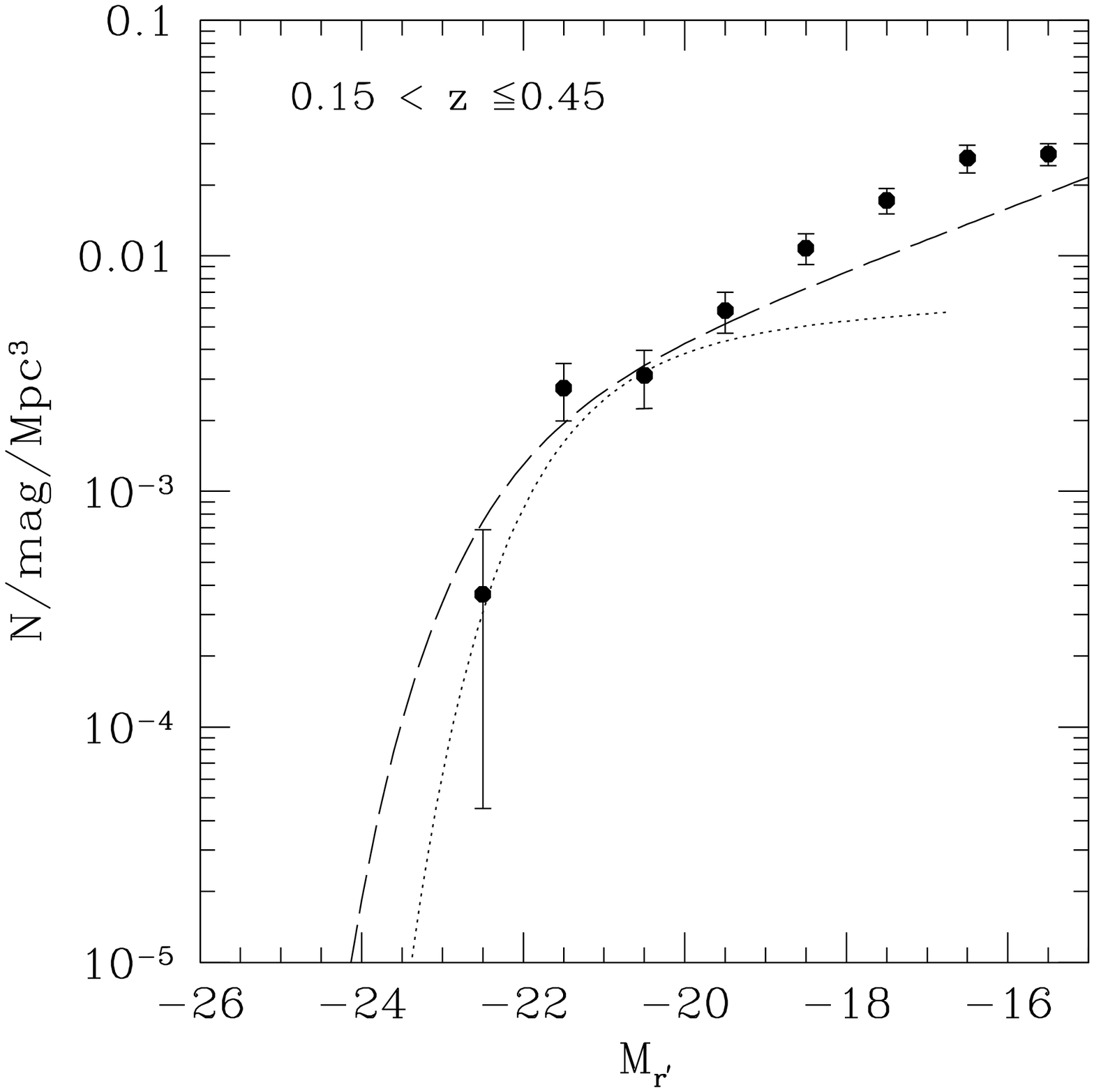}
  \includegraphics[width=0.33\textwidth]{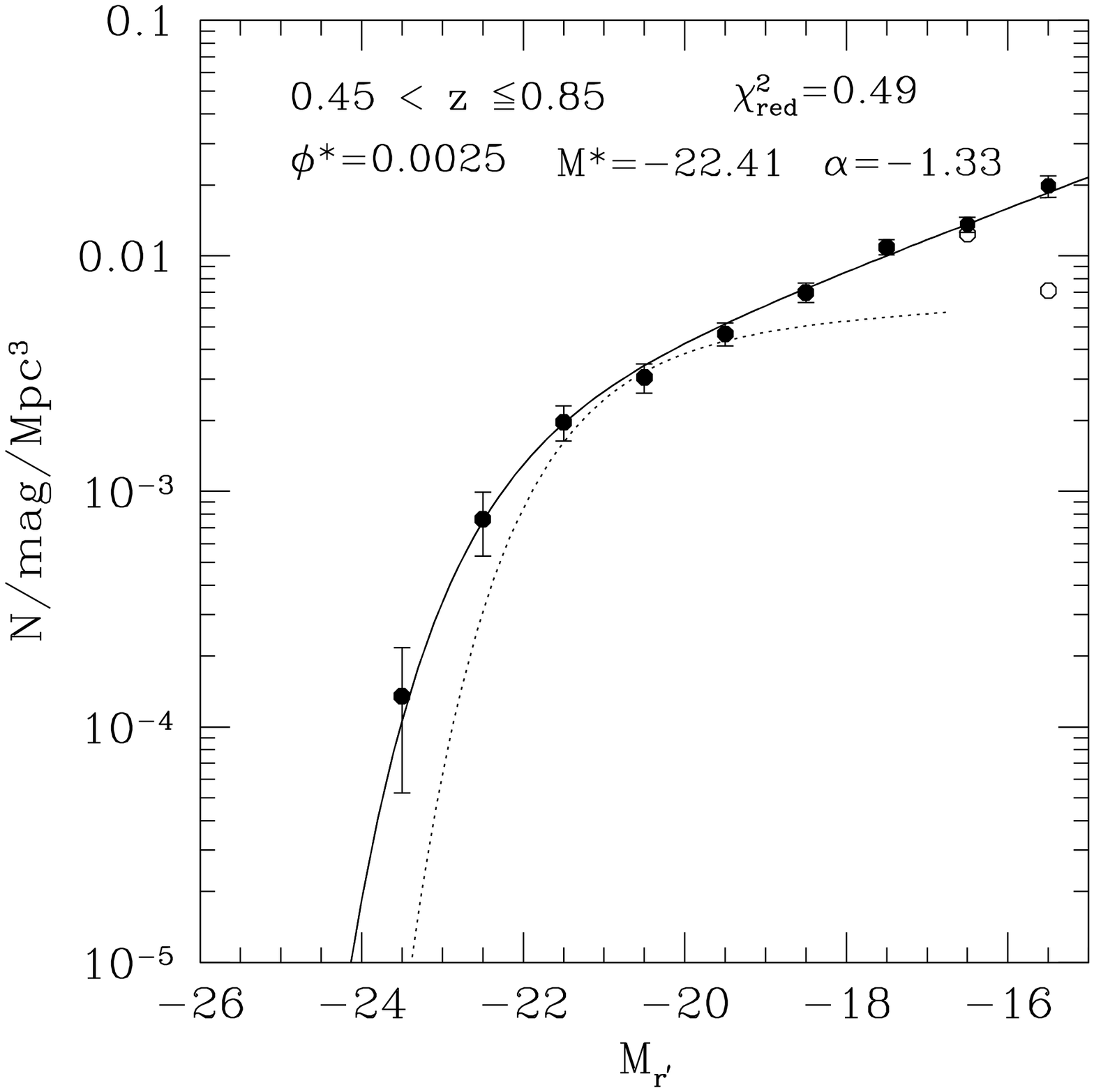}
  \includegraphics[width=0.33\textwidth]{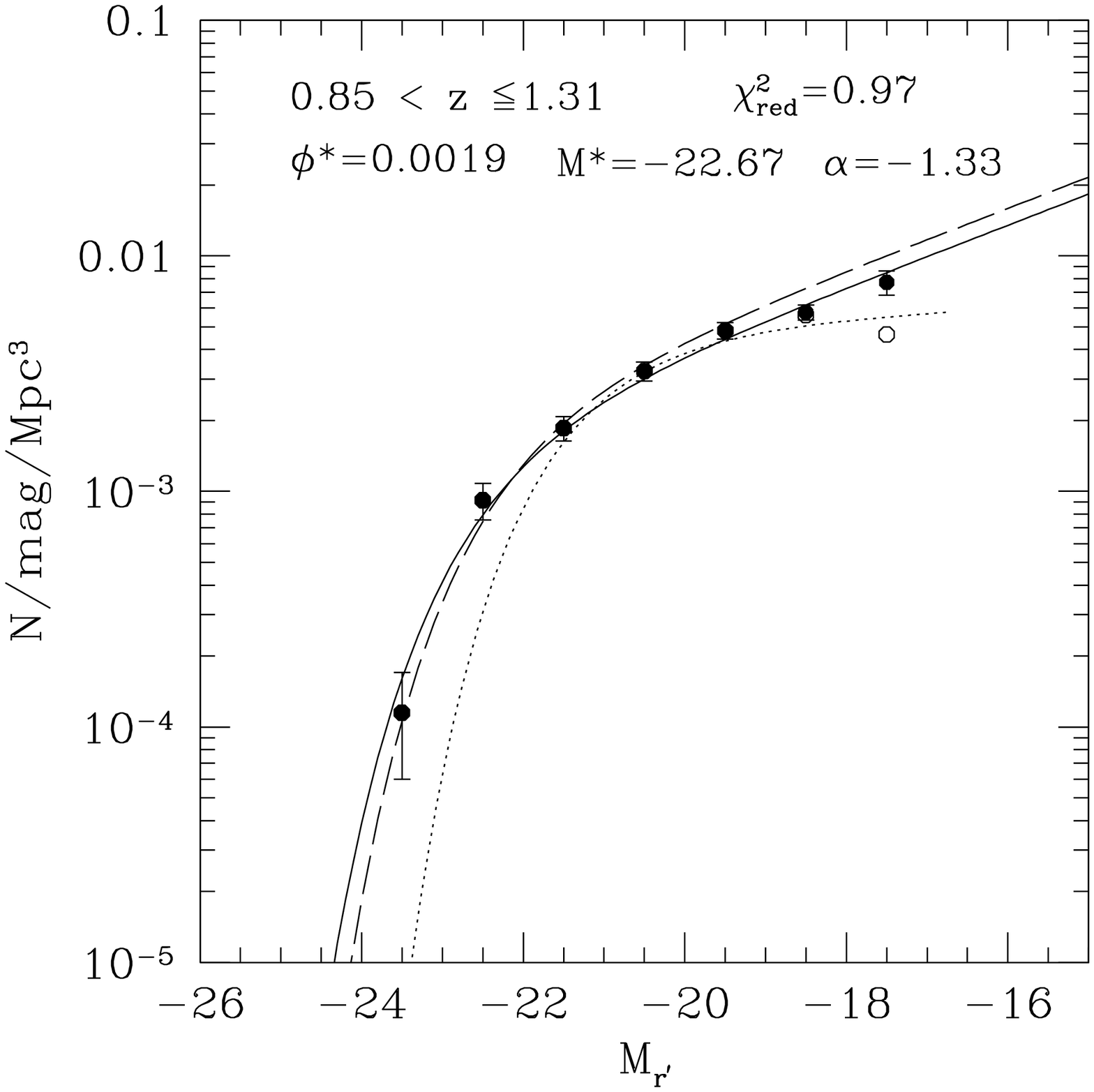}
  \includegraphics[width=0.33\textwidth]{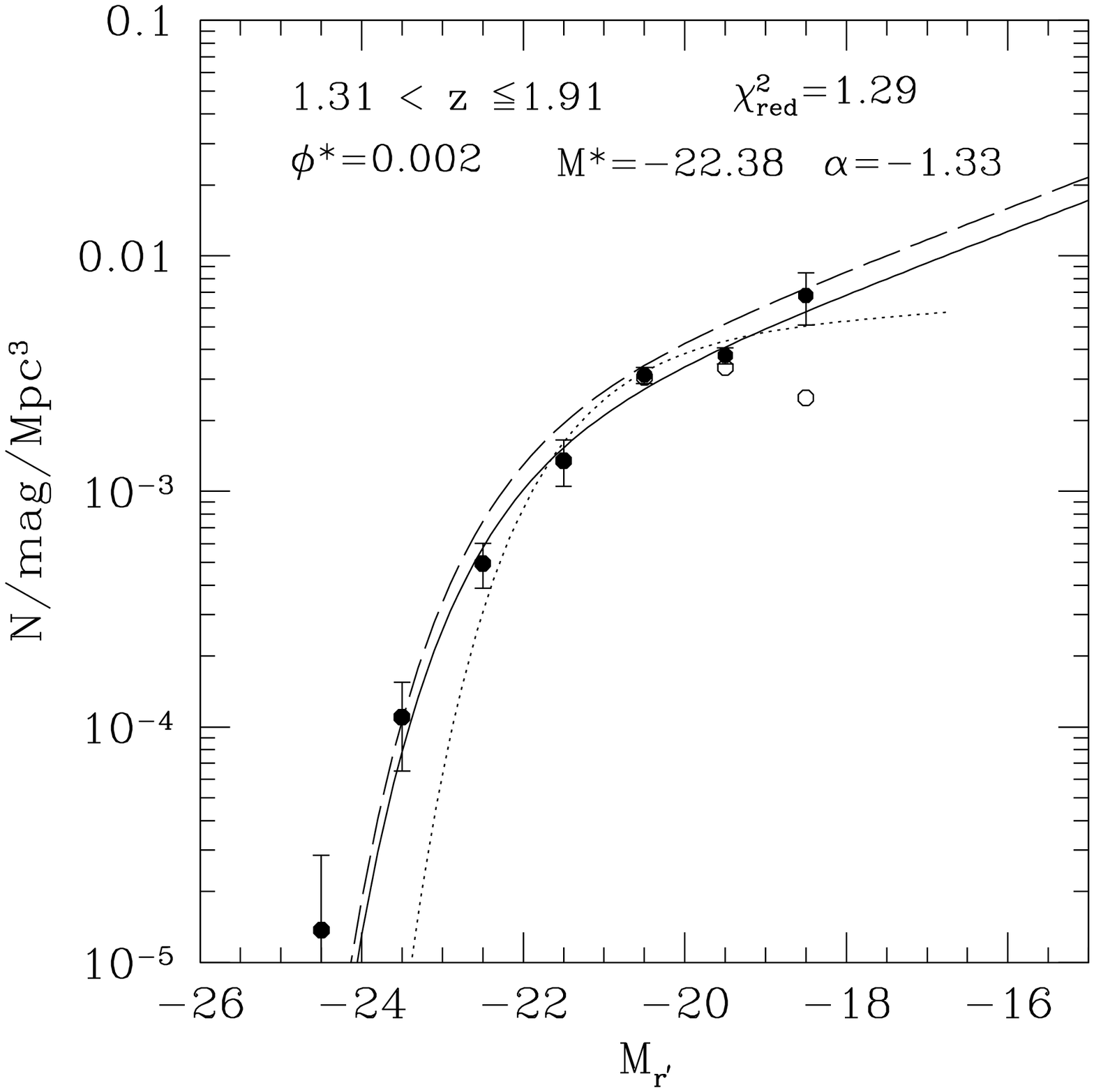}
  \includegraphics[width=0.33\textwidth]{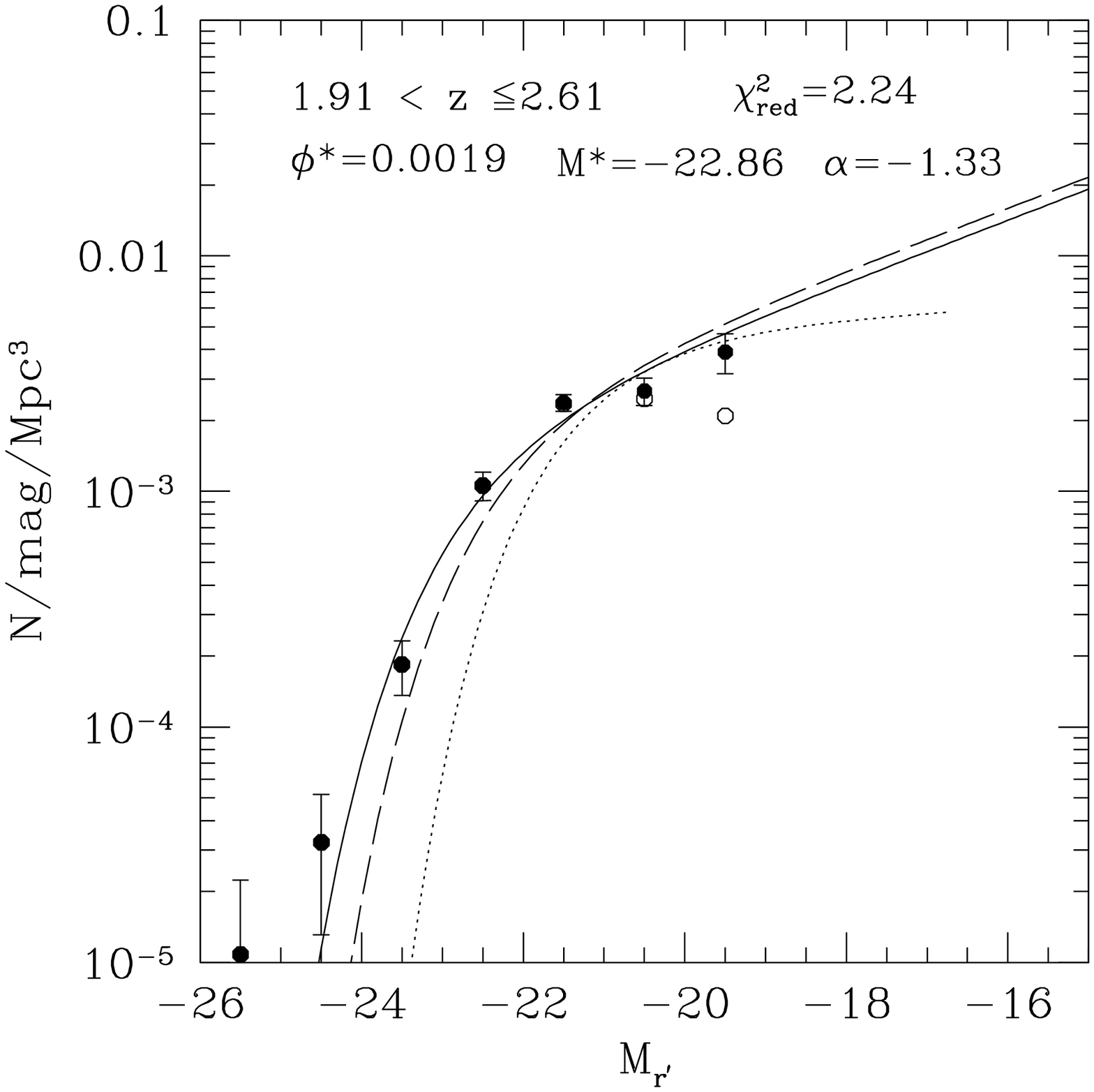}
  \includegraphics[width=0.33\textwidth]{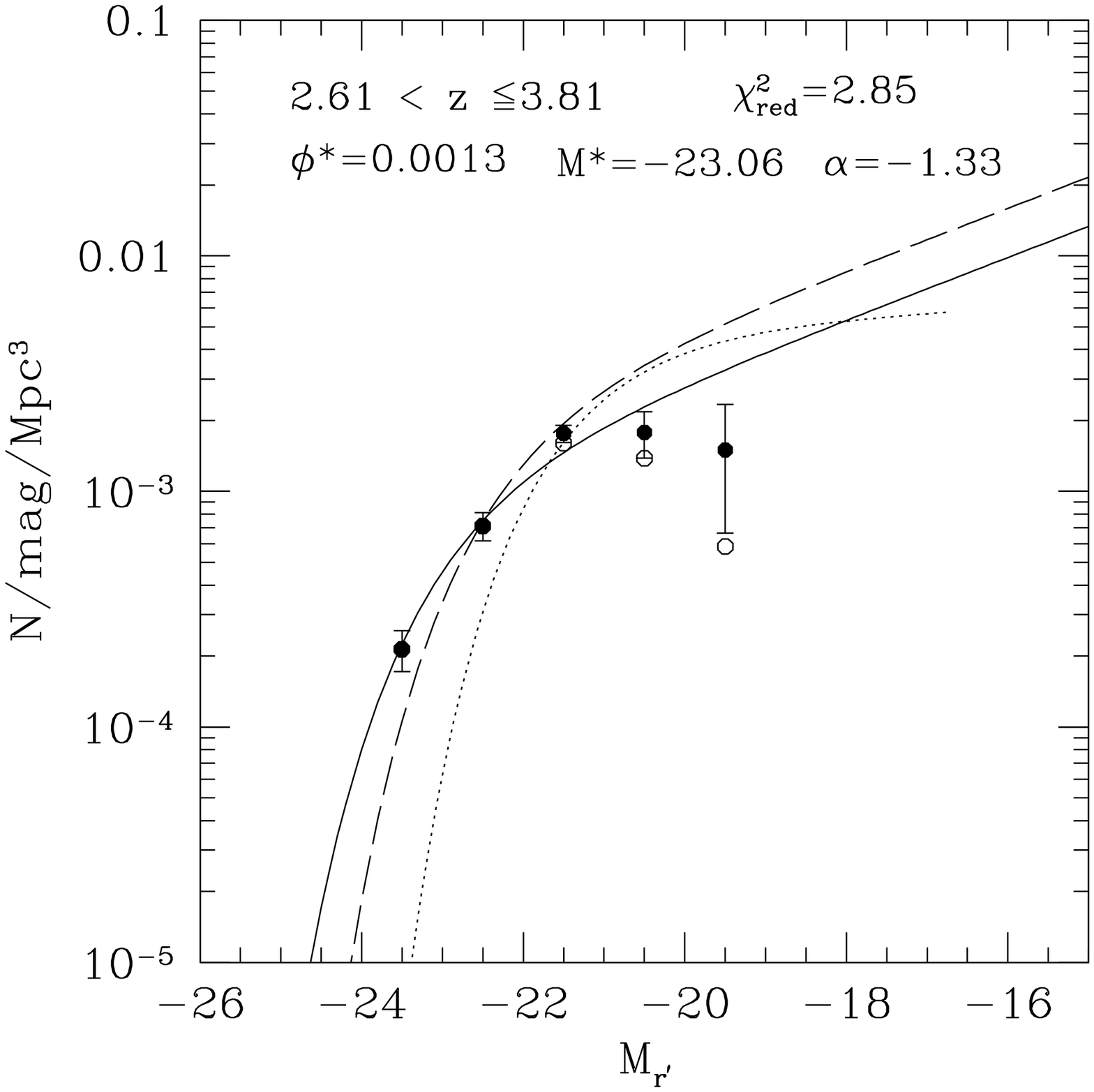}
\caption[LFs in the \textit{r'-band}]
{\label{fig:lfred:lumfkt_fdf_r}
  LFs in the \textit{r'-band} from low redshift
  (\mbox{$\langle z\rangle=0.3$}, upper left panel) to high redshift
  (\mbox{$\langle z\rangle=3.2$}, lower right panel). The filled
  (open) symbols show the LF corrected (uncorrected)
  for $V/V_{max}$. The fitted Schechter functions for a fixed slope
  $\alpha$ are shown as solid lines. Note that we only fit the
  LFs from $\langle z\rangle=0.6$ to $\langle
  z\rangle=3.2$. The parameters of the Schechter functions can be
  found in Table~\ref{tab:lfred:schechter_fit_r}. The dotted line represents
  the local r'-band LF derived from the SDSS
  \citep{blanton:2}. The Schechter fit for redshift $\langle
  z\rangle=0.6$ is indicated as dashed line in all panels.}
\end{figure*}
\begin{table*}[tbh]
\caption[Schechter parameter for the LF in the r'-band]
{\label{tab:lfred:schechter_fit_r}Schechter function fit in the r'-band}
\begin{center}
\begin{tabular}{c|c|c|c}
 redshift interval & M$^\ast$ (mag) & $\phi^\ast$ (Mpc$^{-3}$) & $\alpha$ (fixed)\\
\hline
0.45 -- 0.85 & $-$22.41 +0.23 $-$0.18 & 0.0025 +0.0002 $-$0.0002 &$-$1.33 \\  
0.85 -- 1.31 & $-$22.67 +0.14 $-$0.13 & 0.0019 +0.0001 $-$0.0001 &$-$1.33 \\  
1.31 -- 1.91 & $-$22.38 +0.16 $-$0.16 & 0.0020 +0.0002 $-$0.0002 &$-$1.33 \\  
1.91 -- 2.61 & $-$22.86 +0.13 $-$0.11 & 0.0019 +0.0002 $-$0.0002 &$-$1.33 \\  
2.61 -- 3.81 & $-$23.06 +0.15 $-$0.15 & 0.0013 +0.0002 $-$0.0001 &$-$1.33 \\  
\end{tabular}
\end{center}
\end{table*}

\begin{figure*}[tbp]
\includegraphics[width=0.33\textwidth]{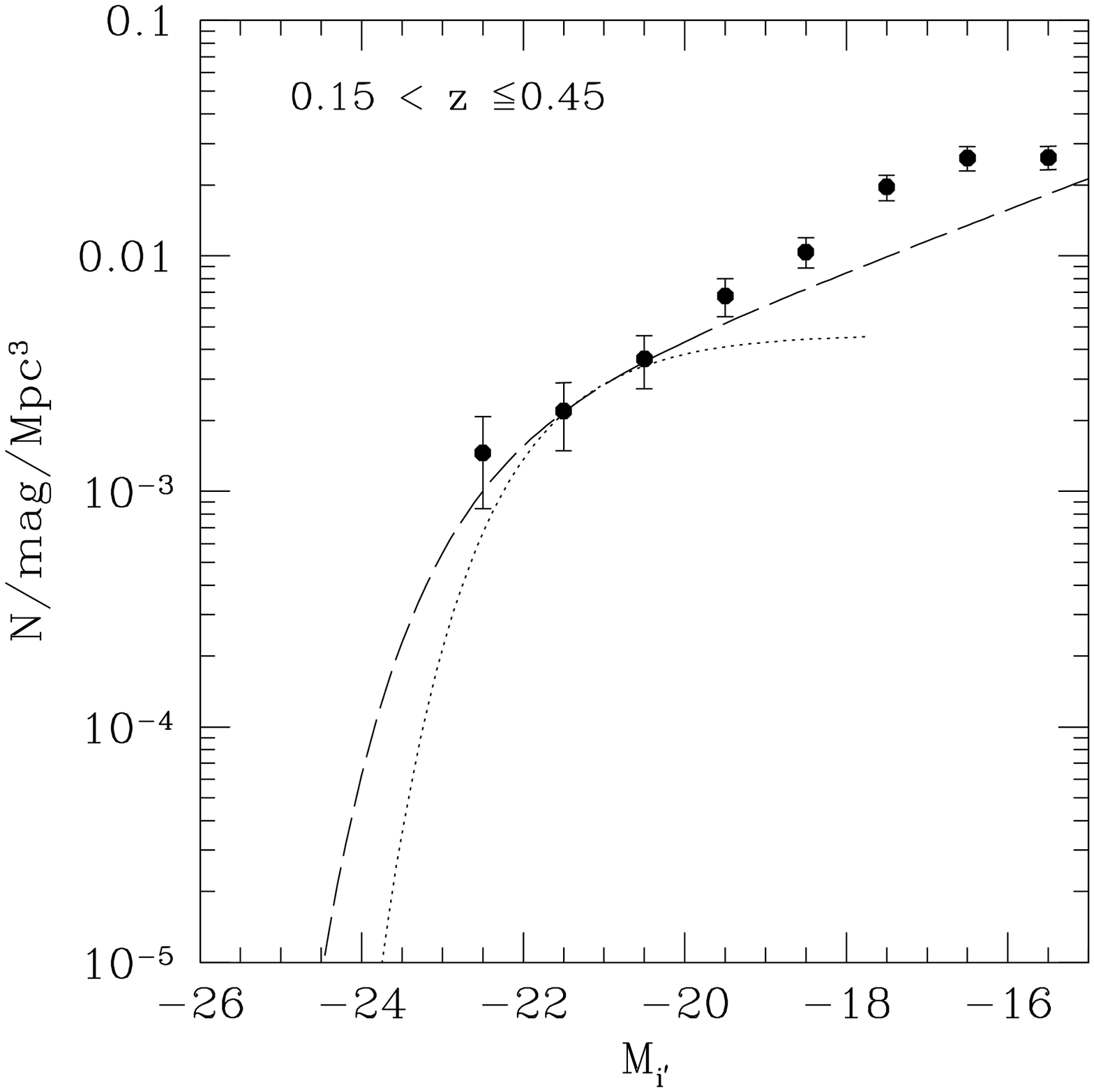}
\includegraphics[width=0.33\textwidth]{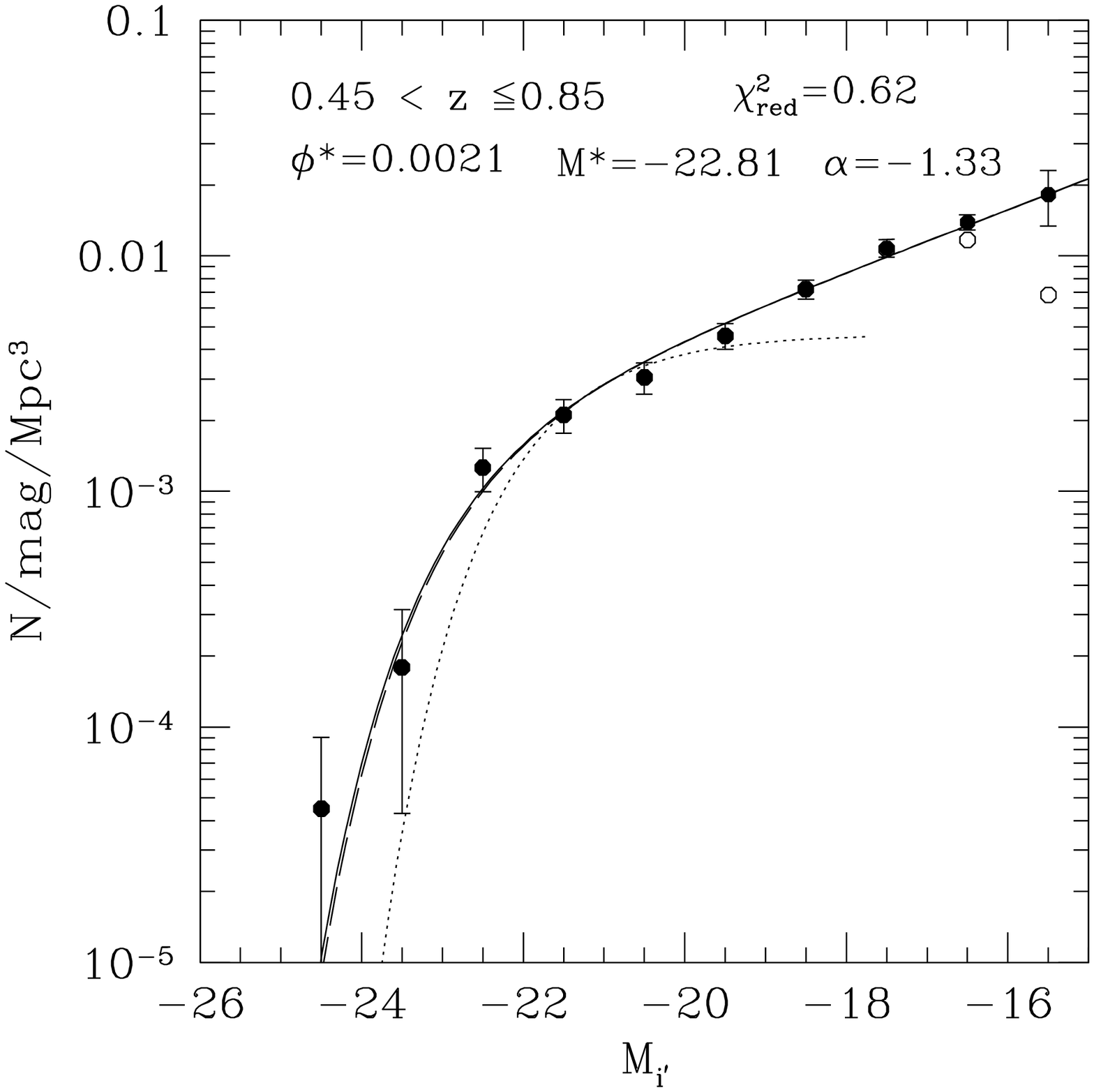}
\includegraphics[width=0.33\textwidth]{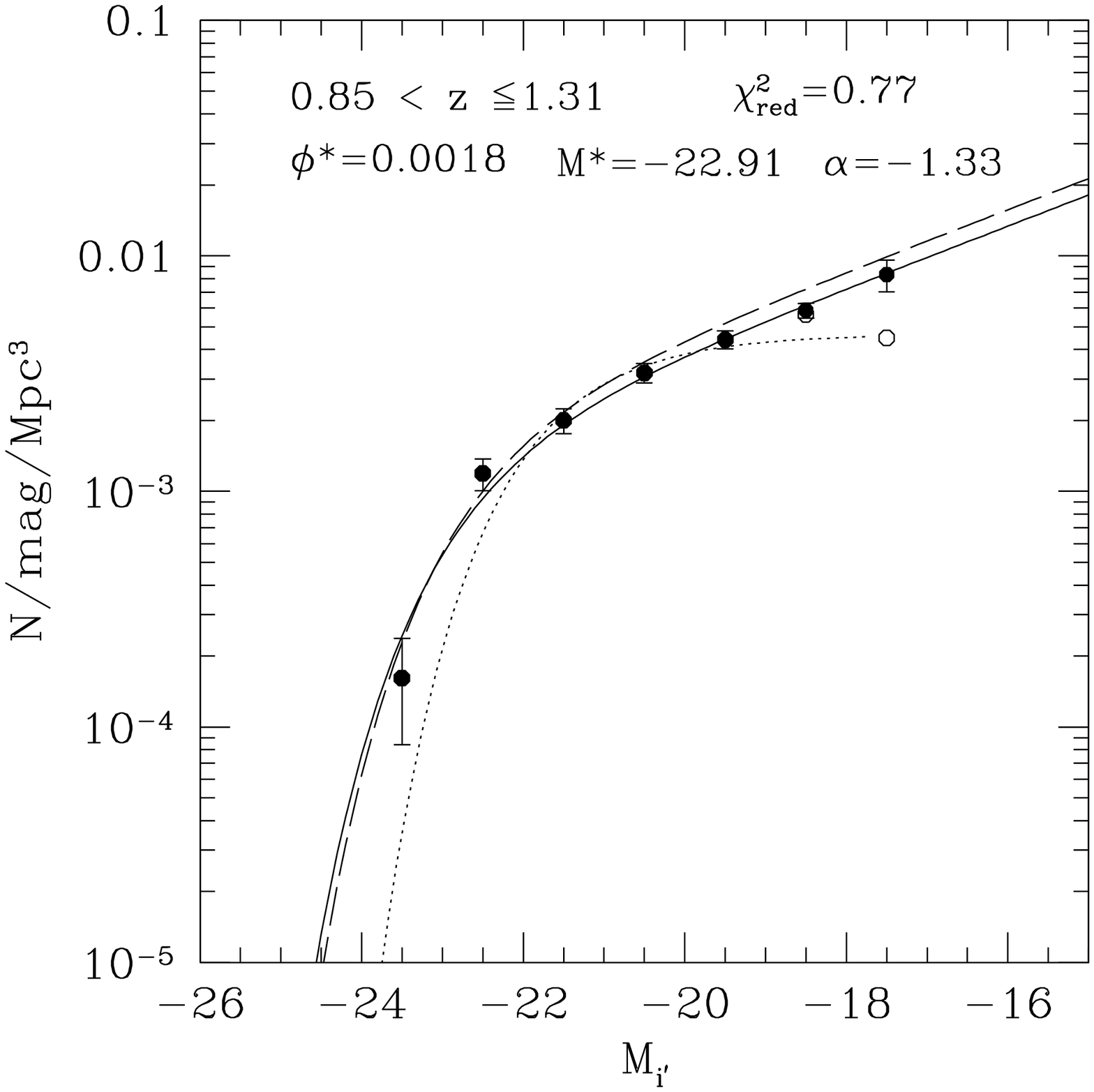}
\includegraphics[width=0.33\textwidth]{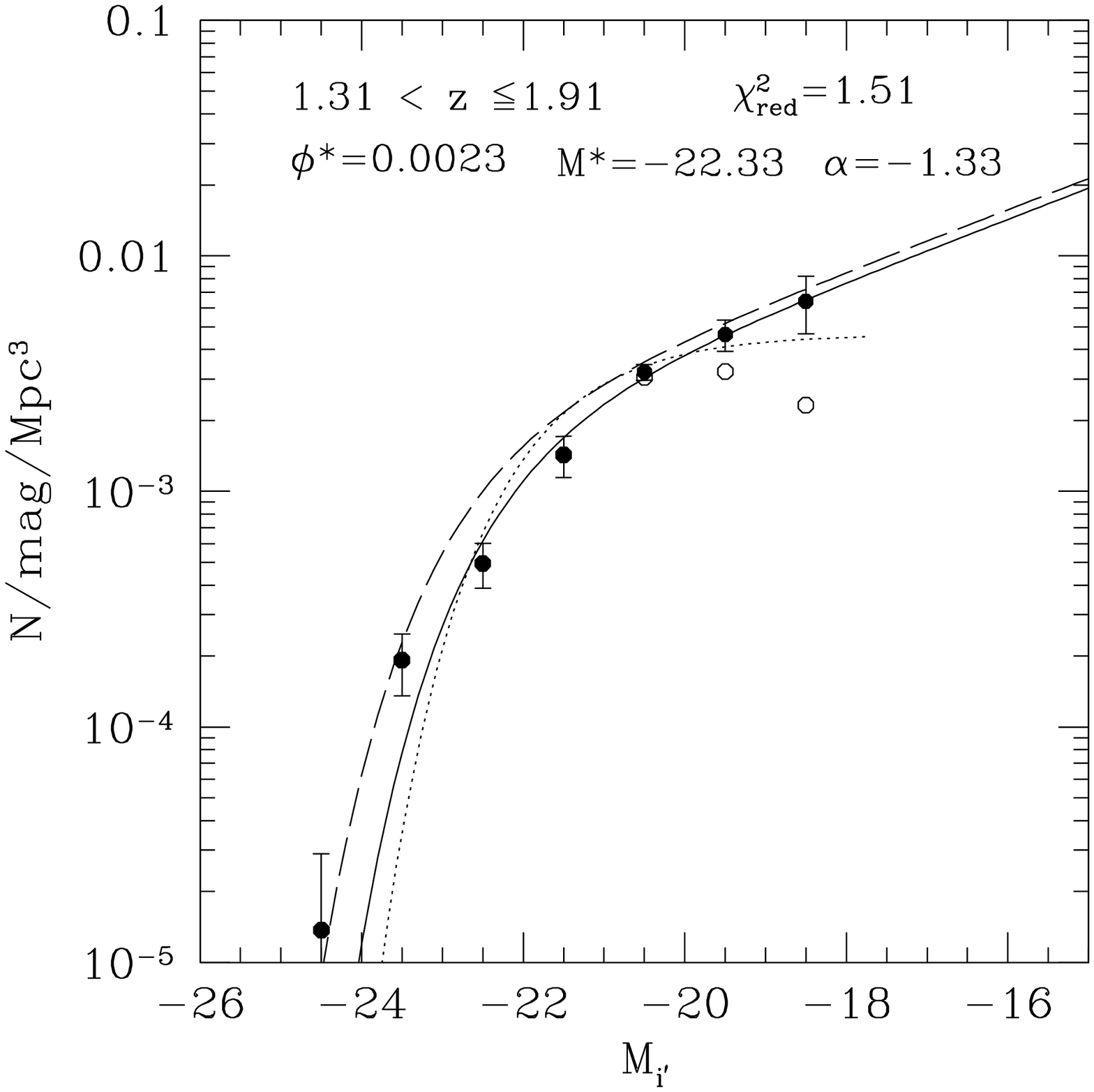}
\includegraphics[width=0.33\textwidth]{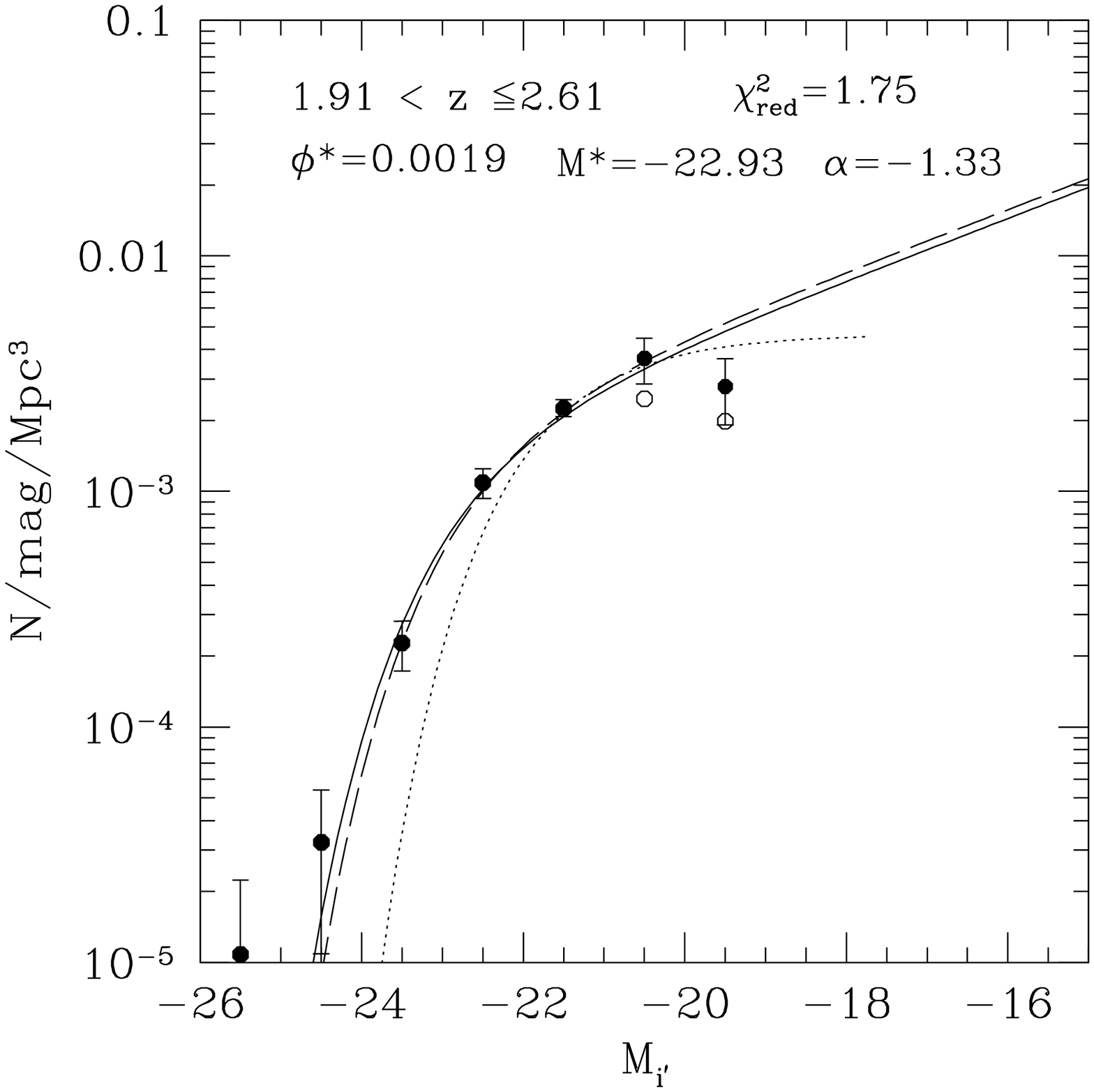}
\includegraphics[width=0.33\textwidth]{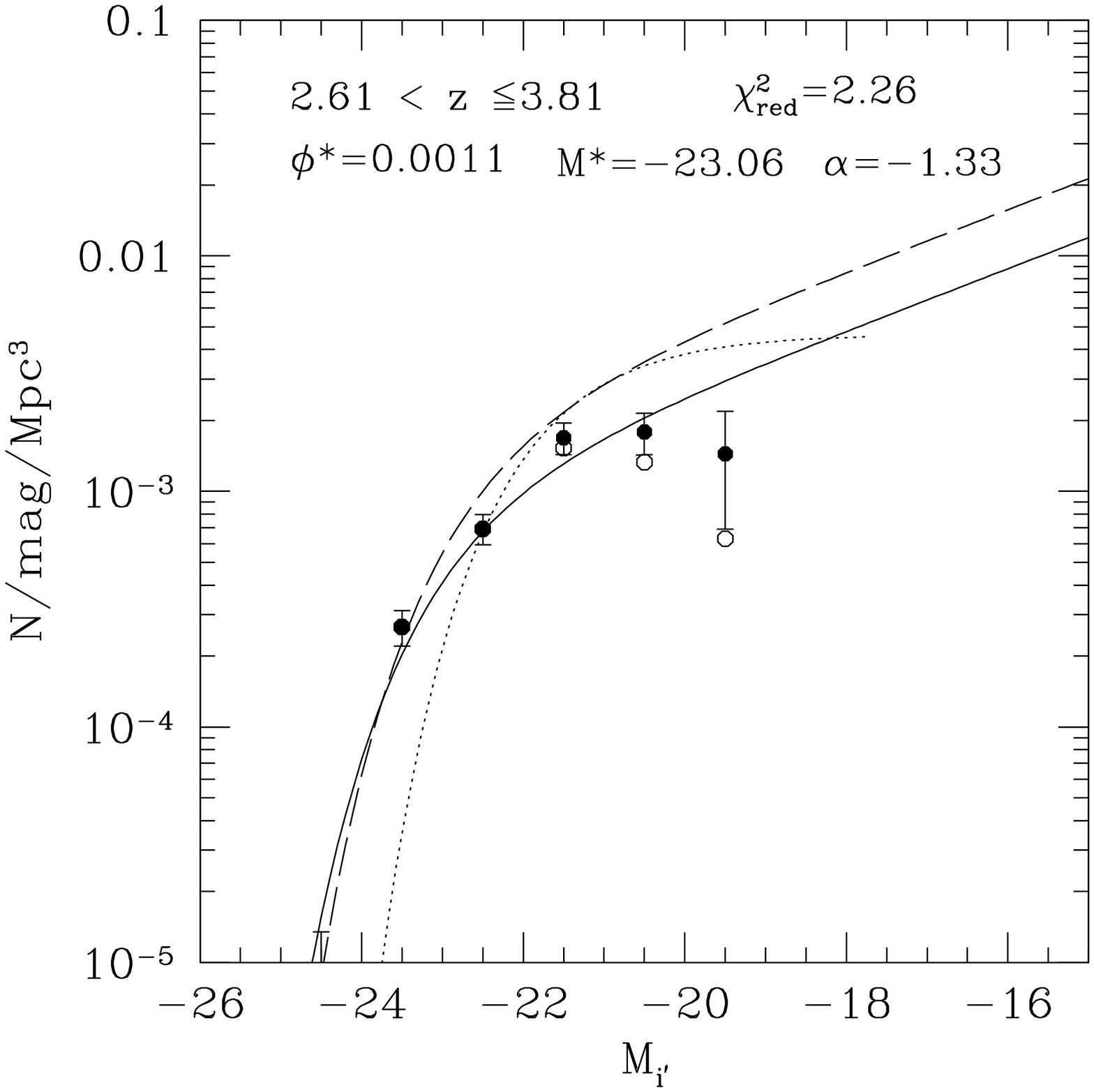}
\caption[LFs in the \textit{i'-band}]
{\label{fig:lfred:lumfkt_fdf_i}
  LFs in the \textit{i'-band} from low redshift
  (\mbox{$\langle z\rangle=0.3$}, upper left panel) to high redshift
  (\mbox{$\langle z\rangle=3.2$}, lower right panel). The filled
  (open) symbols show the LF corrected (uncorrected)
  for $V/V_{max}$. The fitted Schechter functions for a fixed slope
  $\alpha$ are shown as solid lines. Note that we only fit the
  LFs from $\langle z\rangle=0.6$ to $\langle
  z\rangle=3.2$. The parameters of the Schechter functions can be
  found in Table~\ref{tab:lfred:schechter_fit_i}. The dotted line represents
  the local i'-band LF derived from the SDSS
  \citep{blanton:2}. The Schechter fit for redshift $\langle
  z\rangle=0.6$ is indicated as dashed line in all panels.}
\end{figure*}

\begin{table*}[tbh]
\caption[Schechter parameter for the LF in the i'-band]
{\label{tab:lfred:schechter_fit_i}Schechter function fit in the i'-band}
\begin{center}
\begin{tabular}{c|c|c|c}
 redshift interval & M$^\ast$ (mag) & $\phi^\ast$ (Mpc$^{-3}$) & $\alpha$ (fixed)\\
\hline
0.45 -- 0.85 & $-$22.81 +0.23 $-$0.24 & 0.0021 +0.0002 $-$0.0002 &$-$1.33 \\  
0.85 -- 1.31 & $-$22.91 +0.16 $-$0.15 & 0.0018 +0.0001 $-$0.0001 &$-$1.33 \\  
1.31 -- 1.91 & $-$22.33 +0.21 $-$0.18 & 0.0023 +0.0003 $-$0.0003 &$-$1.33 \\  
1.91 -- 2.61 & $-$22.93 +0.14 $-$0.13 & 0.0019 +0.0002 $-$0.0002 &$-$1.33 \\  
2.61 -- 3.81 & $-$23.06 +0.10 $-$0.09 & 0.0011 +0.0001 $-$0.0001 &$-$1.33 \\  
\end{tabular}
\end{center}
\end{table*}

\begin{figure*}[tbp]
\includegraphics[width=0.33\textwidth]{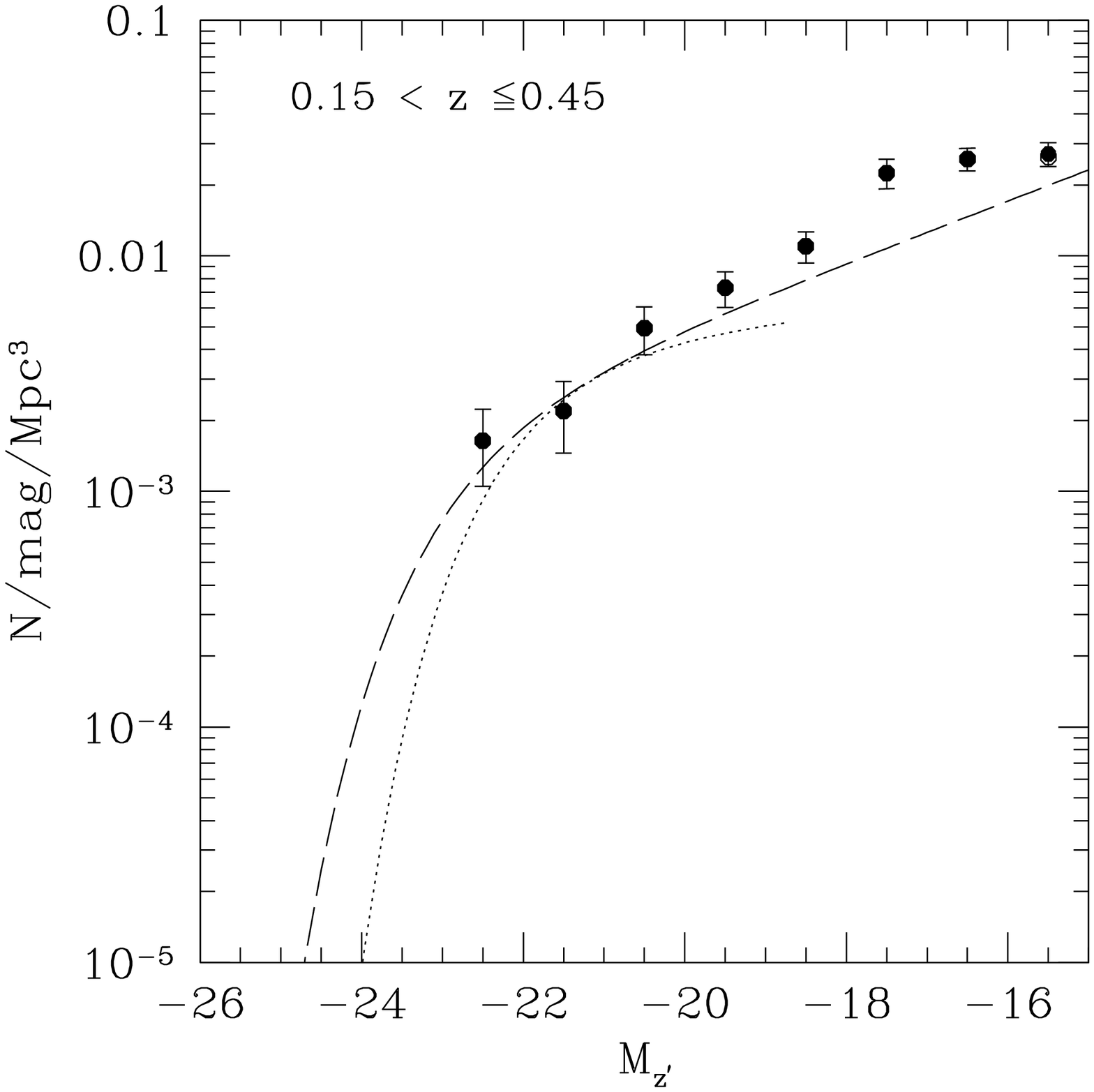}
\includegraphics[width=0.33\textwidth]{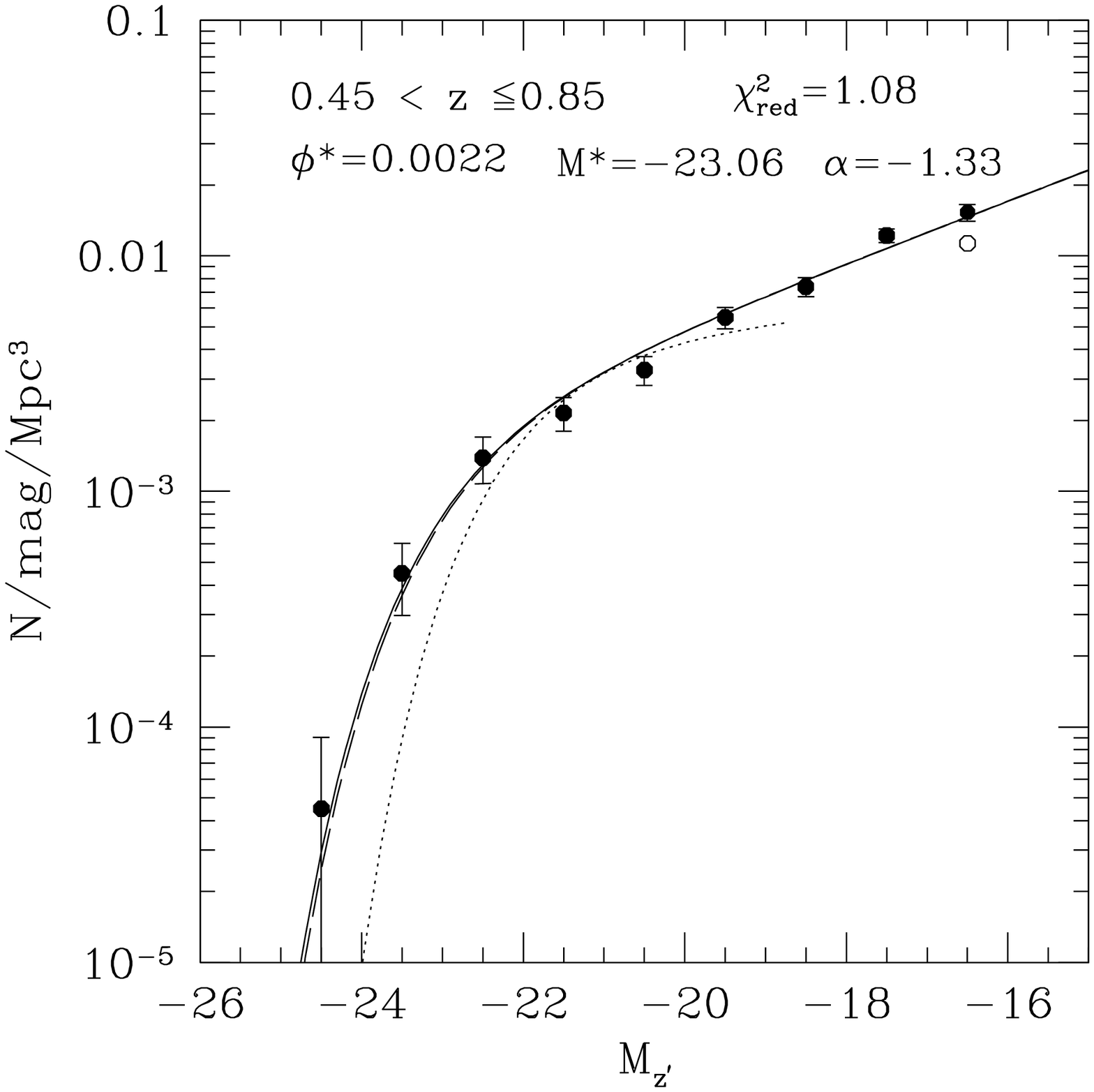}
\includegraphics[width=0.33\textwidth]{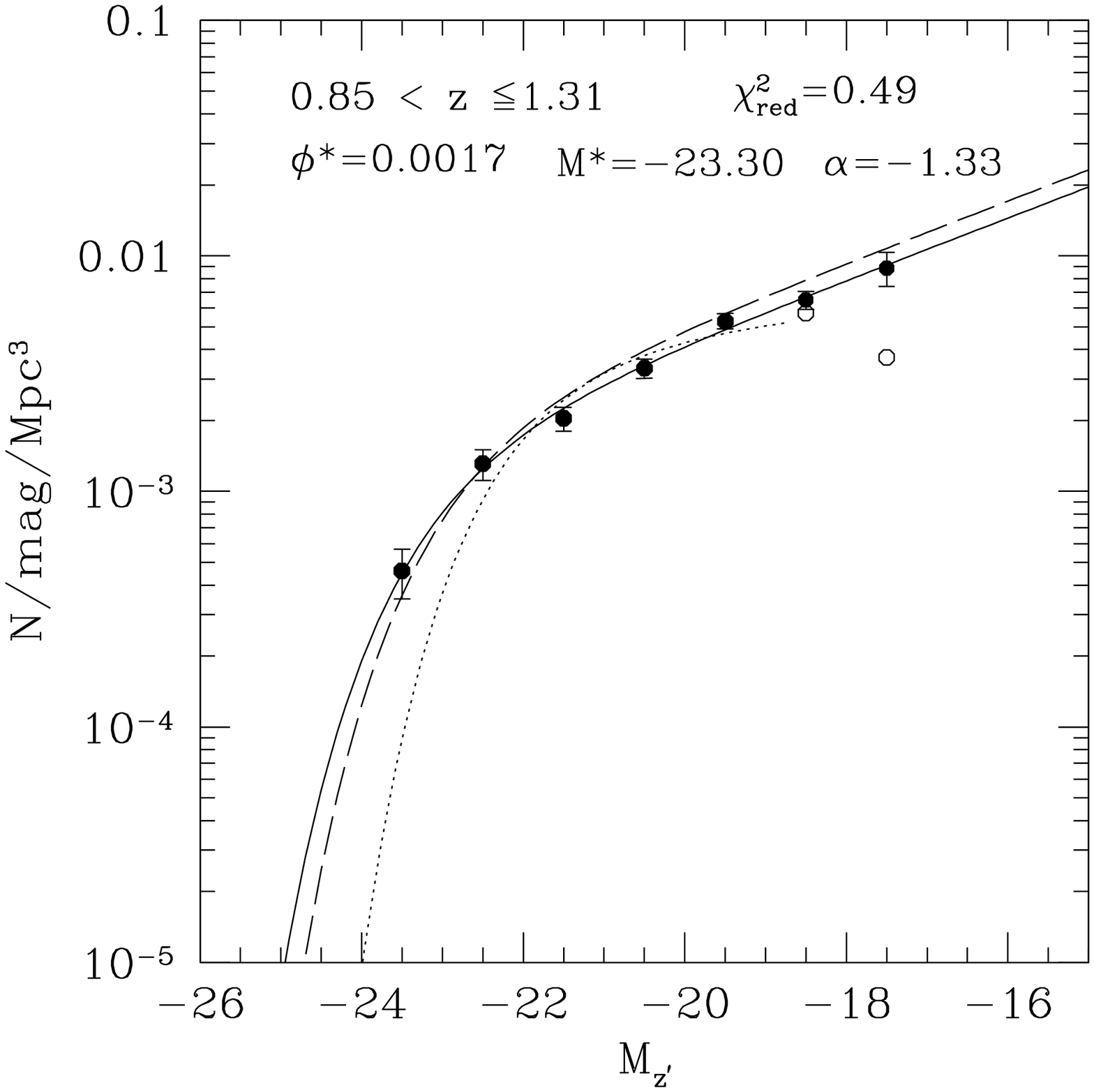}
\includegraphics[width=0.33\textwidth]{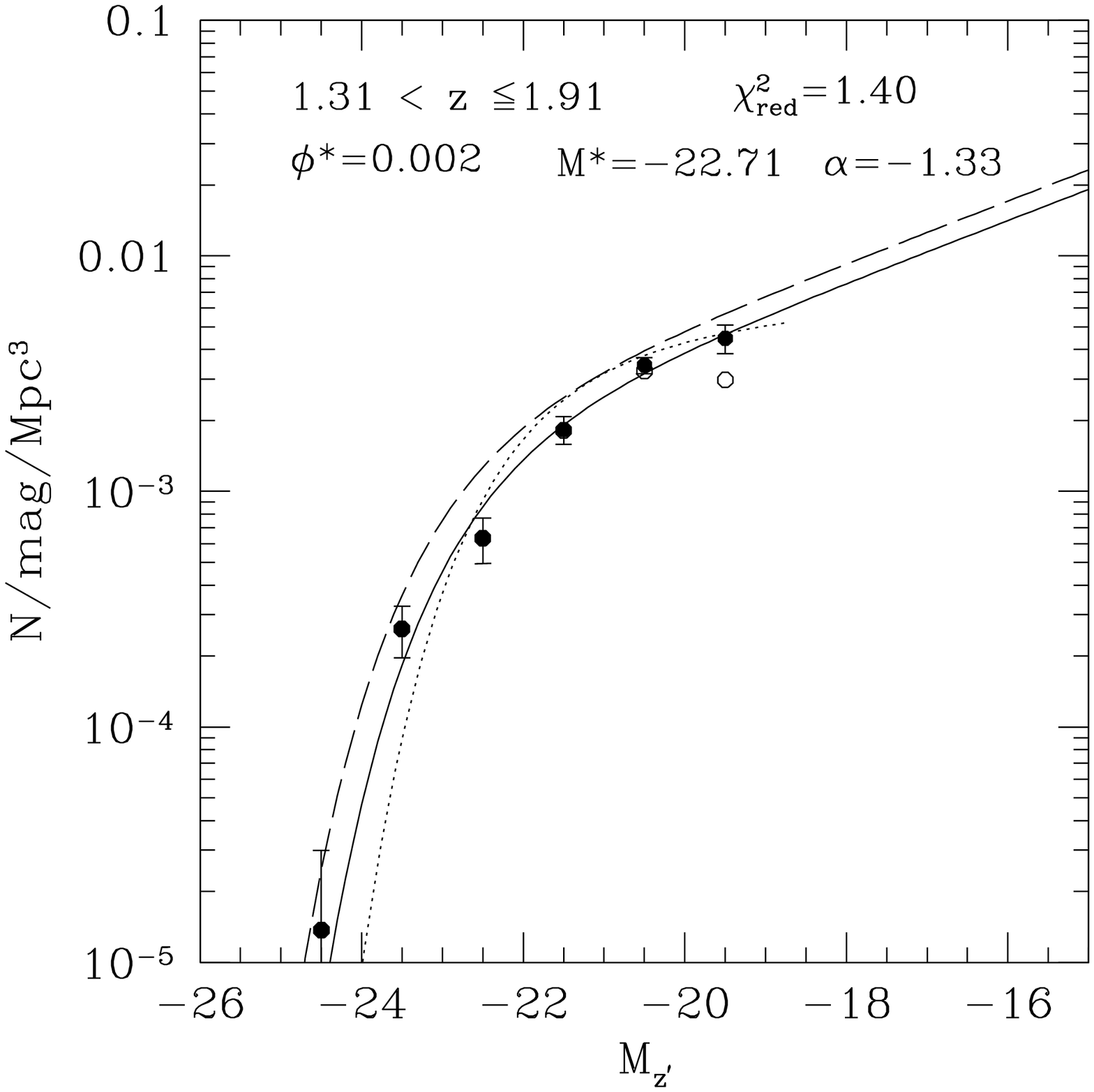}
\includegraphics[width=0.33\textwidth]{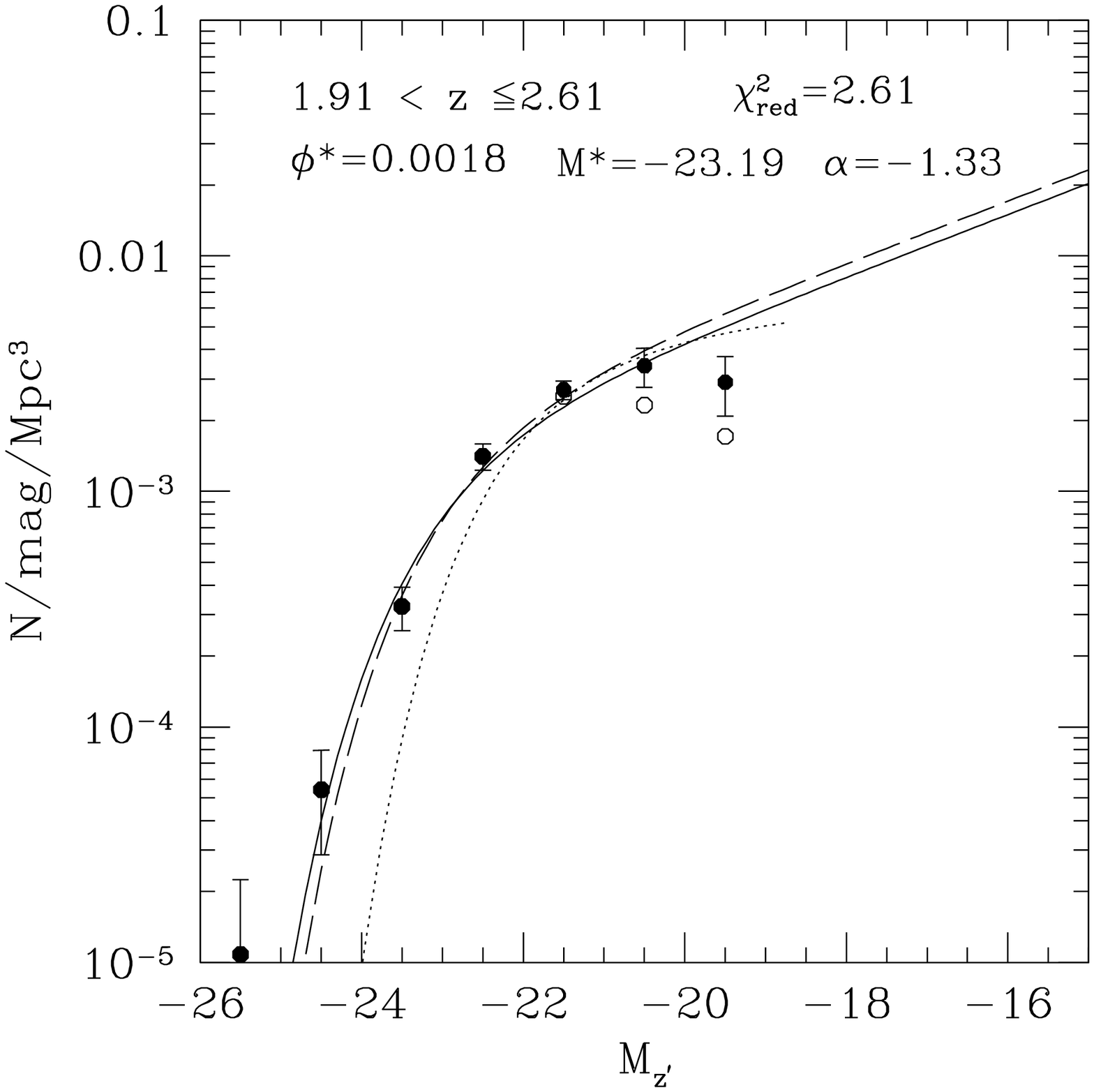}
\includegraphics[width=0.33\textwidth]{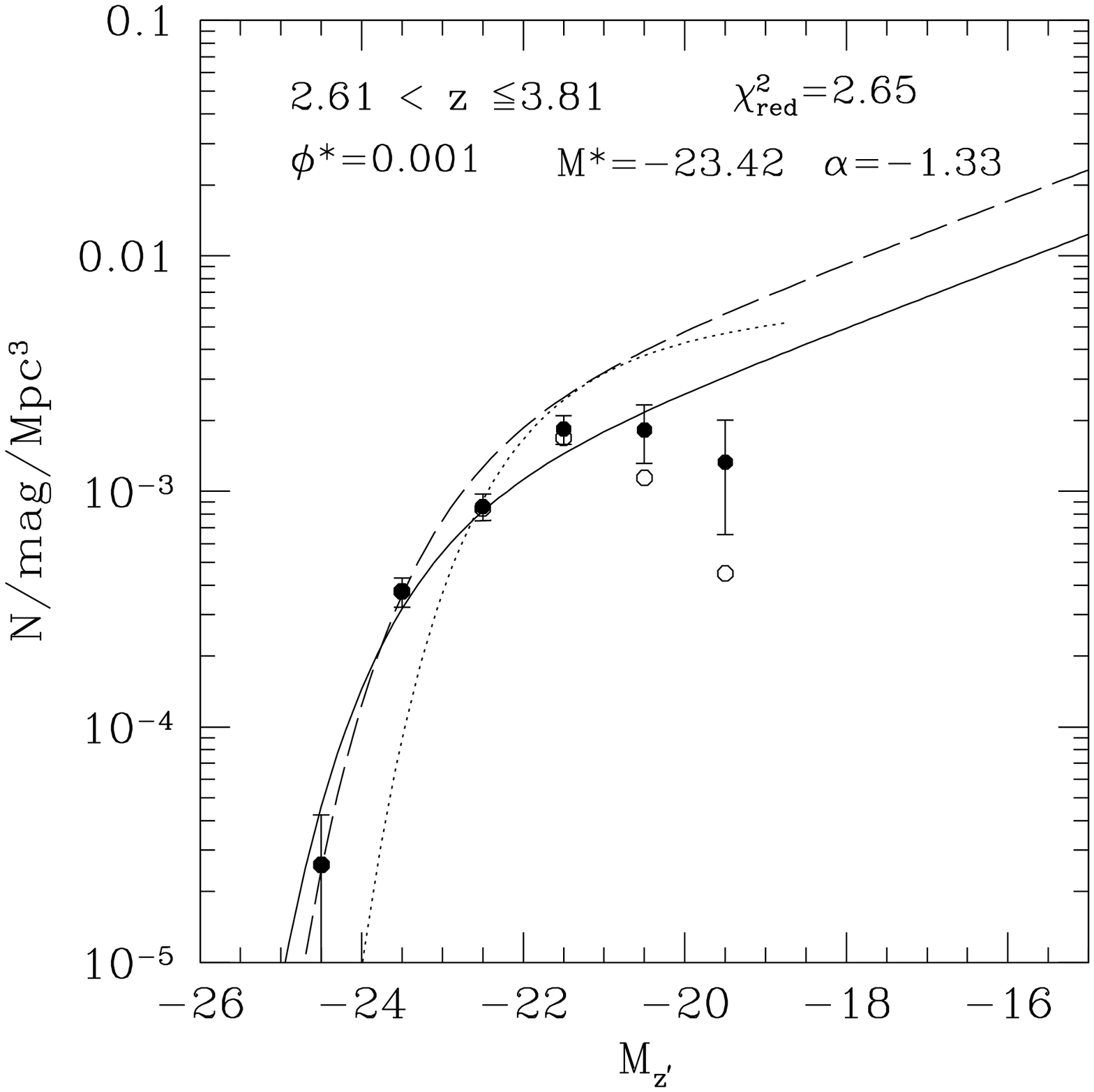}
\caption[LFs in the \textit{z'-band}]
{\label{fig:lfred:lumfkt_fdf_z}
  LFs in the \textit{z'-band} from low redshift
  (\mbox{$\langle z\rangle=0.3$}, upper left panel) to high redshift
  (\mbox{$\langle z\rangle=3.2$}, lower right panel). The filled
  (open) symbols show the LF corrected (uncorrected)
  for $V/V_{max}$. The fitted Schechter functions for a fixed slope
  $\alpha$ are shown as solid lines. Note that we only fit the
  LFs from $\langle z\rangle=0.6$ to $\langle
  z\rangle=3.2$. The parameters of the Schechter functions can be
  found in Table~\ref{tab:lfred:schechter_fit_z}. The dotted line represents
  the local z'-band LF derived from the SDSS
  \citep{blanton:2}. The Schechter fit for redshift $\langle
  z\rangle=0.6$ is indicated as dashed line in all panels.}
\end{figure*}

\begin{table*}[tbh]
\caption[Schechter parameter for the LF in the z'-band]
{\label{tab:lfred:schechter_fit_z}Schechter function fit in the z'-band}
\begin{center}
\begin{tabular}{c|c|c|c}
 redshift interval & M$^\ast$ (mag) & $\phi^\ast$ (Mpc$^{-3}$) & $\alpha$ (fixed)\\
\hline
0.45 -- 0.85 & $-$23.06 +0.25 $-$0.21 & 0.0022 +0.0002 $-$0.0002 &$-$1.33 \\  
0.85 -- 1.31 & $-$23.30 +0.20 $-$0.21 & 0.0017 +0.0002 $-$0.0001 &$-$1.33 \\  
1.31 -- 1.91 & $-$22.71 +0.18 $-$0.17 & 0.0020 +0.0003 $-$0.0002 &$-$1.33 \\  
1.91 -- 2.61 & $-$23.19 +0.13 $-$0.13 & 0.0018 +0.0002 $-$0.0002 &$-$1.33 \\  
2.61 -- 3.81 & $-$23.42 +0.10 $-$0.13 & 0.0010 +0.0001 $-$0.0001 &$-$1.33 \\  
\end{tabular}
\end{center}
\end{table*}

In this section we analyze the LF by means of a Schechter function fit
with a fixed slope of $\alpha =-1.33$.
In Fig.~\ref{fig:lfred:lumfkt_fdf_r} and
Fig.~\ref{fig:lfred:lumfkt_fdf_i} we present the LFs in the r'-band
and in the i'-band, while the results for the z'-band can be found in
Fig.~\ref{fig:lfred:lumfkt_fdf_z}. The filled (open) symbols denote
the LF with (without) completeness correction.  The solid lines show
the Schechter function fitted to the luminosity function.  The best
fitting Schechter parameter, the redshift binning as well as the
reduced $\chi^2$ are also listed in each figure.  The values of the
reduced $\chi^2$ are very good for all redshift bins below $z \sim 2$.
We do not fit our lowest redshift bin data (\mbox{$\langle z
  \rangle\sim 0.3$}) with a Schechter function, because the volume is
too small.  
 For comparison we also show the local LF derived by
  \citet{blanton:2} in the SDSS (see also
  Fig.~\ref{fig:lfred:lumfkt_fdf_sdss}).
The best fitting Schechter
parameters and corresponding $1\sigma$ errors are summarized in
Table~\ref{tab:lfred:schechter_fit_r},
Table~\ref{tab:lfred:schechter_fit_i}, and
Table~\ref{tab:lfred:schechter_fit_z} for the r', i', and z' bands,
respectively.  Even without fitting Schechter functions to the data,
it is obvious that the evolution in characteristic luminosity and
number density between redshifts $\langle z\rangle=0.6$ and $\langle
z\rangle=3.2$ is very moderate if compared to the evolution in the
blue bands.

\section{Parameterizing the evolution of the LFs}
\label{sec:lfred:evol_parameter}

\begin{figure*}[tbp]
\includegraphics[width=0.33\textwidth]{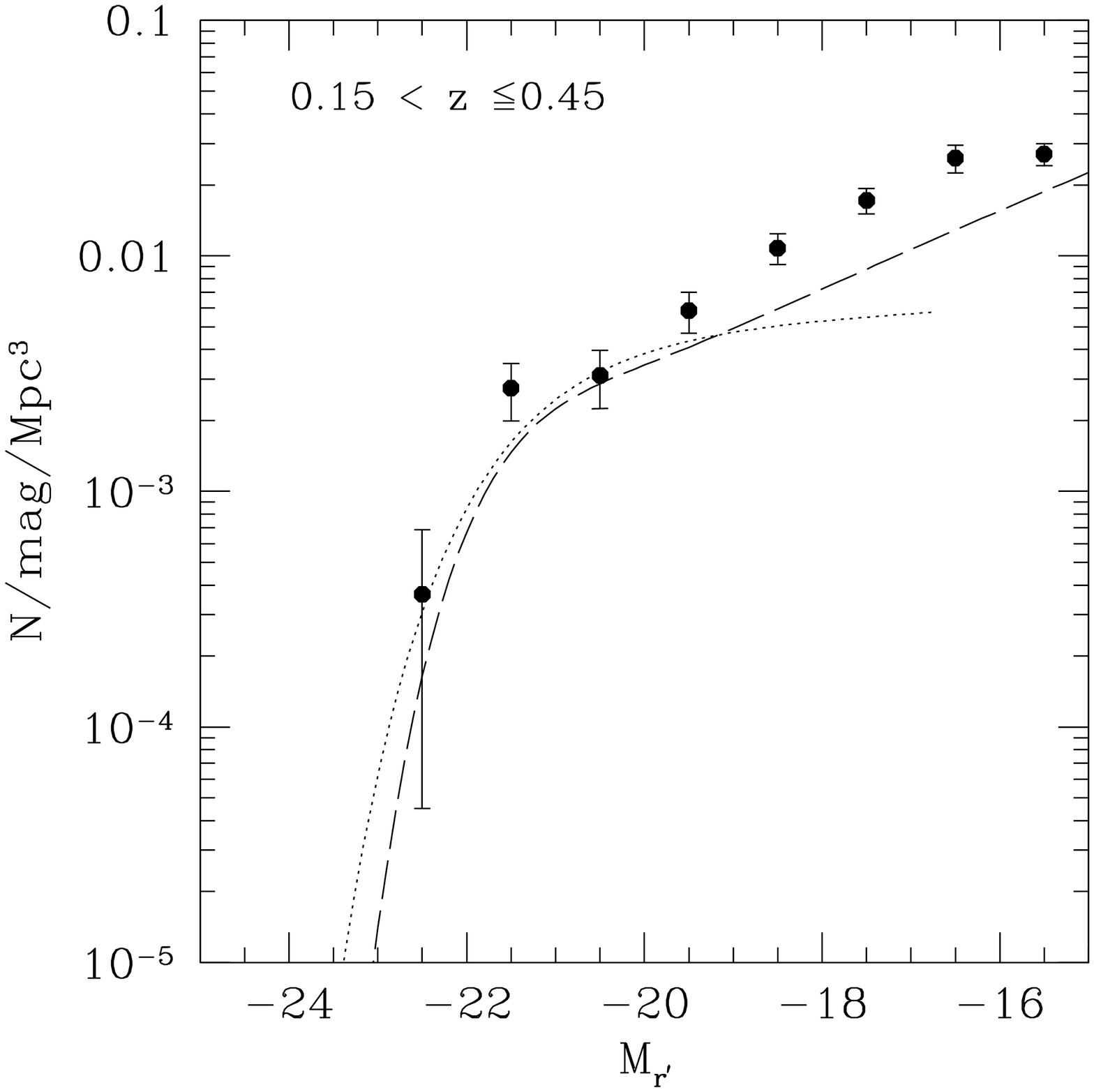}
\includegraphics[width=0.33\textwidth]{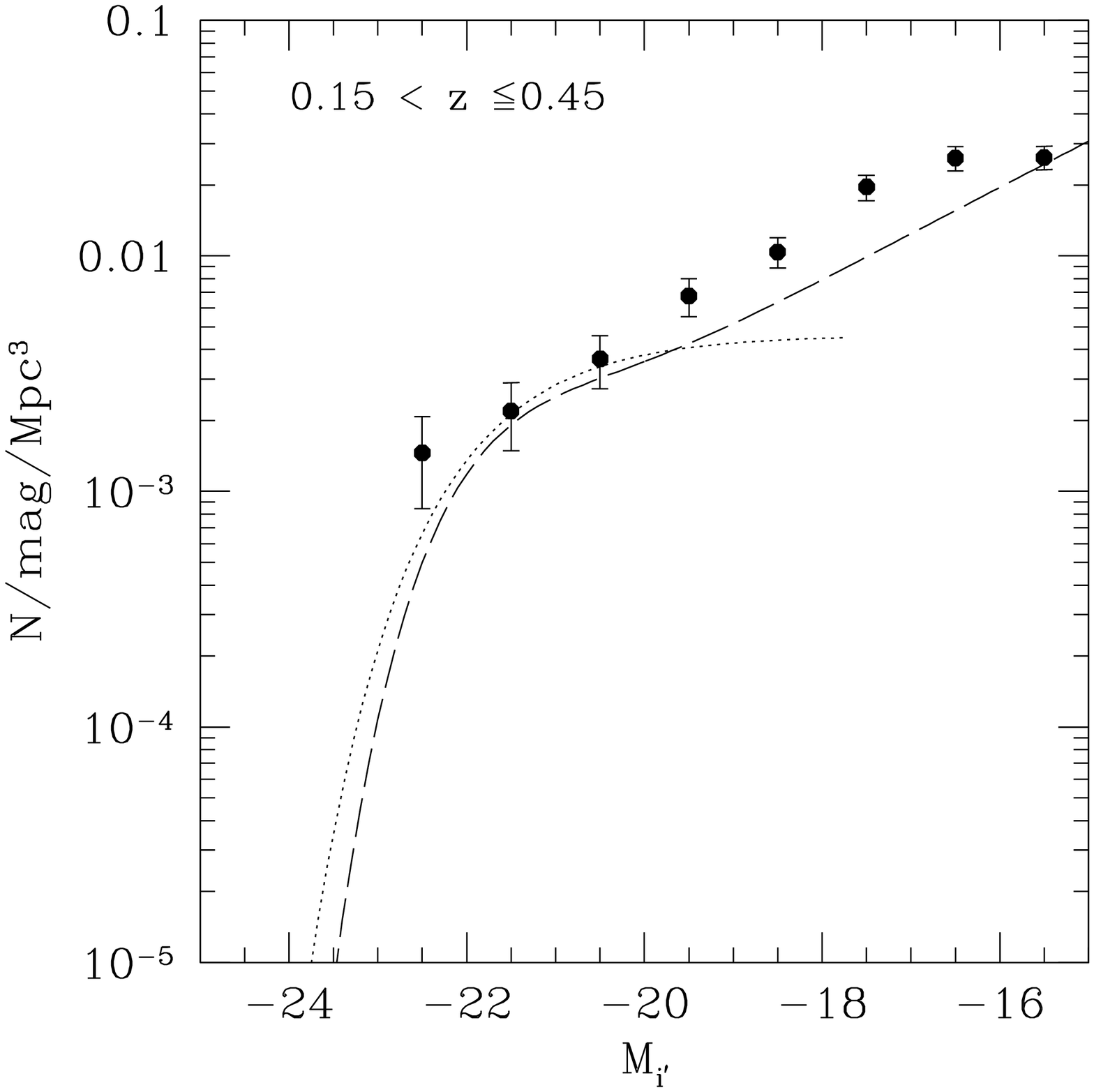}
\includegraphics[width=0.33\textwidth]{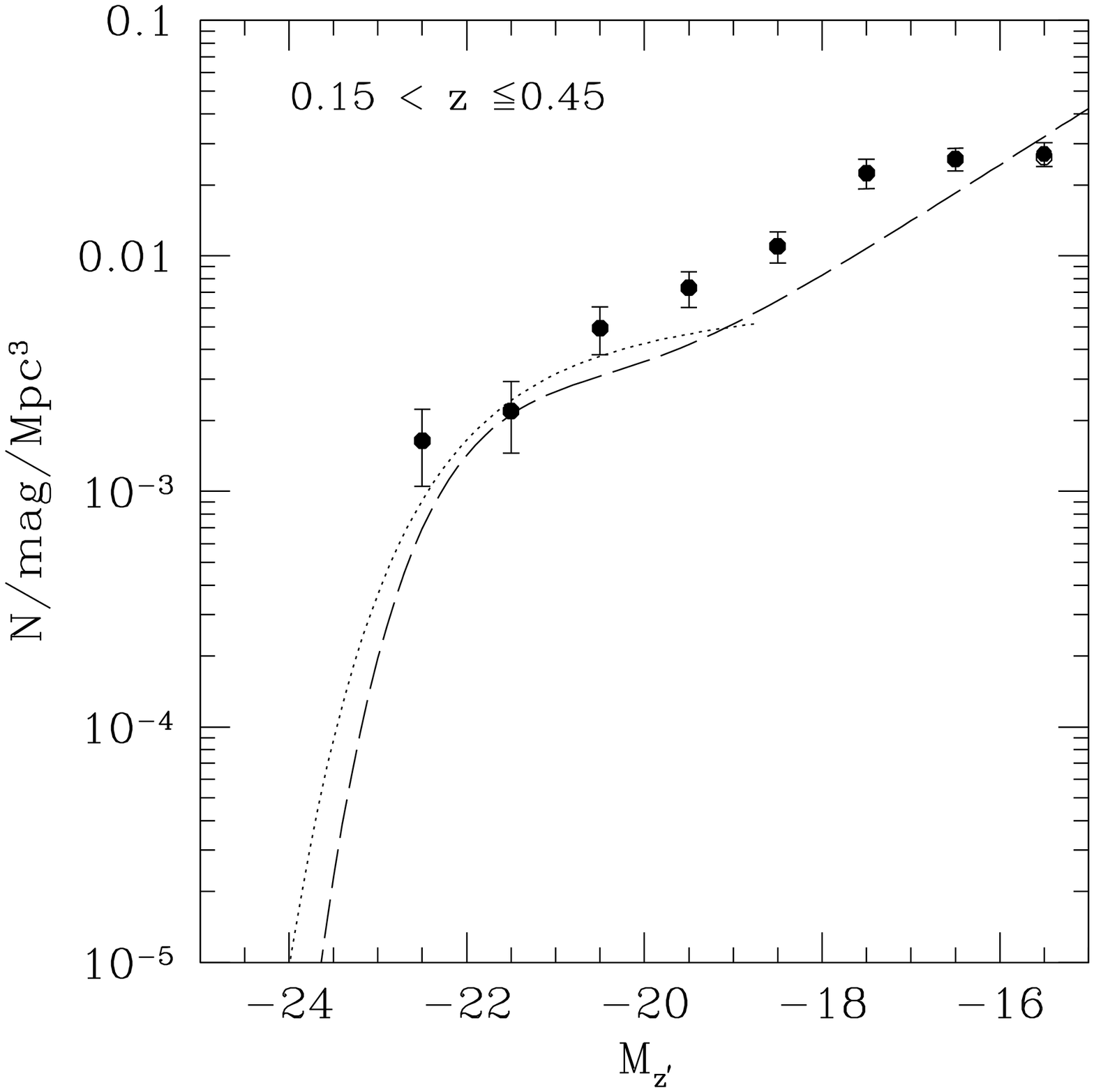}
\includegraphics[width=0.33\textwidth]{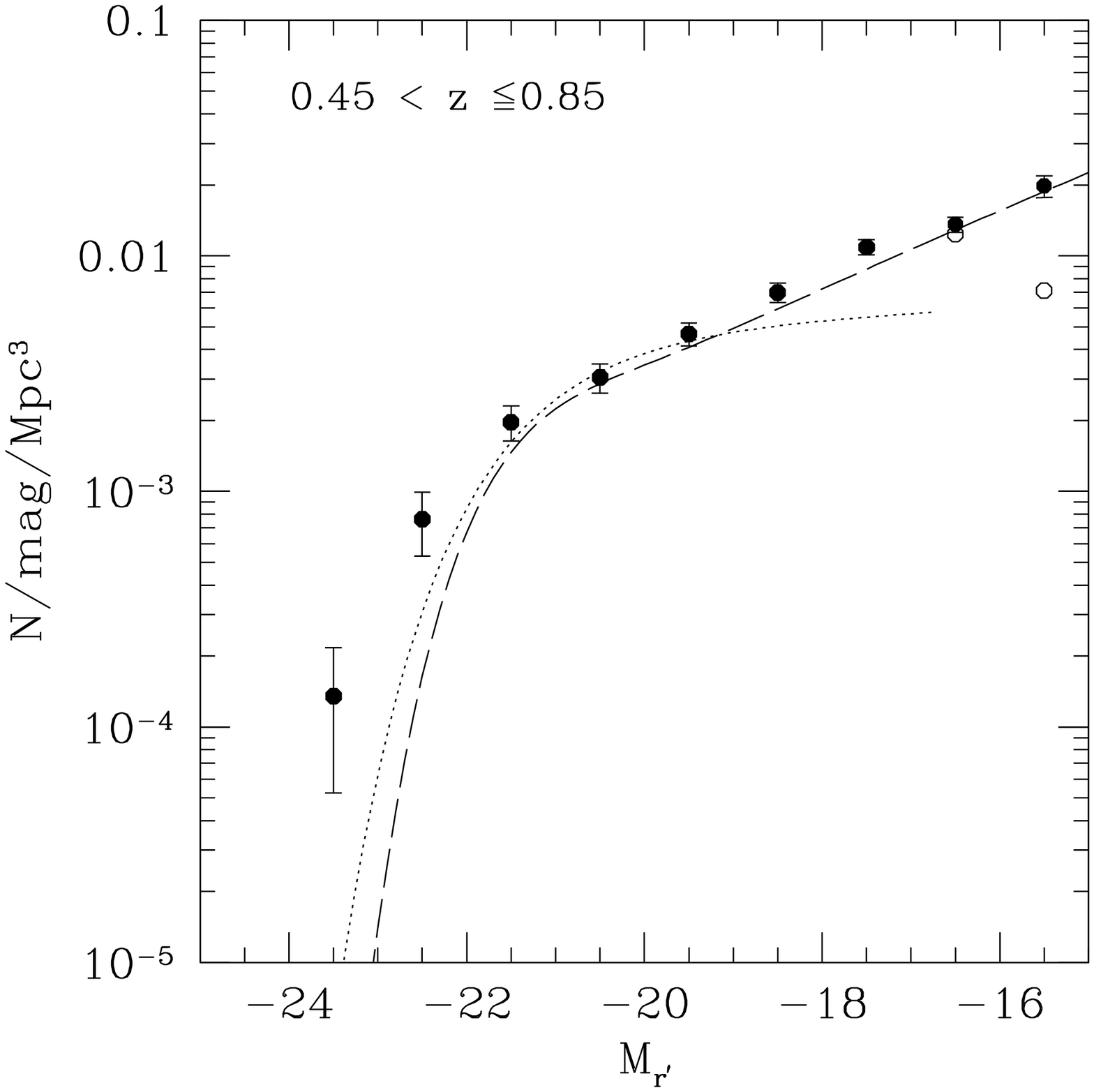}
\includegraphics[width=0.33\textwidth]{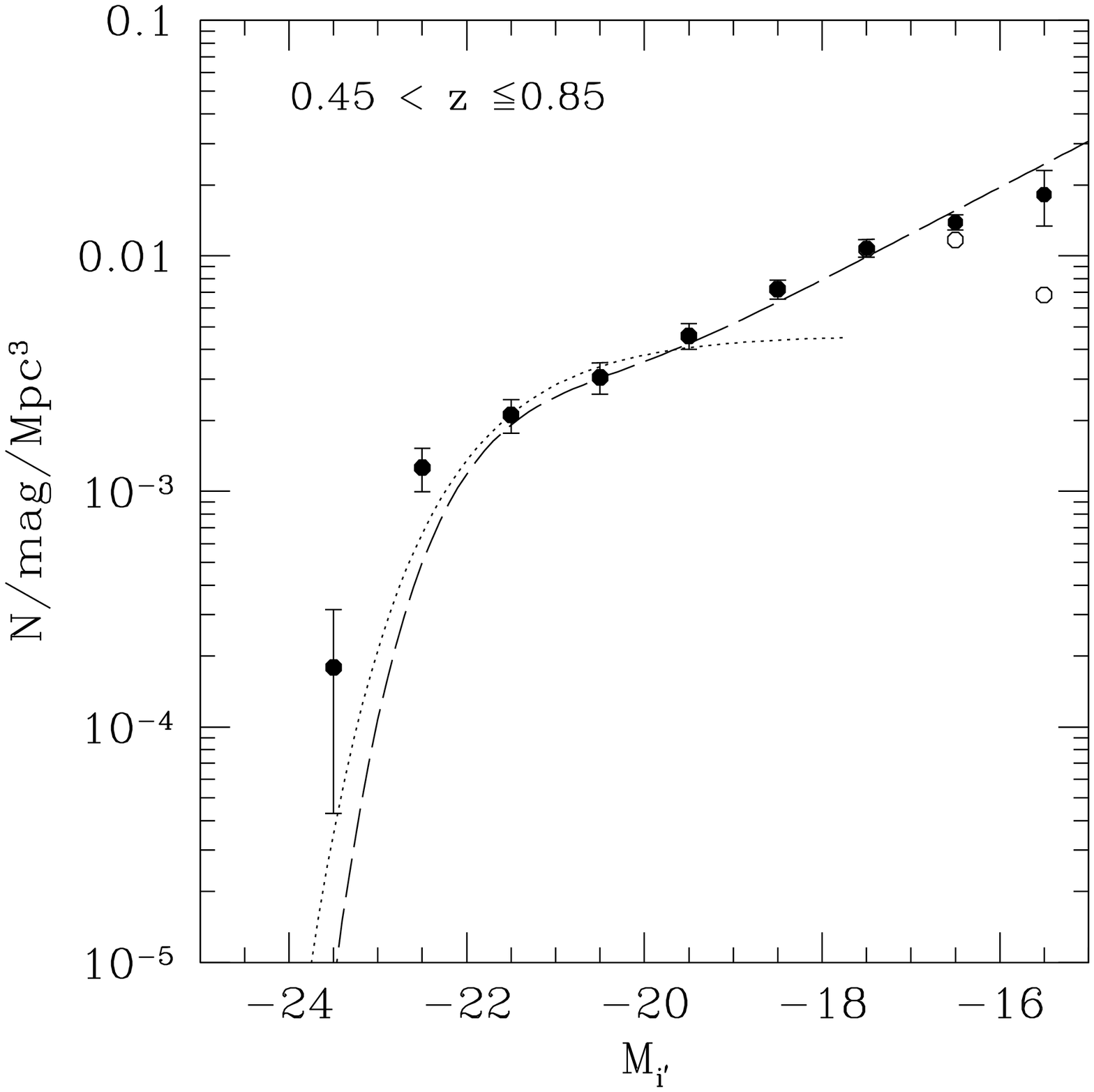}
\includegraphics[width=0.33\textwidth]{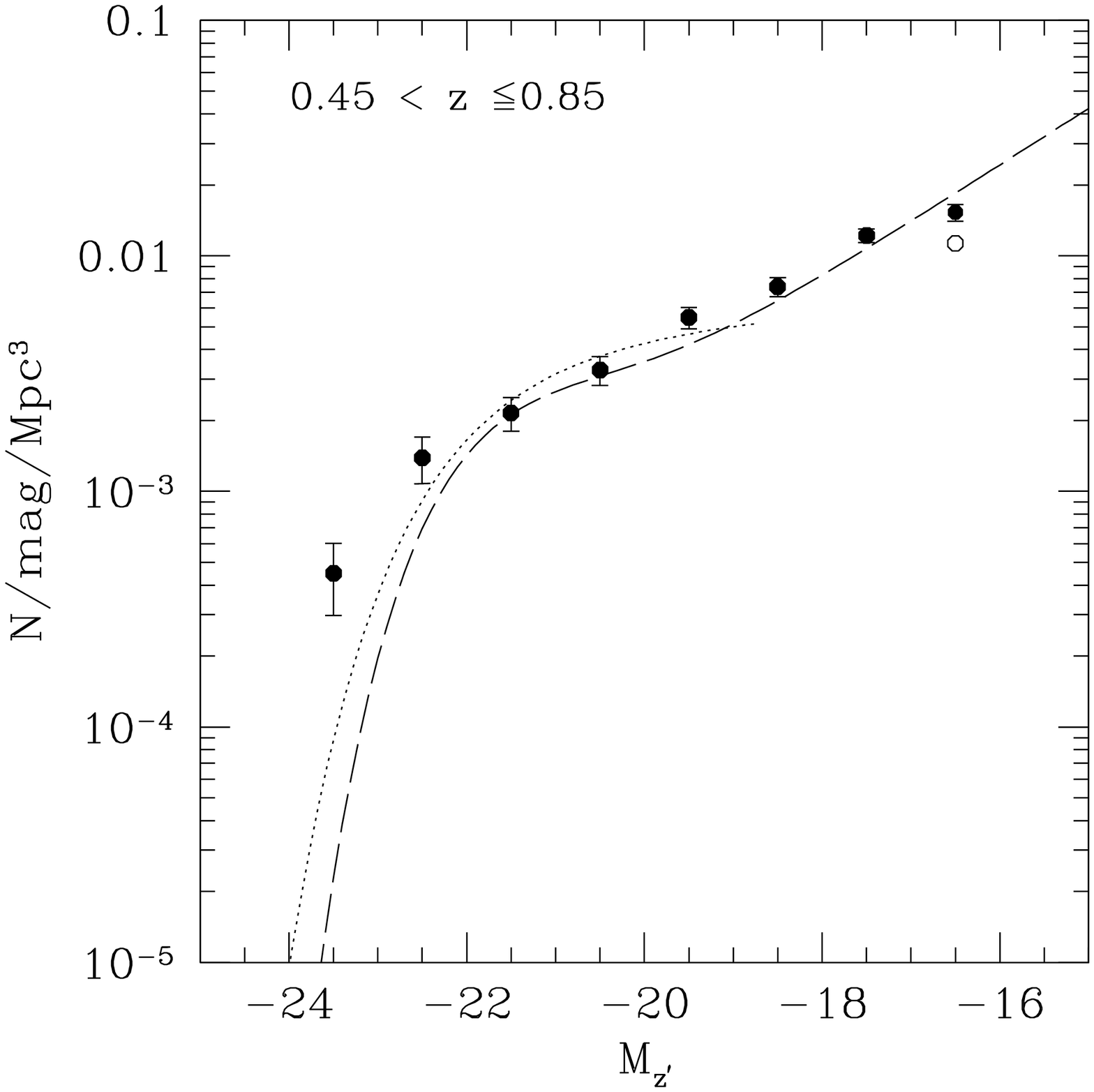}
\caption{\label{fig:lfred:lumfkt_fdf_sdss}
  Comparison of the red FDF LFs in the redshift range \mbox{$\langle
    z\rangle=0.3$} (upper panels) and \mbox{$\langle z\rangle=0.65$}
  (lower panels) with the local Schechter functions as derived in the
  SDSS by
\citet[ dotted
  line]{blanton:2}, and \citet[][ dashed line]{blanton:3}. The filled
  (open) symbols show the FDF LF corrected (uncorrected) for $V/V_{max}$.}
\end{figure*}

To better quantify the redshift evolution of the LFs, we use the
method introduced in FDFLF~I.  We parameterize the evolution of
M$^\ast$ and $\phi^\ast$ with redshift assuming the following simple
relations:
\begin{eqnarray}
  \nonumber
   M^\ast  \, (z) & = & M^\ast_0 \, + \, a \ln (1+z) , \; \\
  \phi^\ast \, (z)& = & \phi^\ast_0 \: \left( 1 \, + \, z
   \right)^b, \; \mathrm{and}   
\label{eqn:lfblue:evol_mstar_phistar}\\
  \nonumber
  \alpha \, (z) & = & \alpha_0 \;\; \equiv \;\; \mathrm{constant} .
\end{eqnarray}
We then derive the best fitting values for the free parameters $a$,
$b$, $M^\ast_0$, and $\phi^\ast_0$ by minimizing the $\chi^2$ of
\begin{eqnarray}
&  \chi^2 & =  \chi^2 \, (a, b, M^\ast_0, \phi^\ast_0)
\label{eqn:lfblue:chi}  \\\nonumber
&   = & \sum_{j=1}^{N_j} \sum_{i=1}^{N_i} \, \frac{\left[
  \phi (M_{ij}) - \Psi (M_{ij}, z_j, a, b, M^\ast_0,\phi^\ast_0)
  \right]^2}{\sigma_{ij}^2} ,
\end{eqnarray}
for the galaxy number densities in all magnitude and redshift bins
simultaneously (for more details see FDFLF~I).
\begin{figure*}[tbp]
\includegraphics[width=0.33\textwidth]{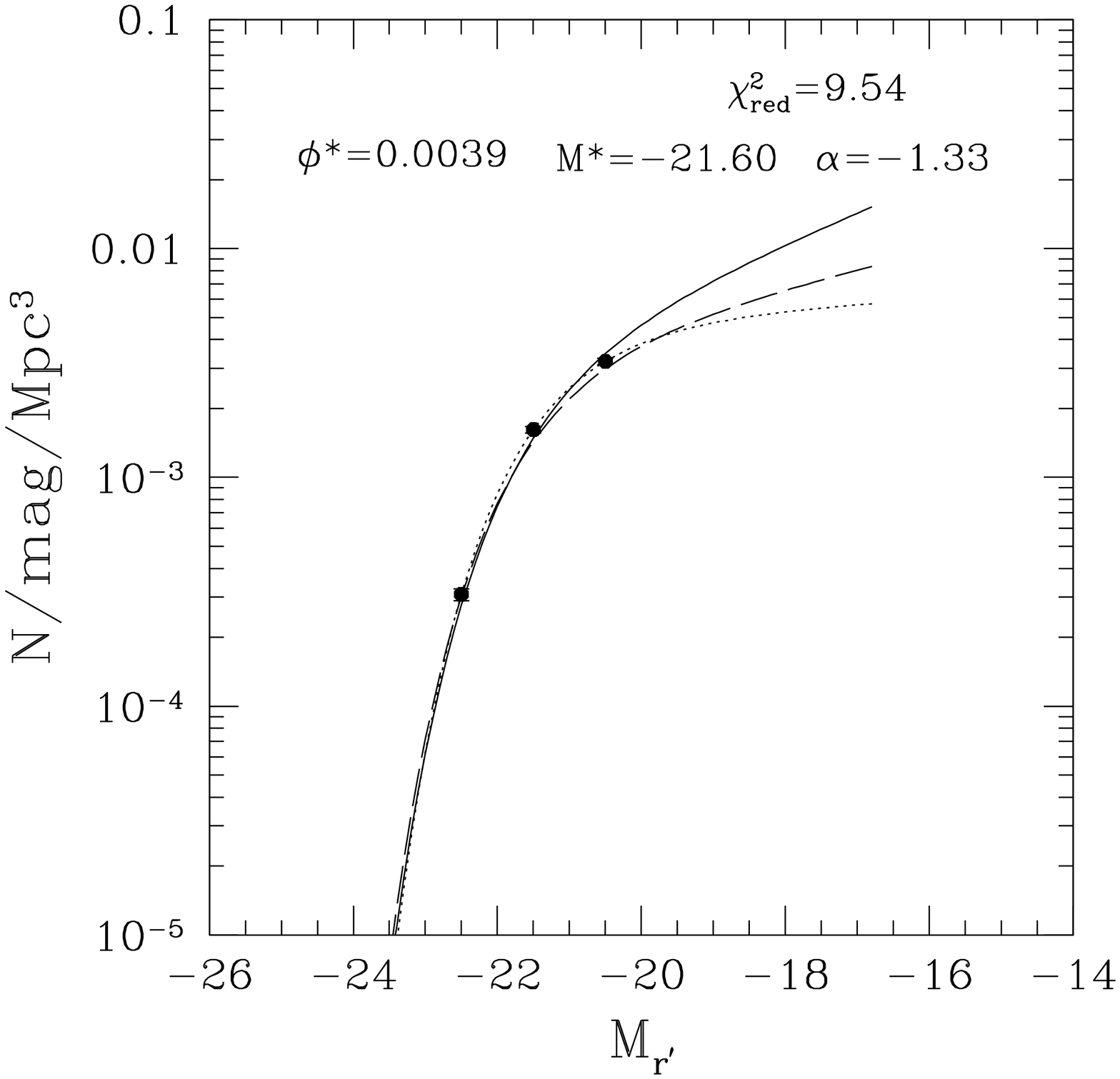}
\includegraphics[width=0.33\textwidth]{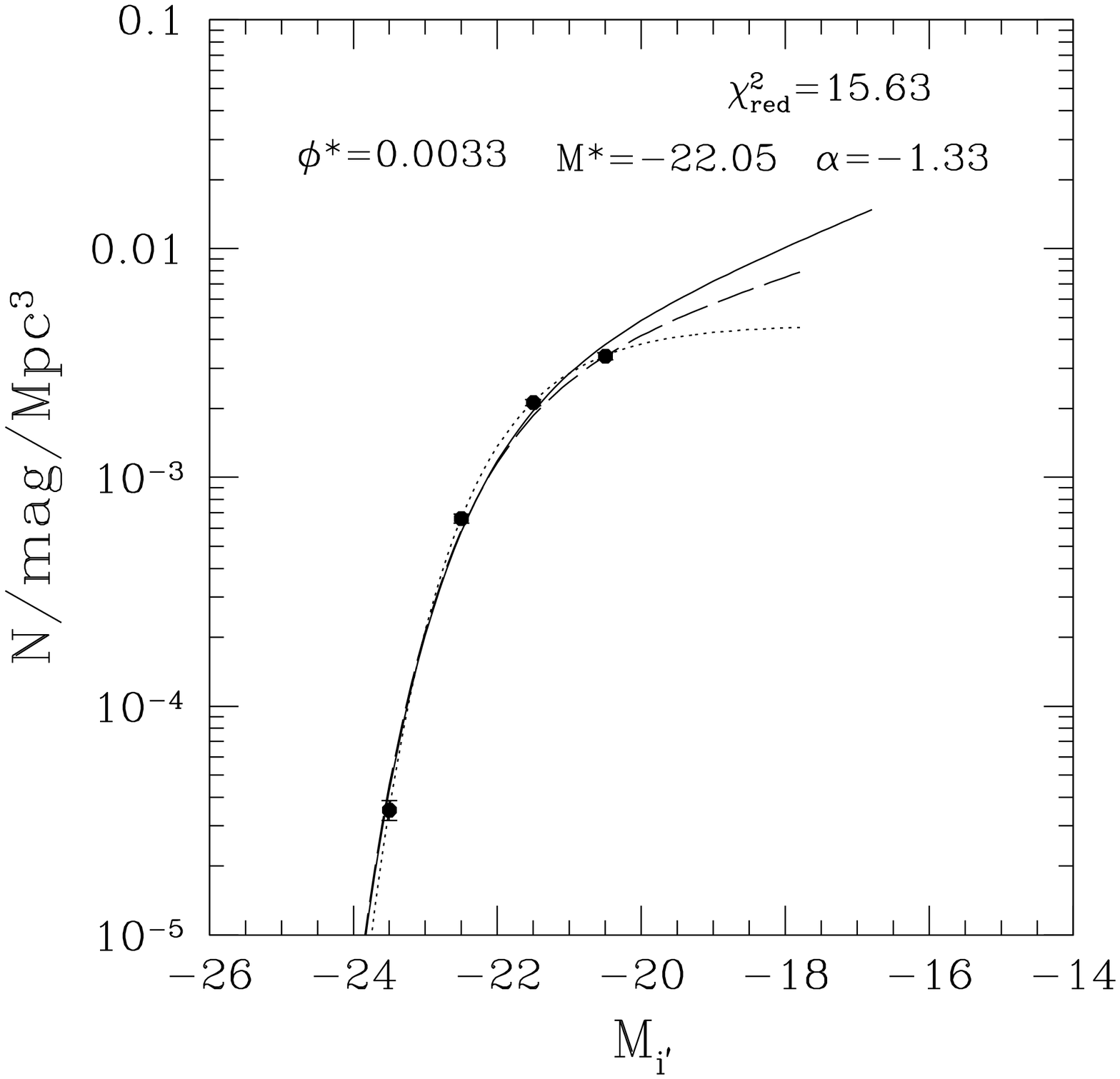}
\includegraphics[width=0.33\textwidth]{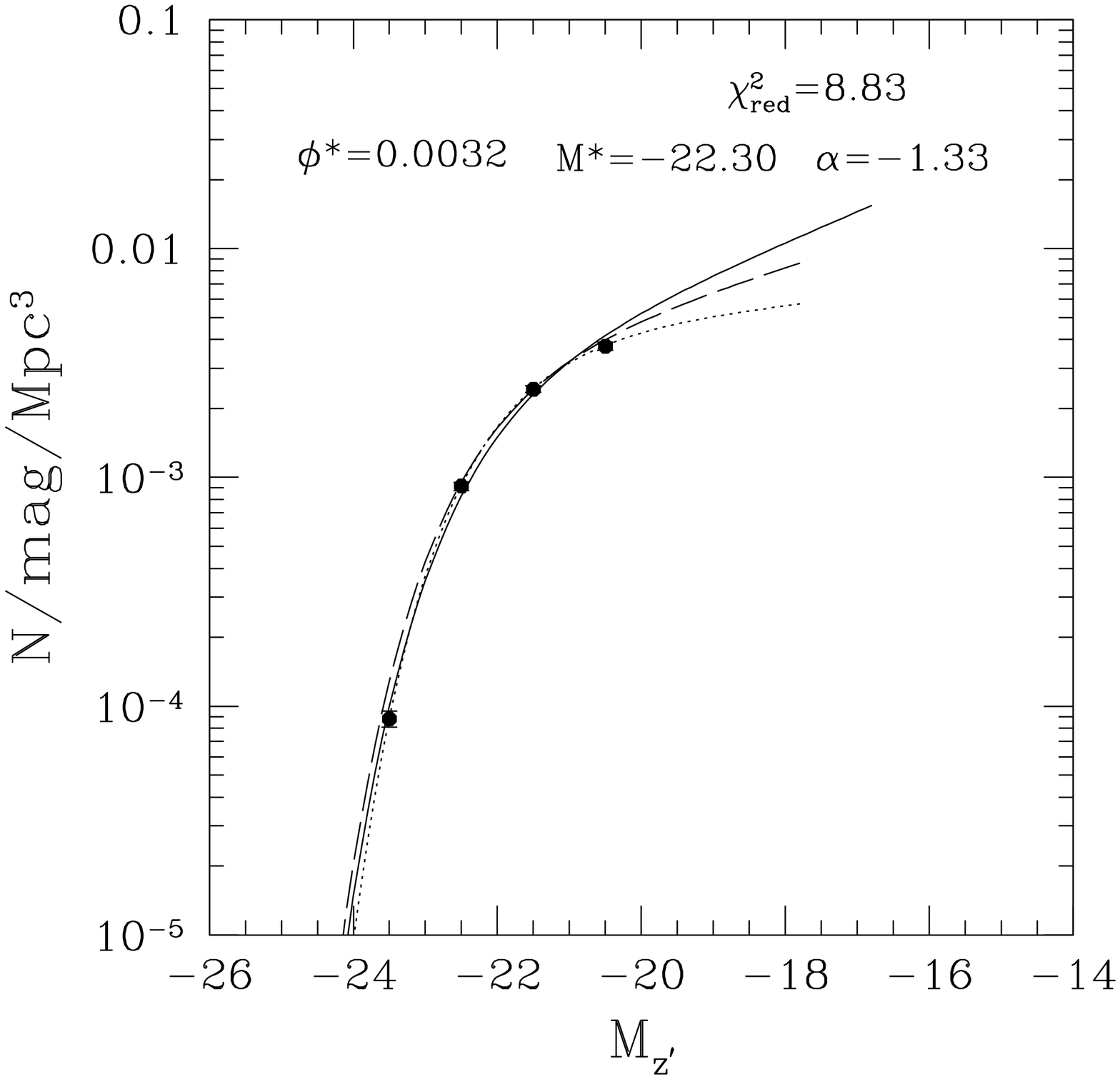}
\caption[LFs as given by \citet{blanton:1} for the 
r', i', and z' bands] {\label{fig:lfred:param:SDSS_orig} The local LFs
  as given by \citet{blanton:2} for the r' (left panel), i' (middle
  panel), and z' (left panel) bands.  The dotted lines in all plots
  represent the best fitting Schechter function of \citet{blanton:2}.
  The solid points and the associated errorbars are derived by the
  Schechter values and corresponding errors of the latter (see text).
  The dashed lines represent the result of \citet{blanton:1} after
  renormalizing $\phi^\ast$ according to \citet{liske:1}. We also
  fit a Schechter function (solid line) with a fixed slope of -1.33 as
  derived from the FDF data do the LF (solid dots). The corresponding
  M$^\ast$, $\phi^\ast$, as well as the reduced $\chi^2$ of the fit is
  also given in the figures.  }
\end{figure*}
The free parameters of the evolutionary model are constrained for
three different cases:
\begin{itemize}
\item \textit{Case~1}: FDF LFs between redshift \mbox{$\langle z
    \rangle\sim 0.65$} and \mbox{$\langle z \rangle\sim 2.26$} are
  used.
\item \textit{Case~2}: FDF LFs between redshift \mbox{$\langle z
    \rangle\sim 0.65$} and \mbox{$\langle z \rangle\sim 3.21$} are
  used.
\item \textit{Case~3}: FDF LFs between redshift \mbox{$\langle z
    \rangle\sim 0.65$} and \mbox{$\langle z \rangle\sim 2.26$} as well
  as the local LF of \citet{blanton:2} are used.
\end{itemize}

As can be seen from Fig.~\ref{fig:lfsedtype:slopes} the slopes
  of the red LFs derived by \citet[][ SDSS Data release 1]{blanton:2}
  are much shallower than those derived in this work and a previous work
  of the same author \citep[][ SDSS EDR]{blanton:1}.
  \citet{blanton:2} argued that the difference in the r-band LF
  (between \citealt{blanton:1} and \citealt{blanton:2}) stems only
  from the inclusion of luminosity evolution within the covered
  redshift range. Very recently \citet{driver:1} showed that the
  B-band LF derived from the Millennium Galaxy Catalogue (MGC, which
  is fully contained within the region of the SDSS) is inconsistent
  with the SDSS z=0 result of \citet{blanton:2} by more than
  $3\sigma$.  On the other hand the $M^\ast$ value of the B-Band LF of
  \citet{driver:1} is consistent with those derived by
  \citet{blanton:1} after that $\phi^\ast$ has been renormalized
  according to \citet{liske:1}. Once corrected, the \citet{blanton:1}
  B-band LF agrees well with the MGC, the 2dFGRS and the ESO Slice
  Project estimates.  \citet{driver:1} conclude, that the discrepancy
  between the MGC and the \citet{blanton:2} LFs is complex, but
  predominantly due to a color bias within the SDSS. They also
  conclude, that the color selection bias might be a general trend
  across all filters.\\
  We compare in Fig.~\ref{fig:lfred:param:SDSS_orig} the local
  Schechter functions as given by \citet{blanton:2} and \citet[][
  $\phi^\ast$ has been renormalized according to
  \citealt{liske:1}]{blanton:1} for the r', i', and z' bands. Although
  there is a reasonable good agreement between the LFs if one focuses
  on the bright part ($M \lsim -20$), they disagree at fainter
  magnitudes.  On the other hand the slope of the LF is strongly
  dependent on the depth of the survey.  The flux limit in the r-band
  selected SDSS survey is about $m_r < 17.79$. A very rough estimate
  of the absolute limiting magnitude at the mean redshift of the
  survey ($\langle z\rangle=0.1$) is therefore $M_{r'} \approx -20$.
  This means that the faint-end of the LF as shown in
  Fig.~\ref{fig:lfred:param:SDSS_orig} depends on the applied
  completeness correction \citep[see also the discussion
  in][]{andreon:1}.  Therefore, we decide to use only the bright part
  ($M \lsim -20$) of the SDSS LFs to constrain the free evolutionary
  parameter of \textit{Case 3}.\\

As the Schechter parameters are coupled, and M$^\ast$ and $\phi^\ast$
of \citet{blanton:2} are derived for a different slope $\alpha$, we
decide not to use M$^\ast$ and $\phi^\ast$ itself, but to reconstruct
a magnitude binned luminosity function out of the Schechter values
M$^\ast$, $\phi^\ast$, and $\alpha$ given in \citet{blanton:2}.
Following the method described in Sect.~\ref{sec:lfred:lit} to
visualize the errors of the literature LFs (shaded regions in the
plots) we derive the 1-magnitude-binned LF
as shown in Fig.~\ref{fig:lfred:param:SDSS_orig} (solid points).\\
Fig.~\ref{fig:lfred:param:SDSS_orig} shows, that a Schechter
  function fit to the SDSS data with a slope of $\alpha=-1.33$ (as
  derived from the FDF data) results in a reduced $\chi_{red}^2 \sim
  10$. This large $\chi_{red}^2$ increases the errorbars of the
  evolutionary parameter since we normalize the result of
  Eq.~(\ref{eqn:lfblue:chi})
  to a $\chi_{red}^2 \sim 1$ before calculating the errors.\\

\begin{figure*}[tbp]
\includegraphics[width=0.33\textwidth]{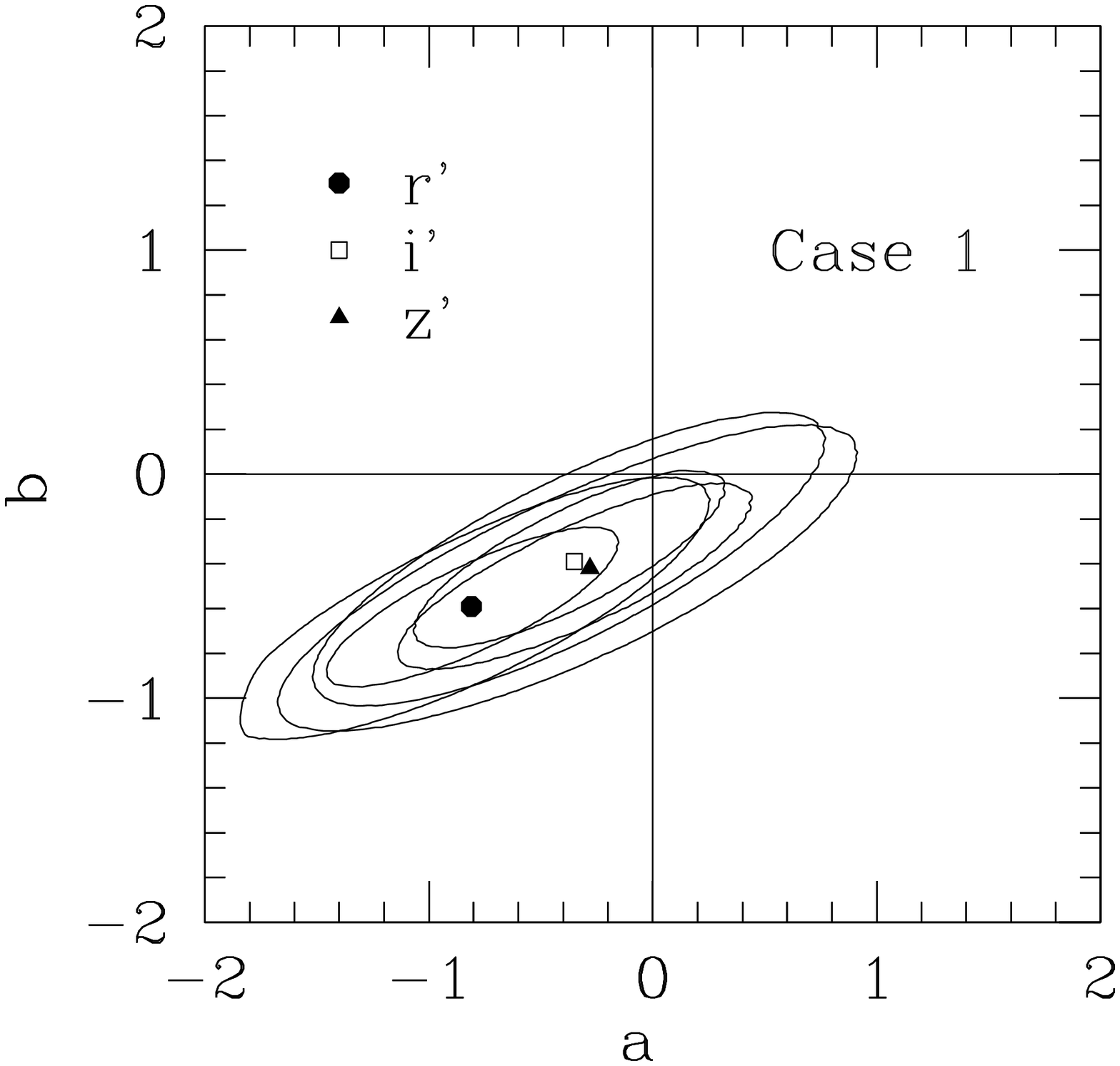}
\includegraphics[width=0.33\textwidth]{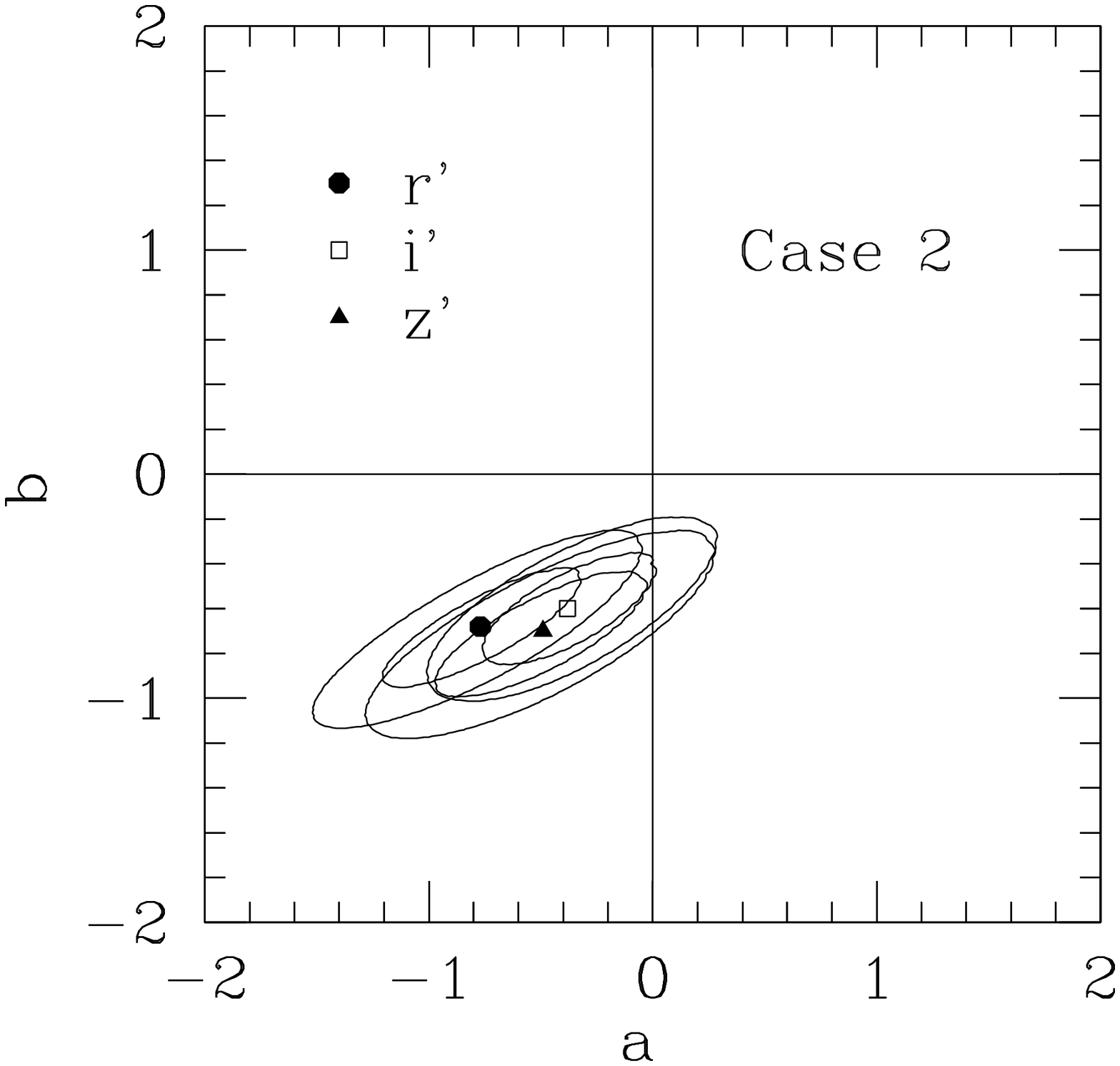}
\includegraphics[width=0.33\textwidth]{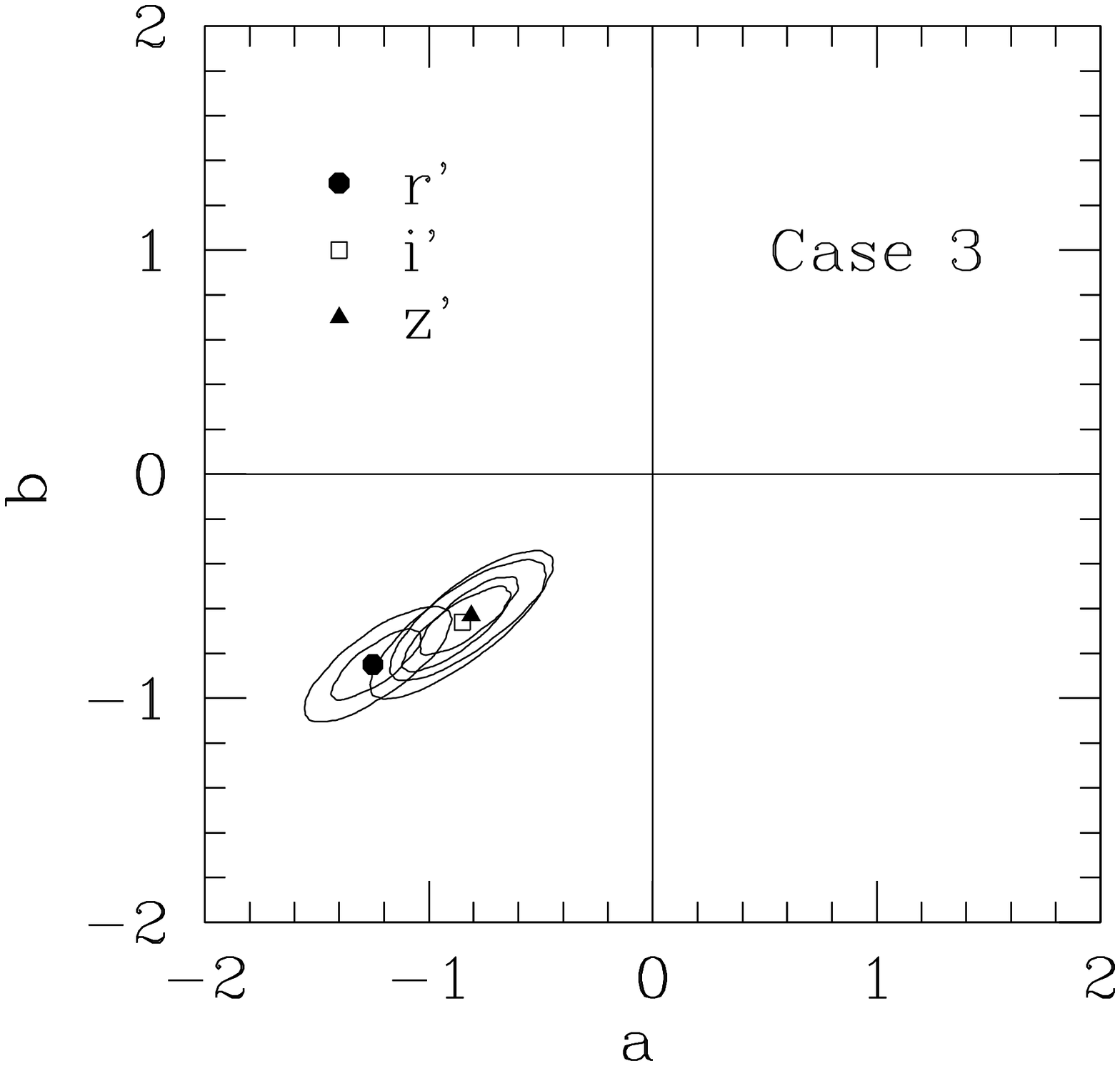}
\caption[Confidence levels of the parameters $a$ and $b$ in the 
  r', i', and z' bands]
{\label{fig:lfred:evol_param} \mbox{$1\sigma$} and \mbox{$2\sigma$}
  confidence levels of the parameters $a$ and $b$ in different bands
  (r', i', and z') for the evolutionary model described in the text.
  Left panel: FDF LFs between redshift
  \mbox{$\langle z \rangle\sim 0.65$} and \mbox{$\langle z \rangle\sim
    2.26$} are used (\textit{Case~1}). The values for $a$ and $b$ can be found in
  Table~\ref{tab:lfred:evo_param:fdf}.
  Middle panel: FDF LFs between redshift
  \mbox{$\langle z \rangle\sim 0.65$} and \mbox{$\langle z \rangle\sim
    3.21$} are used (\textit{Case~2}). The values for $a$ and $b$ can be found in
  Table~\ref{tab:lfred:evo_param:fdf}.
  Right panel: FDF LFs between redshift
  \mbox{$\langle z \rangle\sim 0.65$} and \mbox{$\langle z \rangle\sim
    2.26$} as well as the local LF of
  \citet{blanton:2} are used (\textit{Case~3}).  
  The values for $a$ and $b$ can be found in
  Table~\ref{tab:lfred:evo_param:fdf}. }
\end{figure*}

\begin{figure*}[tbp]
\includegraphics[width=0.33\textwidth]{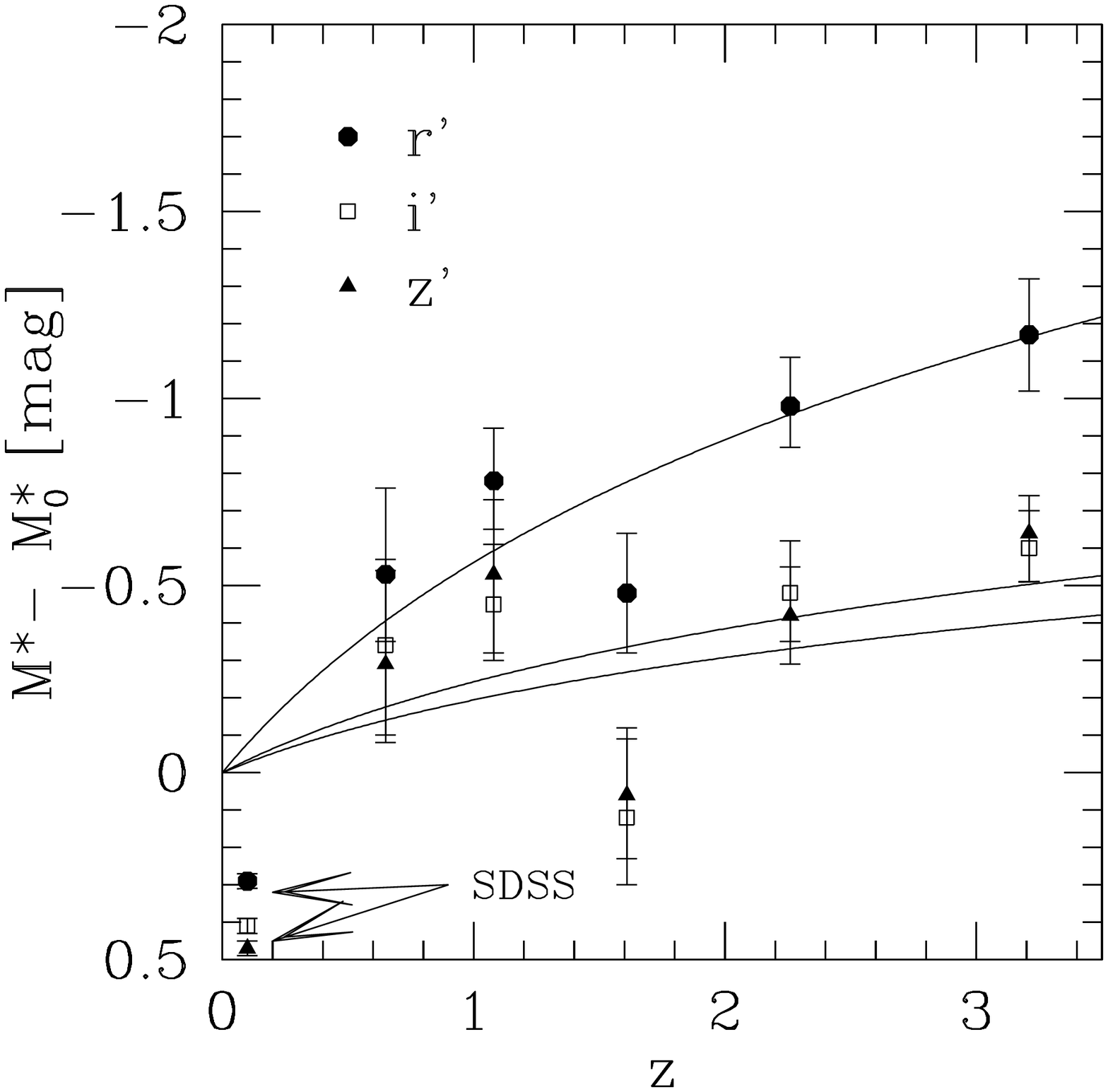}
\includegraphics[width=0.33\textwidth]{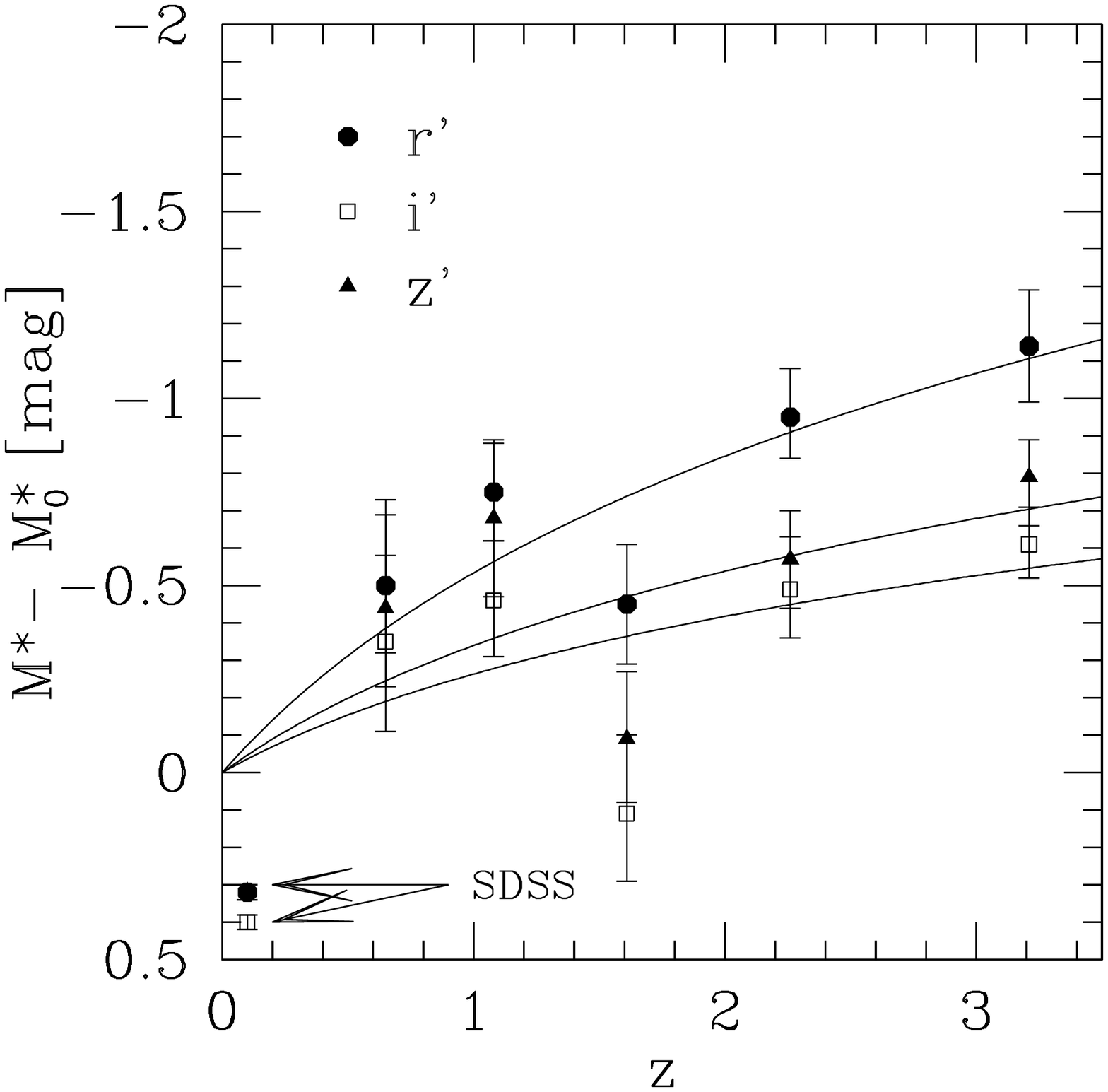}
\includegraphics[width=0.33\textwidth]{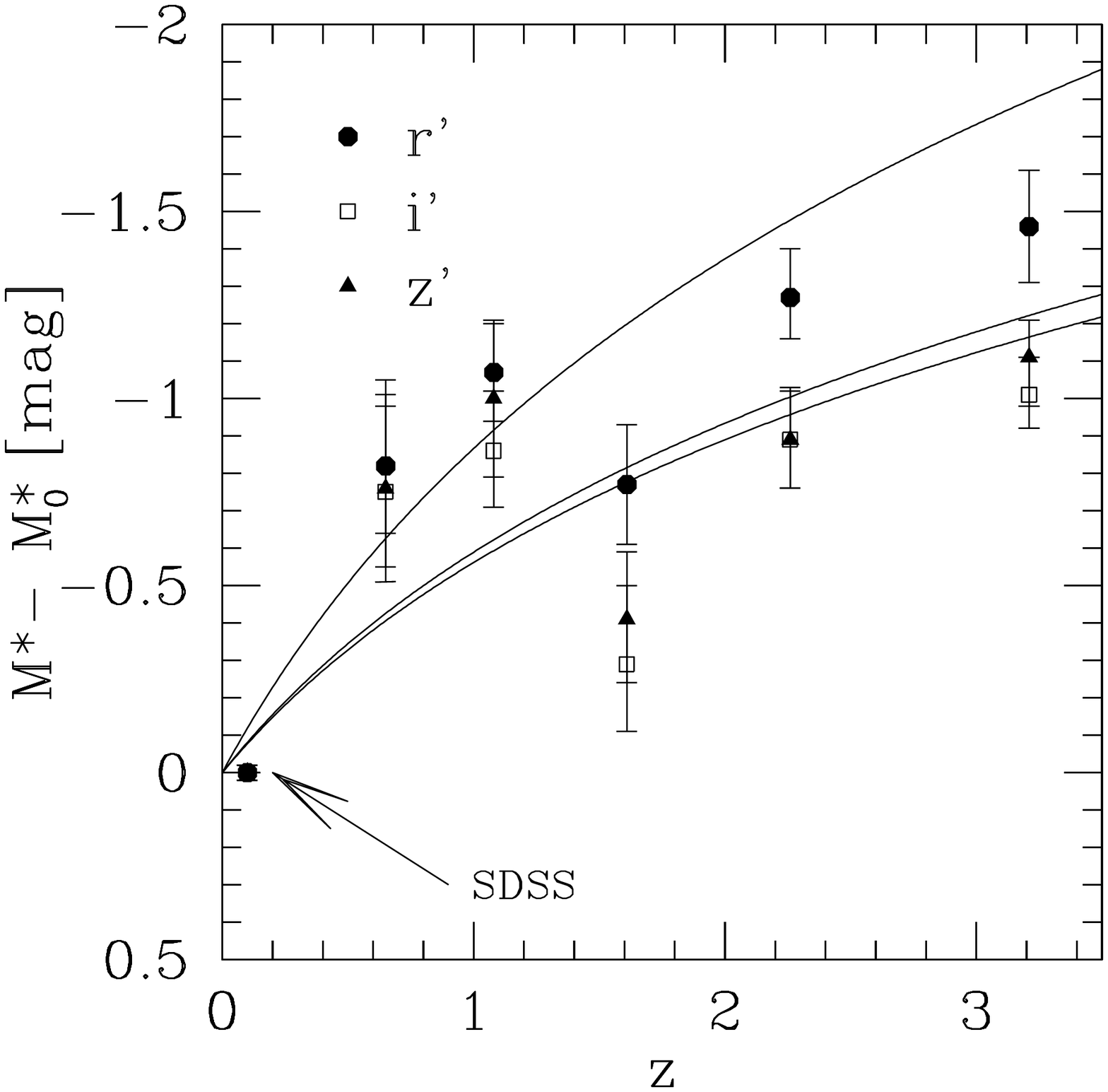}
\caption[Relative evolution of M$^\ast$
with redshift] {\label{fig:lfred:evol_lumfkt_magstar} Relative
  evolution of M$^\ast$ with redshift. The solid line represent the
  best fit of the evolutionary model according to
  Eq.~(\ref{eqn:lfblue:evol_mstar_phistar}).  Left panel: FDF
  LFs between redshift \mbox{$\langle z \rangle\sim
    0.65$} and \mbox{$\langle z \rangle\sim 2.26$} are used to
  constrain the evolutionary model (\textit{Case~1}).  Middle panel:
  FDF LFs between redshift \mbox{$\langle z
    \rangle\sim 0.65$} and \mbox{$\langle z \rangle\sim 3.21$} are
  used to constrain the evolutionary model (\textit{Case~2}).  Right
  panel: FDF LFs between redshift \mbox{$\langle z
    \rangle\sim 0.65$} and \mbox{$\langle z \rangle\sim 2.26$} as well
  as the local LF of \citet{blanton:2} are used to
  constrain the evolutionary model (\textit{Case~3}) (see also text).}
\end{figure*}

\begin{figure*}[tbp]
\includegraphics[width=0.33\textwidth]{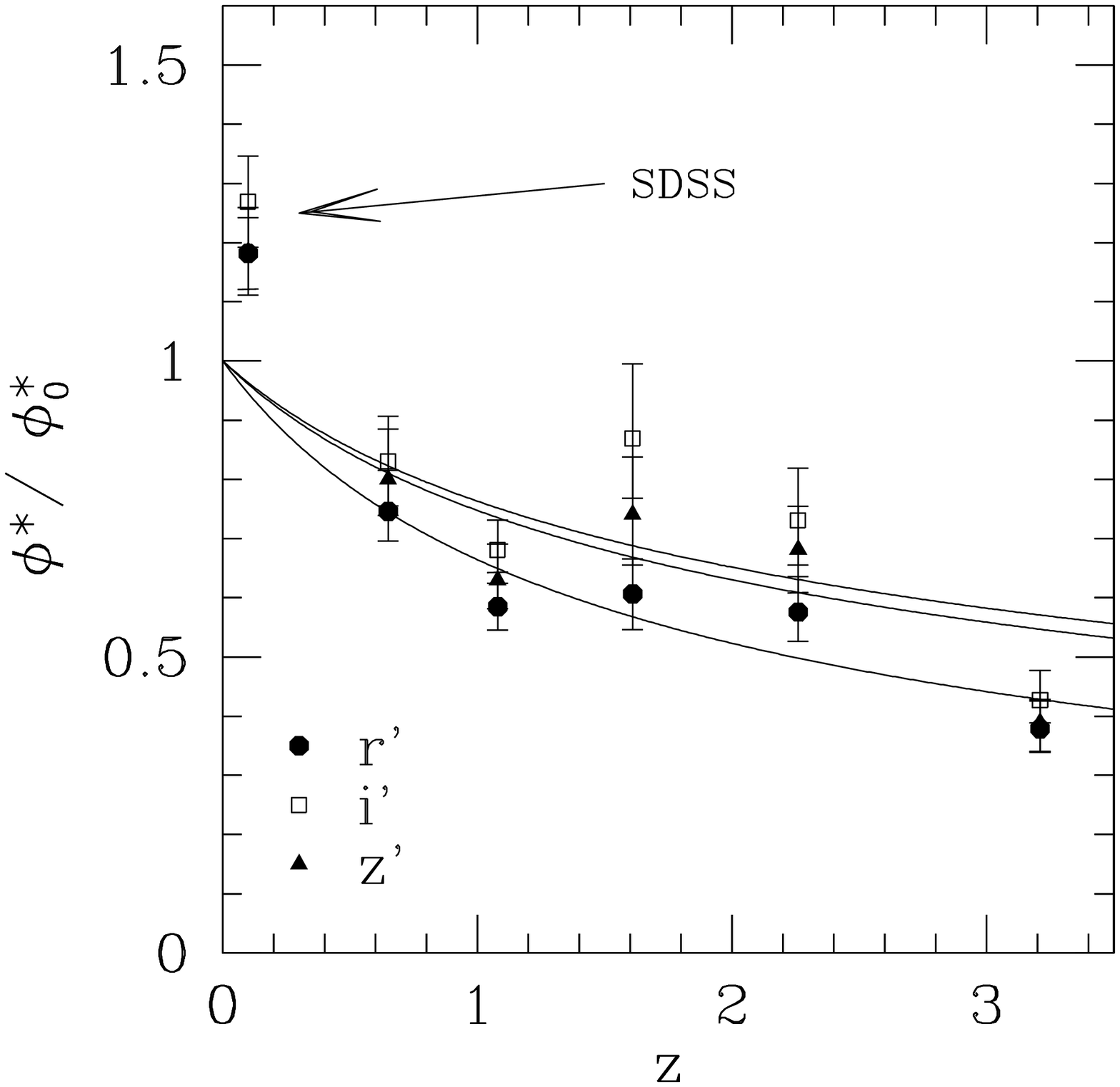}
\includegraphics[width=0.33\textwidth]{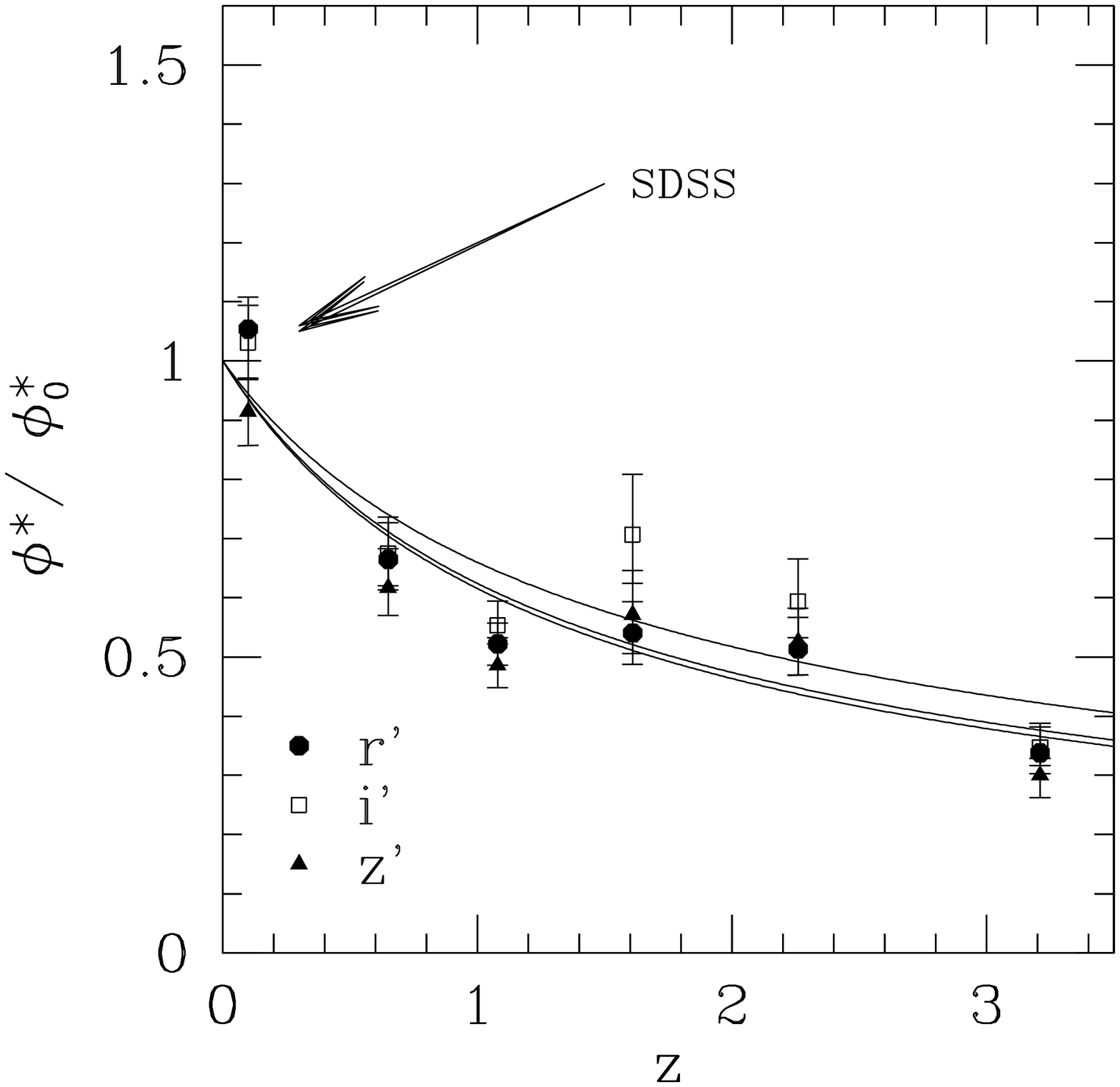}
\includegraphics[width=0.33\textwidth]{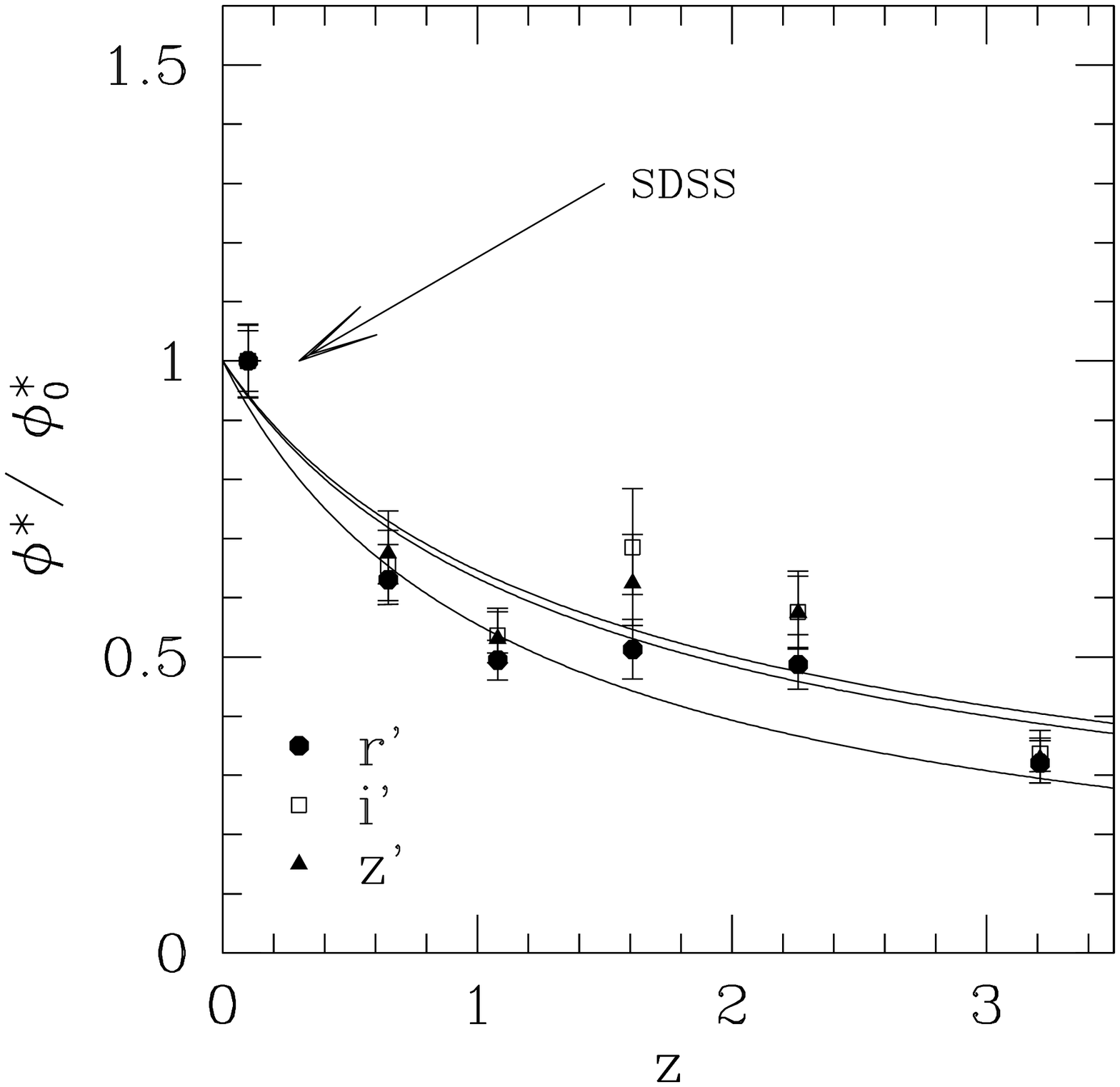}
\caption[Relative evolution of $\phi^\ast$
with redshift] {\label{fig:lfred:evol_lumfkt_phistar} Relative
  evolution of $\phi^\ast$ with redshift. The solid line represent the
  best fit of the evolutionary model according to
  Eq.~(\ref{eqn:lfblue:evol_mstar_phistar}).  Left panel: FDF
  LFs between redshift \mbox{$\langle z \rangle\sim
    0.65$} and \mbox{$\langle z \rangle\sim 2.26$} are used to
  constrain the evolutionary model (\textit{Case~1}).  Middle panel:
  FDF LFs between redshift \mbox{$\langle z
    \rangle\sim 0.65$} and \mbox{$\langle z \rangle\sim 3.21$} are
  used to constrain the evolutionary model (\textit{Case~2}).  Right
  panel: FDF LFs between redshift \mbox{$\langle z
    \rangle\sim 0.65$} and \mbox{$\langle z \rangle\sim 2.26$} as well
  as the local LF of \citet{blanton:2} are used to
  constrain the evolutionary model (\textit{Case~3}) (see also text).}
\end{figure*}

\begin{table*}[tbp]
  \caption{\label{tab:lfred:evo_param:fdf} Evolution parameters according to
    Eq.~(\ref{eqn:lfblue:evol_mstar_phistar}).}
  \begin{center}
    \begin{tabular}{cccccc}
      Filter & \textit{Case}& a  & b & M$^\ast_0$ & $\phi^\ast_0 $   \\
      & &  &  & (mag) & ($\mathrm{Mpc}^{-3}$)   \\
      \hline
      \rule[+3mm]{-1.4mm}{3mm}
      \rule[-3mm]{0mm}{3mm}r'   &\textit{Case 1} &  $-0.81^{+0.43}_{-0.41}$ & $-0.59^{+0.23}_{-0.23}$ & $-21.89^{+0.39}_{-0.42}$ & $0.0033^{+0.0007}_{-0.0005}$\\ 
      \rule[-3mm]{0mm}{3mm}r'   &\textit{Case 2} &  $-0.77^{+0.30}_{-0.28}$ & $-0.68^{+0.17}_{-0.17}$ & $-21.92^{+0.30}_{-0.30}$ & $0.0037^{+0.0005}_{-0.0005}$\\ 
      \rule[-3mm]{0mm}{3mm}r'   &\textit{Case 3} &  $-1.25^{+0.14}_{-0.10}$ & $-0.85^{+0.10}_{-0.08}$ & $-21.49^{+0.03}_{-0.02}$ & $0.0042^{+0.0001}_{-0.0002}$\\ 
 
      \rule[+3mm]{-1.4mm}{3mm}                                                                             
      \rule[-3mm]{0mm}{3mm}i'   &\textit{Case 1} &  $-0.35^{+0.43}_{-0.48}$ & $-0.39^{+0.27}_{-0.24}$ & $-22.46^{+0.44}_{-0.41}$ & $0.0026^{+0.0006}_{-0.0004}$\\ 
      \rule[-3mm]{0mm}{3mm}i'   &\textit{Case 2} &  $-0.38^{+0.26}_{-0.25}$ & $-0.60^{+0.15}_{-0.16}$ & $-22.45^{+0.30}_{-0.30}$ & $0.0032^{+0.0004}_{-0.0004}$\\ 
      \rule[-3mm]{0mm}{3mm}i'   &\textit{Case 3} &  $-0.85^{+0.12}_{-0.18}$ & $-0.66^{+0.08}_{-0.15}$ & $-21.97^{+0.04}_{-0.04}$ & $0.0034^{+0.0002}_{-0.0001}$\\ 
                                                                              
      \rule[+3mm]{-1.4mm}{3mm}
      \rule[-3mm]{0mm}{3mm}z'   &\textit{Case 1} &  $-0.28^{+0.46}_{-0.58}$ & $-0.42^{+0.24}_{-0.30}$ & $-22.77^{+0.56}_{-0.45}$ & $0.0027^{+0.0008}_{-0.0004}$\\ 
      \rule[-3mm]{0mm}{3mm}z'   &\textit{Case 2} &  $-0.49^{+0.29}_{-0.31}$ & $-0.70^{+0.17}_{-0.19}$ & $-22.62^{+0.38}_{-0.32}$ & $0.0035^{+0.0006}_{-0.0006}$\\ 
      \rule[-3mm]{0mm}{3mm}z'   &\textit{Case 3} &  $-0.81^{+0.11}_{-0.16}$ & $-0.63^{+0.11}_{-0.12}$ & $-22.22^{+0.04}_{-0.05}$ & $0.0033^{+0.0002}_{-0.0001}$\\ 

    \end{tabular}
  \end{center}
\end{table*}

The \mbox{$1\sigma$} and \mbox{$2\sigma$} confidence levels of the
evolution parameters $a$ and $b$ for the different filters and cases
are shown in Fig.~\ref{fig:lfred:evol_param}. These contours were
derived by projecting the four-dimensional $\chi^2$ distribution to
the $a$-$b$ plane, i.e.\ for given $a$ and $b$ we use the value of
$M^\ast_0$ and $\phi^\ast_0$ which minimizes the $\chi^2(a,b)$.  For
\textit{Case~1} (left panel) the errorbars of $a$ and $b$ are rather
large and although the best fitting values suggest a redshift
evolution we are also compatible (within $2\sigma$) with no evolution
of M$^\ast$ and $\phi^\ast$.  The error ellipses for \textit{Case~2}
(middle panel) are smaller than in \textit{Case~1} and for the r'-band
LF we see a luminosity and a density evolution on a $2\sigma$ level.
For the i'-band and z'-band LFs we see only a density evolution on a
$2\sigma$ level.  Including also the local LF of \citet{blanton:2} in
the evolution analysis as in \textit{Case~3} (left panel) we are able
to derive $a$ and $b$ with higher precision since $M^\ast_0$ and
$\phi^\ast_0$ are more restricted.  The luminosity and density
evolution is clearly visible on more than $2\sigma$ level.  Please
note that combining different datasets like the FDF and the SDSS can
introduce systematic errors due to different selection techniques and
calibration differences not fully taken into account
(see also discussion above).  
Nevertheless, a comparison of the FDF LFs with the SDSS Schechter
functions in Fig.~\ref{fig:lfred:lumfkt_fdf_sdss} shows a relatively
good agreement 
(at the bright end).
Furthermore, a detailed comparison of the UV LFs of the
FDF with the LF derived in large surveys e.g.  \citet[][ based on
COMBO-17]{combo17:1}, \citet[][ based on LBG analysis]{steidel:1},
\citet[][ based on Subaru Deep Field/Survey]{iwata:1,ouchi:3} or
pencil beam surveys e.g \citet[][ based on both HDFs]{poli:1}
presented in FDFLF~I shows good agreement in the overlapping magnitude
range at all redshifts.  We are thus confident, that remaining
systematic differences (e.g. due to the influence of large scale
structure; LSS) must be small.

The values for the free parameters $a$, $b$, $M^\ast_0$, and
$\phi^\ast_0$ as well as the associated errors can be found in
Table~\ref{tab:lfred:evo_param:fdf}. The evolution parameters $a$,
$b$, $M^\ast_0$, and $\phi^\ast_0$ derived in \textit{Case~1},
\textit{Case~2}, and \textit{Case~3} agree \textit{all} within
$2\sigma$. Most of the values differ only by $1\sigma$ or less.

In Fig.~\ref{fig:lfred:evol_lumfkt_magstar} we illustrate the relative
redshift evolution of $M^\ast$ for the different filters and different
cases, whereas the relative redshift evolution of $\phi^\ast$ is shown
in Fig.~\ref{fig:lfred:evol_lumfkt_phistar}.  Note that $a$, $b$,
$M^\ast_0$, and $\phi^\ast_0$ were derived by minimizing
Eq.~(\ref{eqn:lfblue:chi}) and not the differences between the (best
fitting) lines and the data points in
Fig.~\ref{fig:lfred:evol_lumfkt_magstar} and
Fig.~\ref{fig:lfred:evol_lumfkt_phistar}. As for the blue bands
(FDFLF~I) the simple parametrization of
Eq.~(\ref{eqn:lfblue:evol_mstar_phistar}) is able to describe the
evolution of the galaxy LFs also in the red bands very well.\\

Recently \citet{blanton:3} used the data of the SDSS Data Release 2 to
analyze the very local (\mbox{$0.00 < z < 0.05$}) LF (corrected
for surface-brightness incompleteness) down to extremely low
luminosity galaxies. They found, that a Schechter function is an
insufficient parametrization of the LF as there is an upturn in the
slope of the LF for $M_r-5\log(h_{100}) > -18$.  We therefore compare
in Fig.~\ref{fig:lfred:lumfkt_fdf_sdss} the red FDF LFs in two
redshift ranges (\mbox{$\langle z\rangle=0.3$} and \mbox{$\langle
  z\rangle=0.65$}) with the local Schechter functions as derived in
the SDSS by \citet{blanton:2}, and \citet{blanton:3}.  Considering the
small volume covered by the FDF in the redshift bin \mbox{$\langle
  z\rangle=0.3$} and the fact, that we see clustered {\em
  spectroscopic} redshifts at $z=0.22$, $z=0.33$, and $z=0.39$, the
agreement between the LFs and the Schechter functions is relatively
good for $M < -19$. For the fainter part, the measured number density
disagree with \citet{blanton:2} and \citet{blanton:3} in all three
analyzed bands.  If we do the same comparison at \mbox{$\langle
  z\rangle=0.65$} where the FDF covers a relatively large volume
minimizing the influence of LSS, the measured LFs follow the very
local Schechter function of \citet{blanton:3} also in the faint
magnitude regime. Moreover, the upturn of the faint-end of the LF as
found by \citet{blanton:3} in the SDSS or by \citet{popesso:1} in the
RASS-SDSS Galaxy Cluster Survey \citep[see also ][]{gonzales:1}, is
visible also in the FDF data (at least at \mbox{$\langle
  z\rangle=0.65$}).\\
This upturn seems to be less pronounced in the UV (FDFLF~I). A
possible reason for this could again be the different contribution of
the SED-type LFs presented in Fig.~\ref{fig:lfsedtype:uv_i_schechter}.
In the red bands, the difference between the characteristic
luminosities between the LFs for types 1, 2, 3 and type 4 together
with the dominance of the type-4 LF at the faint end results in a dip
at $M\sim -20$.

Although a Schechter function is an insufficient parametrization of
the LF derived by \citet{blanton:3} we used their results as local
reference point to calculate the evolution of the LF in the various
bands by minimizing Eq.~(\ref{eqn:lfblue:chi}). Due to the upturn of
the faint-end of the local LF and the fact that our evolutionary model
assumes a normal Schechter function, the reduced $\chi^2$ of
Eq.~(\ref{eqn:lfblue:chi}) is of the order of 9. As we do not want to
increase the number of free parameters by using a double Schechter
function (at higher redshifts the data are not able to constrain a
possible upturn in the LF), we increase the errors of $a$, $b$,
$M^\ast_0$, and $\phi^\ast_0$.  We do this by an appropriate scaling
of the errors $\sigma_{ij}$ of Eq.~(\ref{eqn:lfblue:chi}) to obtain a
reduced $\chi^2$ of unity.  
 A comparison of the evolution parameter $a$ and $b$ with
  those derived in \textit{Case~3} shows, that the evolution in the
  characteristic luminosity agrees with \textit{Case~3}, but the
  evolution of the characteristic density decreases from $b \sim -0.7$
  to $b \sim -0.5$ being closer to \textit{Case~1} and
  \textit{Case~2}.
However, a no-evolution
hypothesis can be excluded on a \mbox{$2\sigma$} level in all three
bands if the results of \citet{blanton:3} are used as local reference
points.

If we compare the evolutionary parameters $a$ and $b$ of the red bands
with those derived in the blue bands (FDFLF~I), the following trend
can be seen: with increasing waveband the redshift evolution of
$M^\star$ and $\phi^\star$ decreases. Furthermore, if we include in
our analysis also the results obtained in the SDSS \citep{blanton:2}
the brightening of $M^\star$ and the decrease in $\phi^\star$ for
increasing redshift is still visible in the red bands at more than
$3\sigma$.

\section{Comparison with observational results from literature}
\label{sec:lfred:lit}

To put the FDF results on the evolution of the LFs into perspective,
we compare them to other surveys using the following
approach:\\
\textit{First} we convert results from the literature to our cosmology
(\mbox{$\Omega_M=0.3$}, \mbox{$\Omega_\Lambda=0.7$}, and \mbox{$H_0=70
  \, \mathrm{km} \, \mathrm{s}^{-1} \, \mathrm{Mpc}^{-1}$}). Although
this conversion may not be perfect (we can only transform number
densities and magnitudes but lack the knowledge of the individual
magnitudes and redshifts of the galaxies), the errors introduced in
this way are not large and the method is suitable for our purpose.
\textit{Second}, in order to avoid uncertainties due to conversion
between different filter bands, we always convert our data to the same
band as the survey we want to compare with.  \textit{Third}, we try to
use the same redshift binning as in the literature.

To visualize the errors of the literature LFs we perform Monte-Carlo
simulations using the $\Delta$M$^\ast$, $\Delta\phi^\ast$, and
$\Delta\alpha$ given in the papers. In cases where not all of these
values could be found in the paper, this is mentioned in the figure
caption.  We do not take into account any correlation between the
Schechter parameters and assume a Gaussian distribution of the errors
$\Delta$M$^\ast$, $\Delta\phi^\ast$, and $\Delta\alpha$.  From 1000
simulated Schechter functions we derive the region where \mbox{68.8
  \%} of the realizations lie.  The resulting region, roughly
corresponding to 1$\sigma$ errors, is shaded in the figures.  The LFs
derived in the FDF are also shown as filled and open circles. The
filled circles are completeness corrected whereas the open circles are
not corrected. The redshift binning used to derive the LF in the FDF
as well as the literature redshift binning is given in the upper part
of every figure.  Moreover, the limiting magnitude of the respective
survey is indicated by the low-luminosity cut-off of the shaded region
in all figures.  If the limiting magnitude was not explicitly given it
was estimated from the figures
in the literature.\\
A comparison of our FDF results with LFs based on spectroscopic
distance determinations
\citep{blanton:2,blanton:3,lin:2,lin:1,brown:1,shapley:1,ilbert:2}
as well as with LFs based on photometric redshifts
\citep{combo17:1,chen:1,dahlen:1} follows:\\

\noindent\textit{\citet{blanton:2,blanton:3}:}\\
In Fig.~\ref{fig:lfred:lumfkt_fdf_sdss} we compare the red FDF LFs in
two redshift regimes (\mbox{$\langle z\rangle=0.3$} and \mbox{$\langle
  z\rangle=0.65$}) with the local Schechter functions as derived in
the SDSS by \citet{blanton:2}, and \citet{blanton:3}. As previously
discussed, the agreement between the LFs and the Schechter functions
is relatively good for $M < -19$. For the fainter part, the measured
number density disagree with \citet{blanton:2} and \citet{blanton:3}.
If we do the same comparison at \mbox{$\langle z\rangle=0.65$} where
the FDF covers a relatively large volume minimizing the influence of
LSS, the measured LFs follow the very local Schechter function of
\citet{blanton:3} also in the faint magnitude regime.
Note that \citet{blanton:3} explicitly corrected for
  surface-brightness incompleteness when deriving the very local LFs.\\

\begin{figure*}[tbp]
\includegraphics[width=0.50\textwidth]{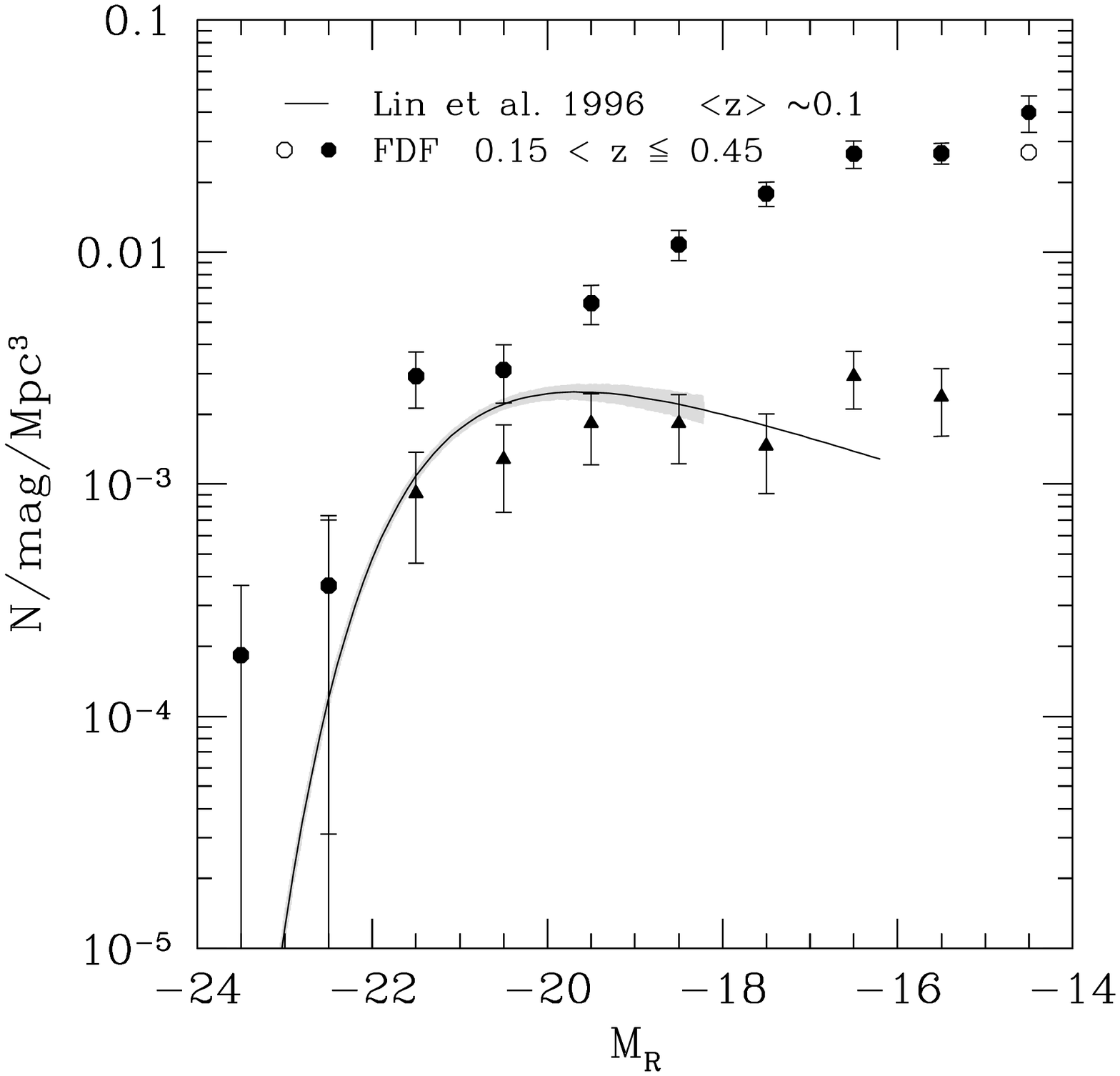}
\includegraphics[width=0.50\textwidth]{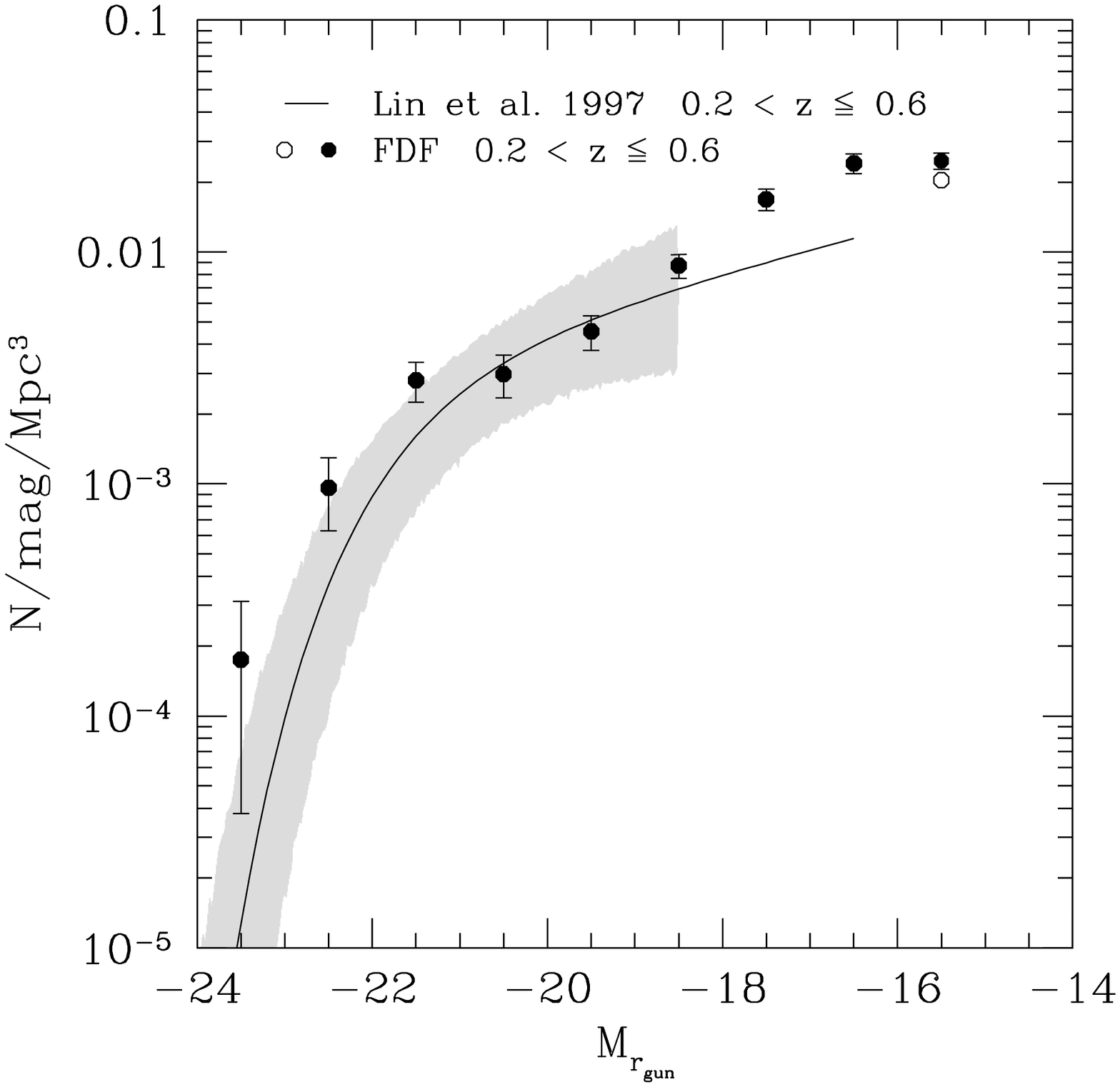}
\caption[Comparison of the FDF LF with \citet{lin:2} and \citet{lin:1}]
{\label{fig:lfred:lin} Left panel: Comparison of the R-band luminosity
  function of the FDF (filled/open circles, \mbox{$\langle z
    \rangle\sim 0.3$}) with the Schechter function derived in
  \textit{\citet{lin:2}} (\mbox{$\langle z \rangle\sim 0.1$}). The
  shaded region is based on $\Delta$M$^\ast$, $\Delta\phi^\ast$, and
  $\Delta\alpha$. The triangles show the r'-band FDF LF at
  \mbox{$\langle z \rangle\sim 0.3$} for SED type 1 galaxies.  Right
  panel: Comparison of the Gunn-r-band LF of the FDF with the
  Schechter function derived in \citet{lin:1} ($z=0.2$--$0.6$). The
  shaded region is based on $\Delta$M$^\ast$, $\Delta\phi^\ast$, and
  $\Delta\alpha$.}
\end{figure*}

\noindent\textit{\citet{lin:2}:}\\
Despite the small volume covered by the FDF at low redshift
we compare in Fig.~\ref{fig:lfred:lin} (left panel) our LF with the LF
derived by \citet{lin:2} in the Las Campanas Redshift Survey (LCRS).
Their sample contains 18678 sources selected from CCD photometry in a
``hybrid'' red Kron-Cousins R-band with a mean redshift of
\mbox{$\langle z \rangle\sim 0.1$}. The solid line in
Fig.~\ref{fig:lfred:lin} (left panel) represents the LF in the R-band
from \citet{lin:2} whereas the filled circles show our $V/V_{max}$
corrected LF derived at $0.15 <z\le 0.45$.  There is a rather large
disagreement between the LF in the FDF and in the LCRS, which is
mainly due to the different slope ($\alpha=-0.7$ for the LCRS) but
also the FDF galaxy number density at the bright end seems to be
slightly higher than in the LCRS. This might be partly attributed to
cosmic variance and/or to the selection method. The difference at the
faint-end is a well known LCRS feature related to their selection
method which biases LCRS towards early type systems.  Indeed, our LF
for SED type 1 galaxies (triangles in Fig.~\ref{fig:lfred:lin}) shows
a very good agreement with \citet{lin:2}.\\

\noindent\textit{\citet{lin:1}:}\\
Based on 389 field galaxies from the Canadian Network for
Observational Cosmology cluster redshift survey (CNOC1) selected in
the Gunn-r-band \citet{lin:1} derived the LF in the restframe
Gunn-r-band.  In Fig.~\ref{fig:lfred:lin} (right panel) we compare our
luminosity function with the LF derived by \citet{lin:1} in the
redshift range $z=0.2$--$0.6$.  There is a very good agreement between
the FDF data and the CNOC1 survey concerning the LF, if we compare
only the magnitude range in common to both surveys (shaded region).
Also the slope derived in \citet{lin:1} ($\alpha=-1.25 \pm 0.19 $,
Table 2 of the paper) is compatible with the slope in the FDF.\\

\begin{figure*}
\includegraphics[width=0.50\textwidth]{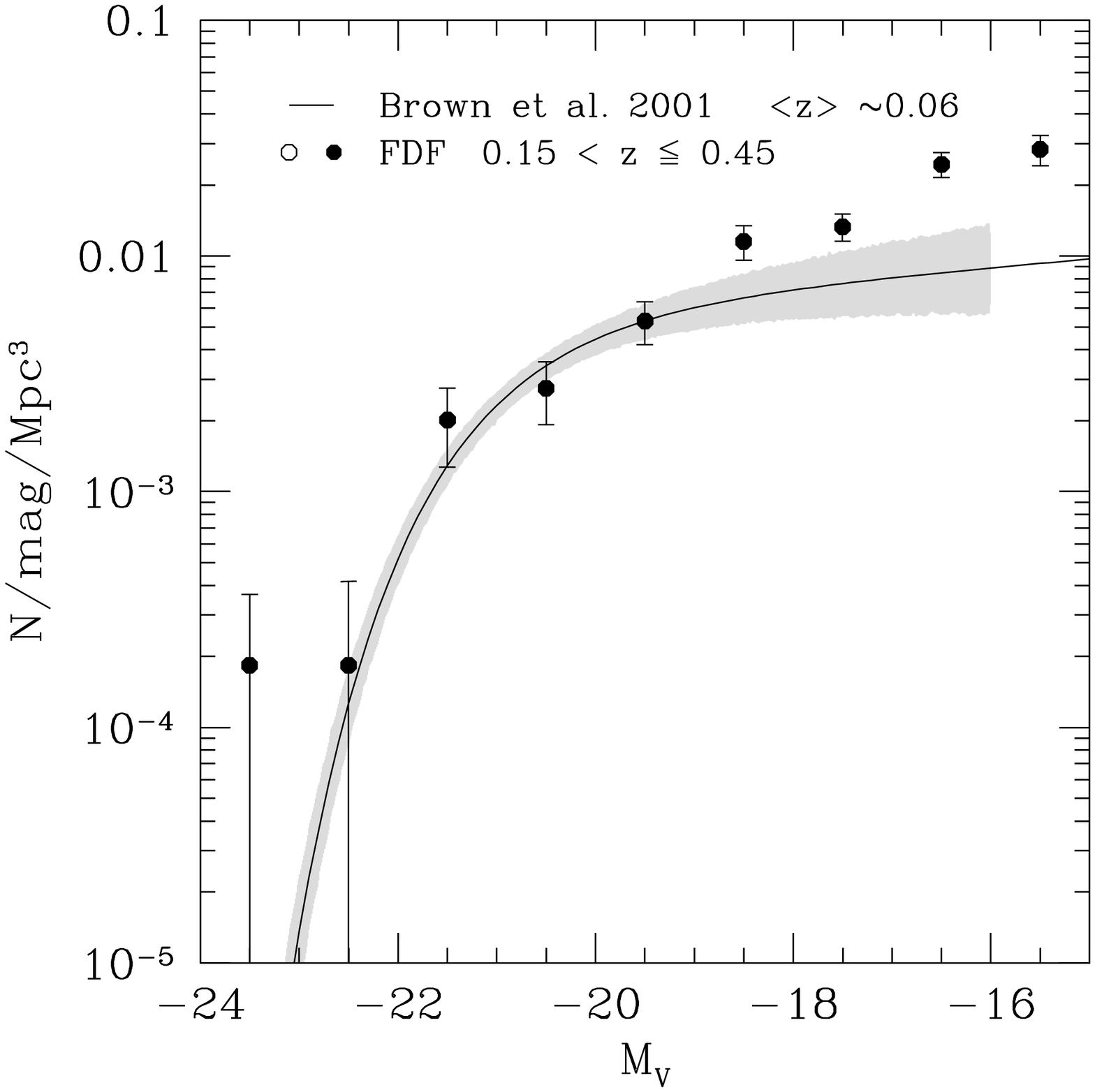}
\includegraphics[width=0.50\textwidth]{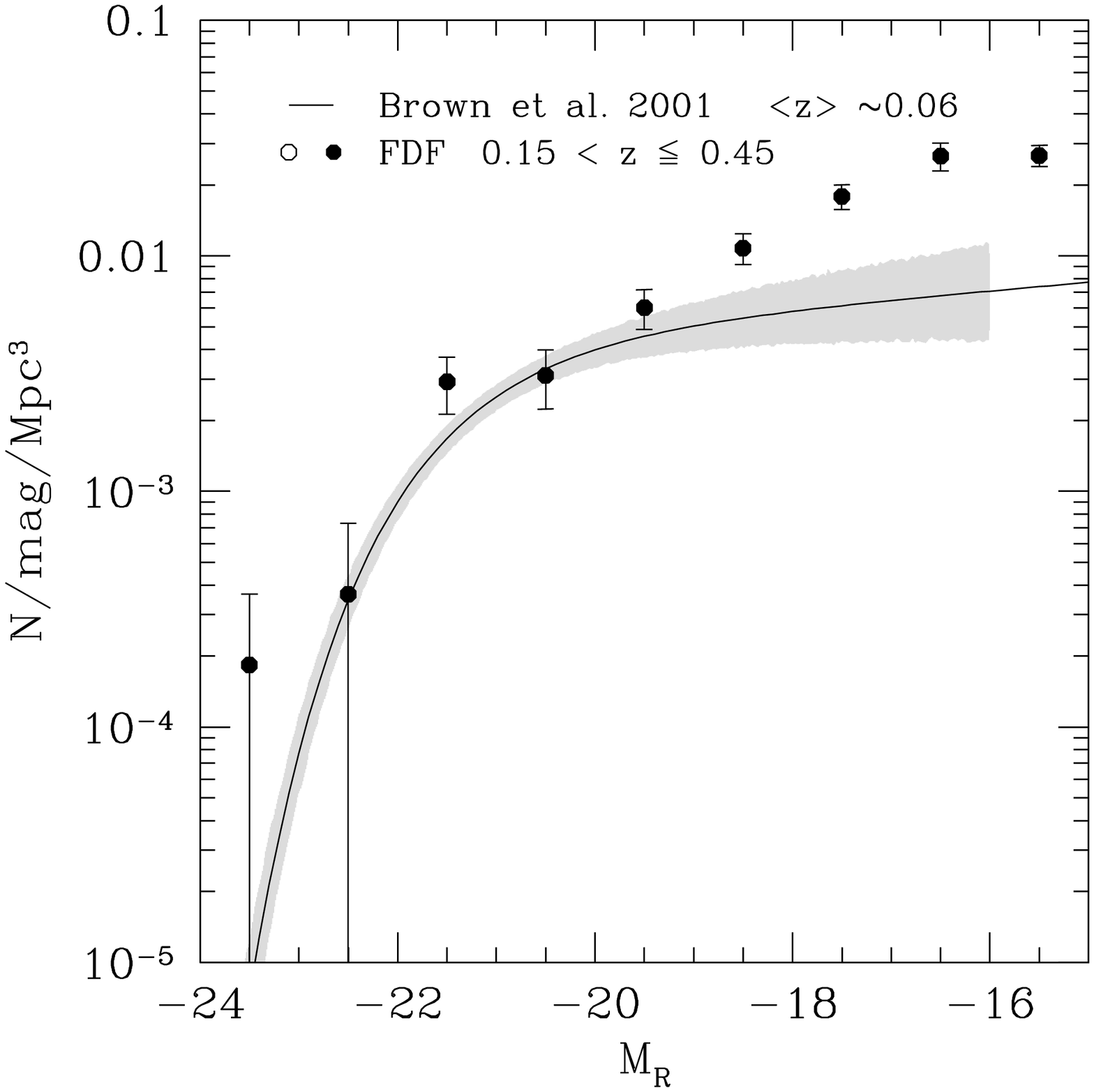}
\caption[Comparison of the FDF LF with \textit{\citet{brown:1}}]
{\label{fig:lfred:brown} Comparison of the V-band (left panel) and
  R-band (right panel) LF of the FDF with the local
  (\mbox{$\langle z \rangle\sim 0.06$}) Schechter function derived in
  \textit{\citet{brown:1}}. The shaded region is based on
  $\Delta$M$^\ast$, $\Delta\phi^\ast$, and $\Delta\alpha$.  }
\end{figure*}

\noindent\textit{\citet{brown:1}:}\\
\citet{brown:1} use 64 deg$^2$ of V and R images to measure the local
V- and R-band LF. They analyzed about 1250 V \& R selected galaxies
from the Century Survey \citep{geller:1} with a mean spectroscopic
redshift of \mbox{$\langle z \rangle\sim 0.06$}.\\
A comparison between the LF of \citet{brown:1} and the FDF is shown in
Fig.~\ref{fig:lfred:brown} for the V-band (left panel) and the R-band
(right panel). Although the agreement is quite good for the bright
end, the number density of the faint-end is substantially higher in
the FDF (while the slope of the LF derived in the FDF is
$\alpha=-1.25$, the slope derived by \citet{brown:1} is
$\alpha=-1.09 \pm 0.09$ in the V- as well as in the R-band).\\

\begin{figure}[btp]
\centering
\includegraphics[width=0.50\textwidth]{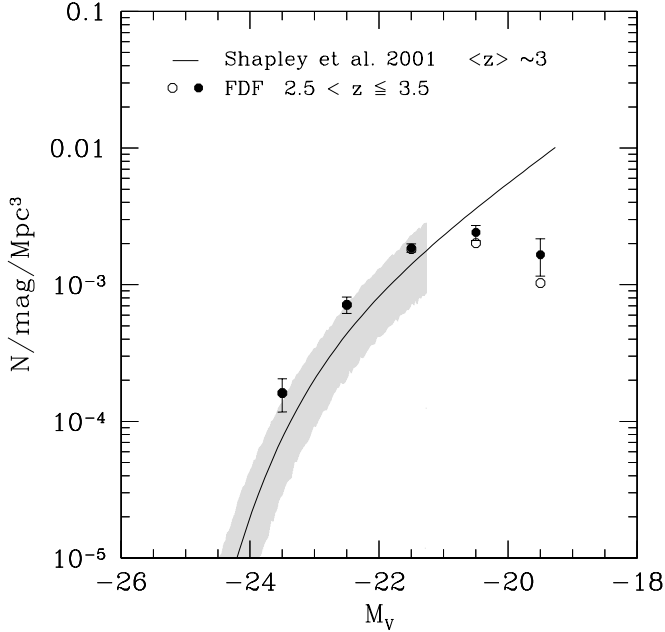}
\caption[Comparison of the FDF luminosity
  function with \textit{\citet{shapley:1}}]
{\label{fig:lfred:shapley} Comparison of the V-band luminosity
  function of the FDF with the Schechter function derived in
  \textit{\citet{shapley:1}} at \mbox{$\langle z \rangle\sim 3.0$}.
  The shaded region is based on $\Delta$M$^\ast$, $\Delta\phi^\ast$,
  and $\Delta\alpha$, where the cut-off at low luminosity indicates
  the limiting magnitude of the sample.}
\end{figure}

\noindent\textit{\citet{shapley:1}:}\\
\citet{shapley:1} analyzed 118 photometrically selected LBGs with
K$_{s}$-band measurements covering an area of 30 arcmin$^2$.  63 galaxies
have additional J-band measurements and 81 galaxies are
spectroscopically confirmed. Using this sample \citet{shapley:1}
derived the luminosity function in the restframe V-band at redshift of
\mbox{$\langle z \rangle\sim 3.0$}. Fig.~\ref{fig:lfred:shapley} shows
a comparison of the V-band LF derived by \citet{shapley:1} with the LF
in the FDF at \mbox{$\langle z \rangle\sim 3.0$}.  The agreement is
very good if we again concentrate on the shaded region. On the other
hand, because of the depth of the FDF we can trace the LF two
magnitudes deeper and therefore give better constraints on the slope
of the Schechter function.  Comparing the faint-end of the FDF LF with
the extrapolated Schechter function of \citet{shapley:1} clearly
shows, that the very steep slope of $\alpha=-1.85$ is not seen
in the FDF dataset.\\

\begin{figure*}[tbp]
\includegraphics[width=0.30\textwidth]{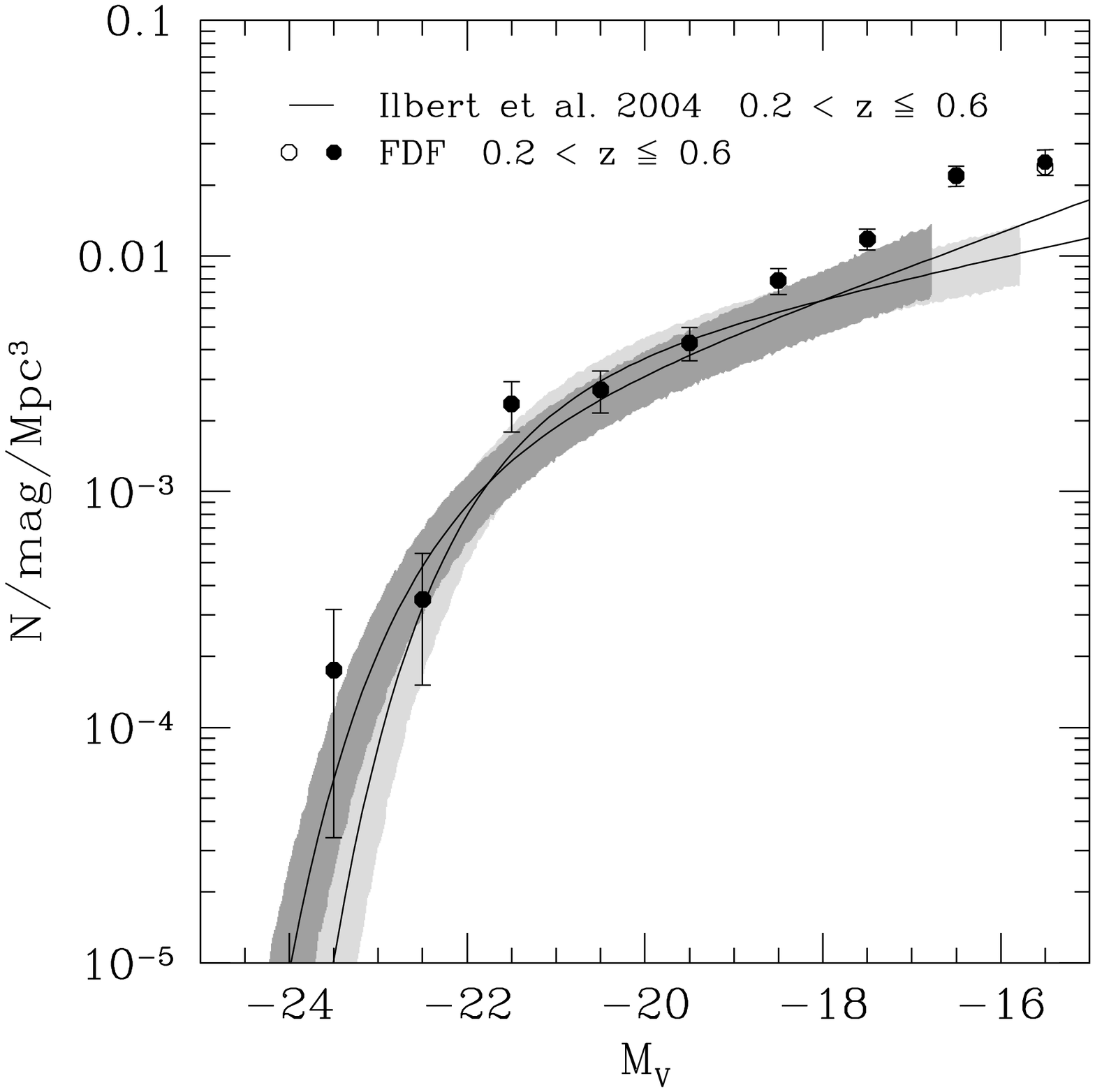}
\includegraphics[width=0.30\textwidth]{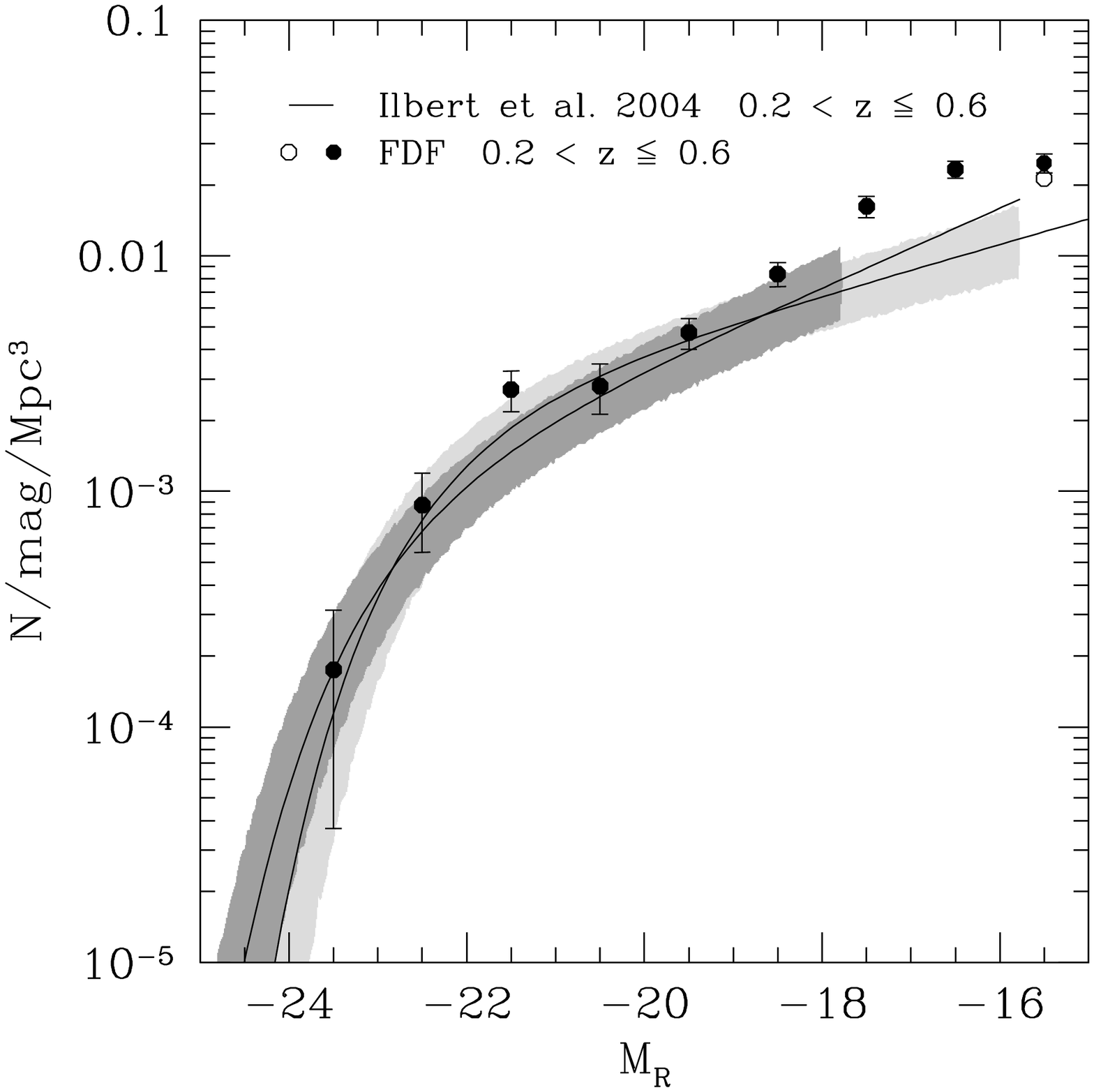}
\includegraphics[width=0.30\textwidth]{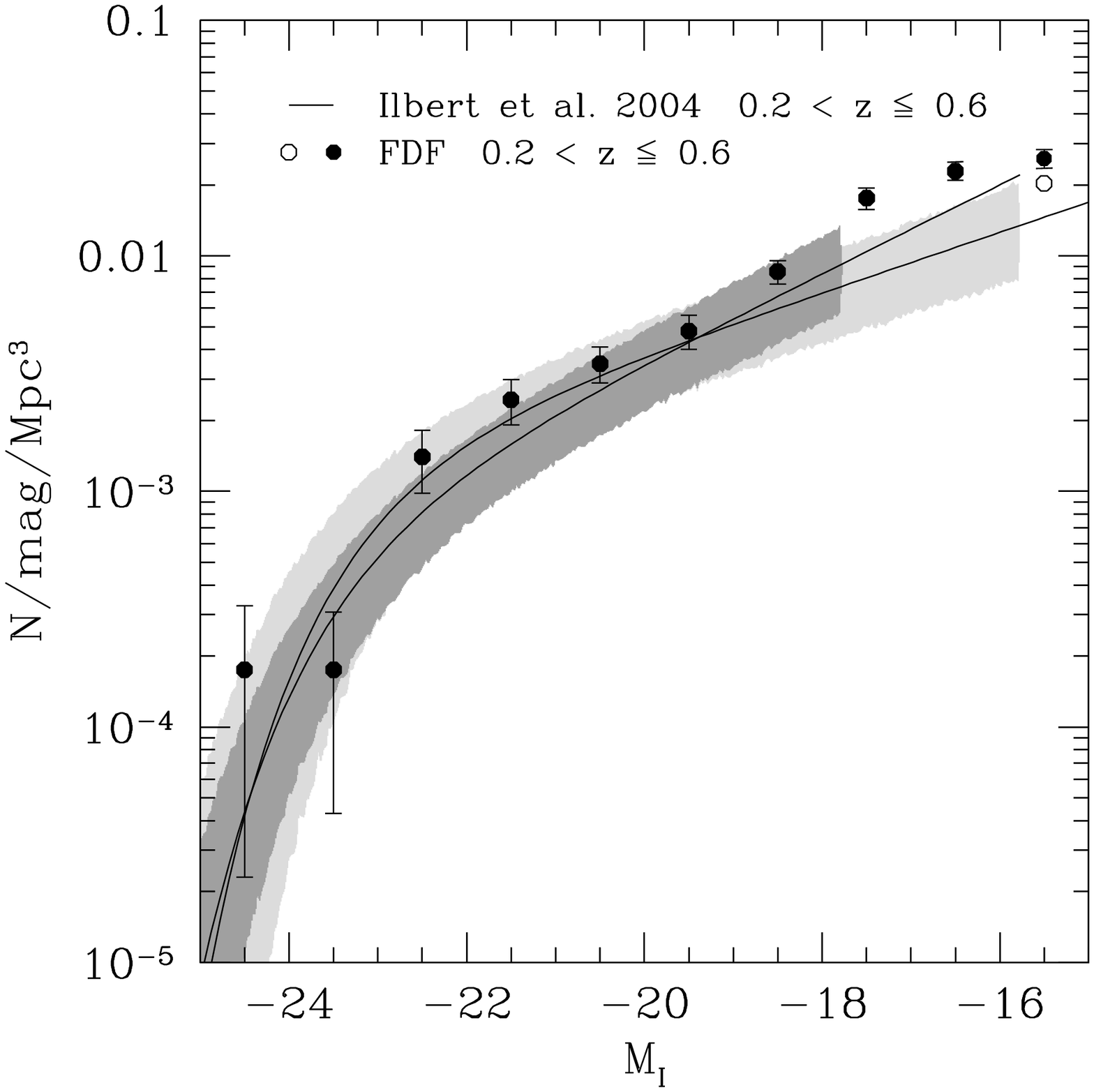}
\includegraphics[width=0.30\textwidth]{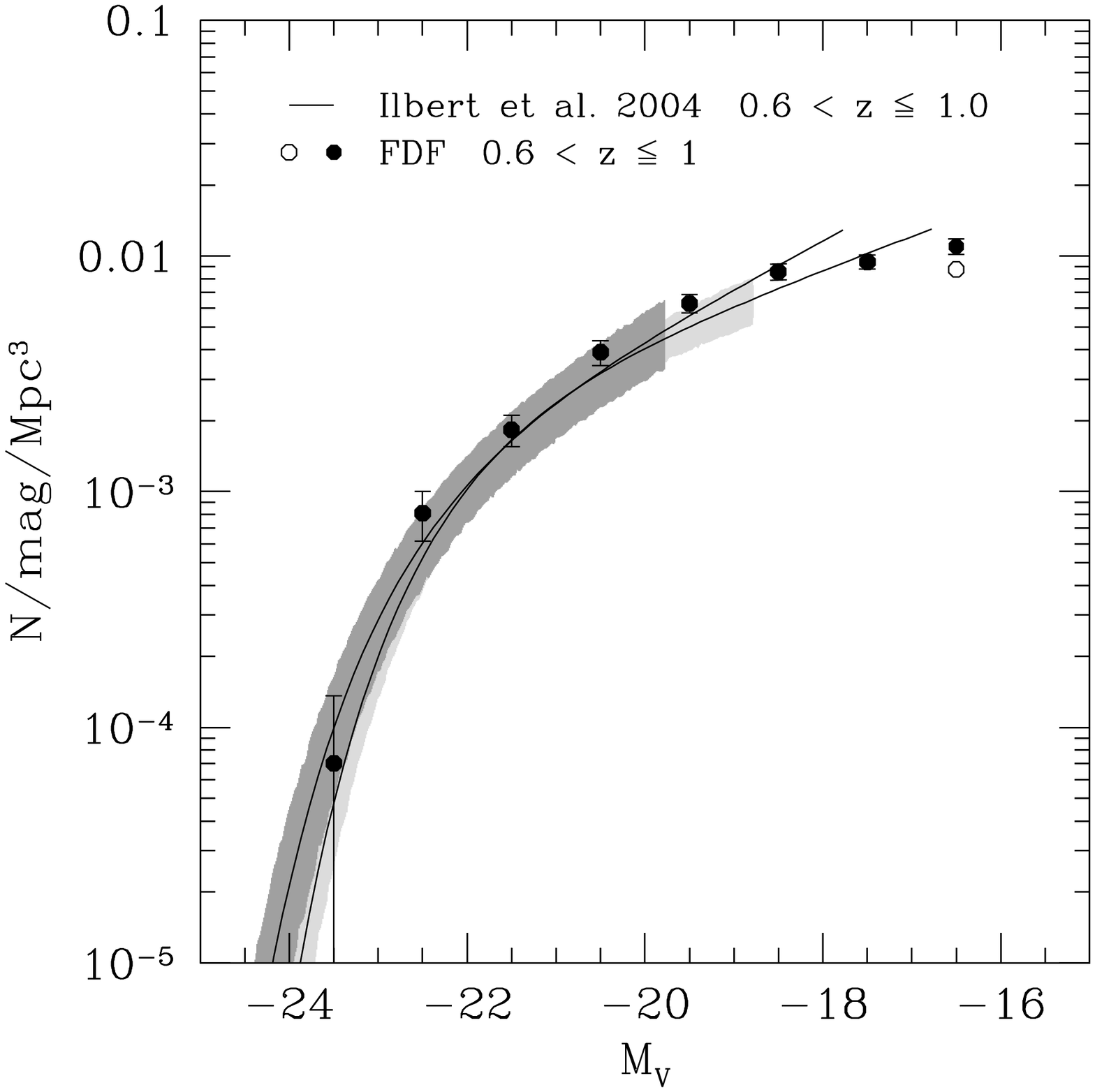}
\includegraphics[width=0.30\textwidth]{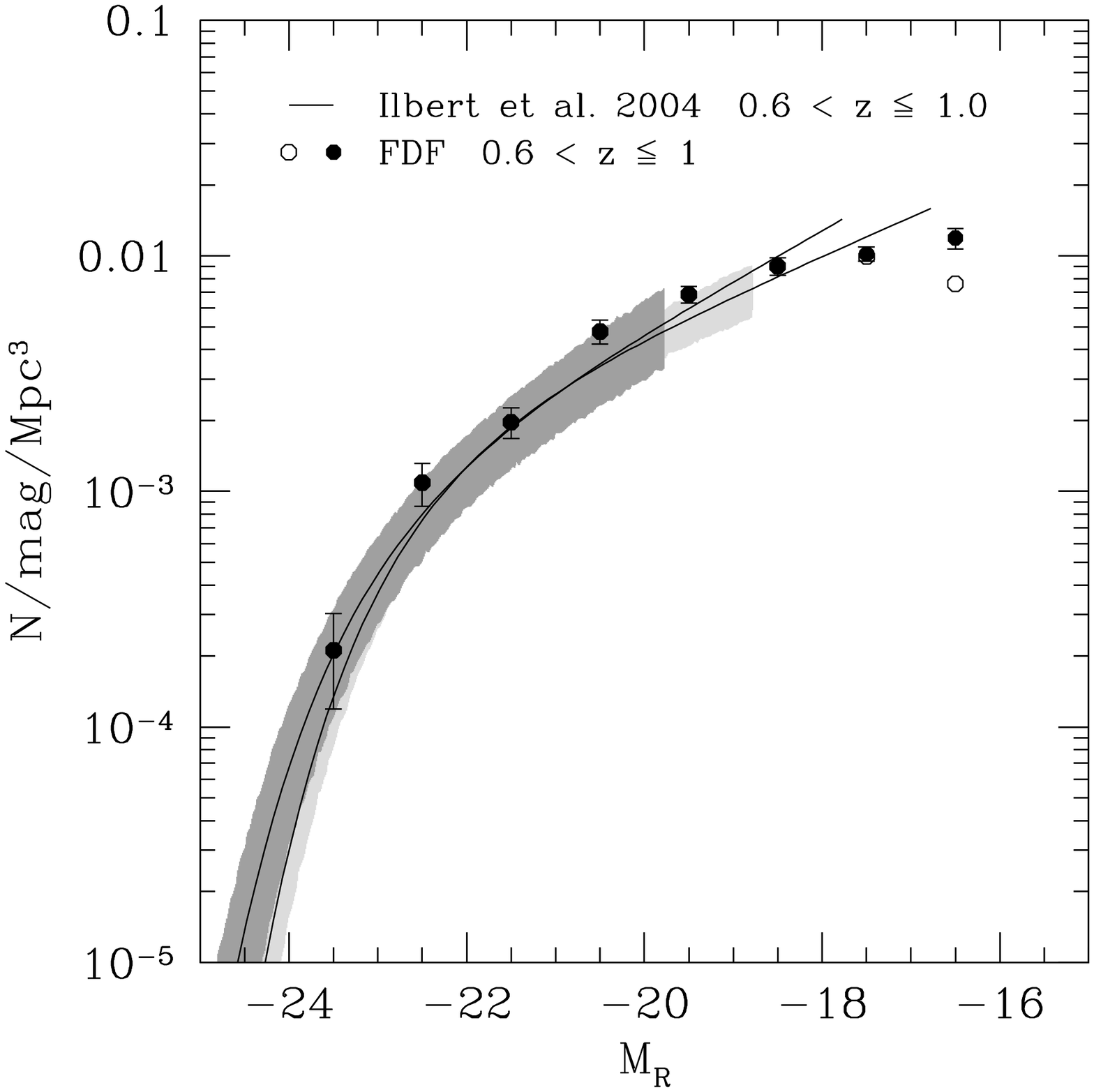}
\includegraphics[width=0.30\textwidth]{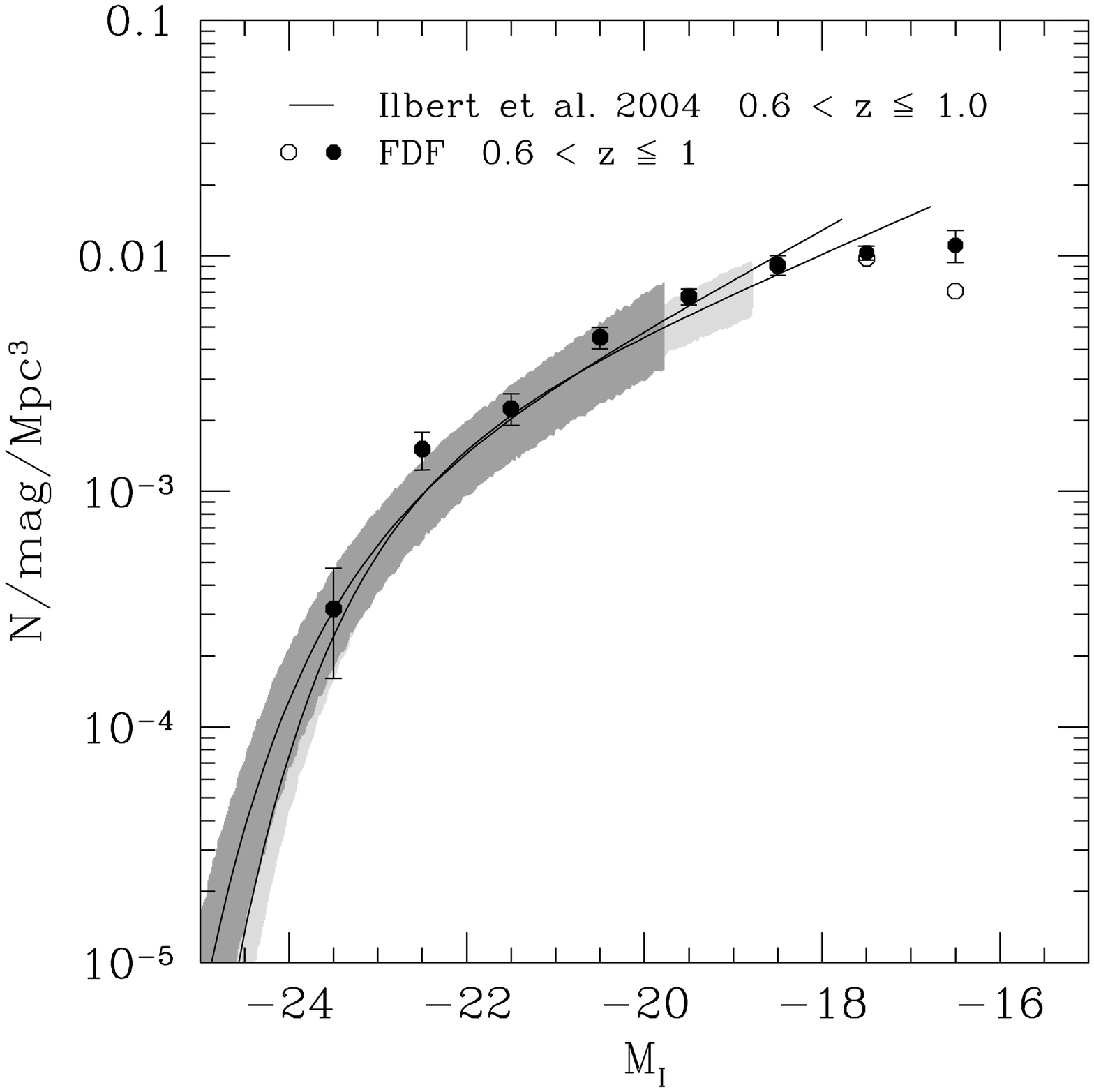}
\includegraphics[width=0.30\textwidth]{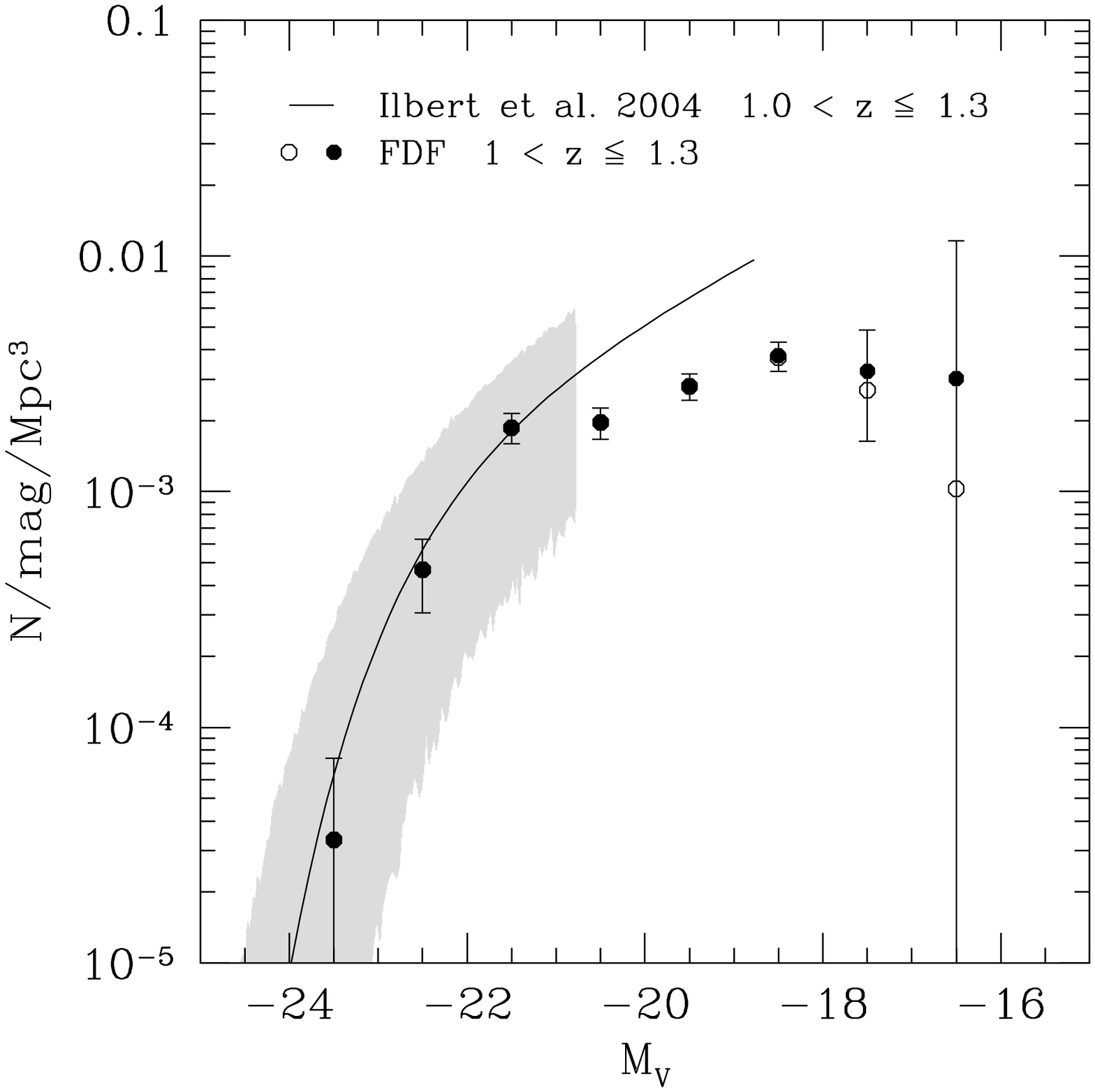}
\includegraphics[width=0.30\textwidth]{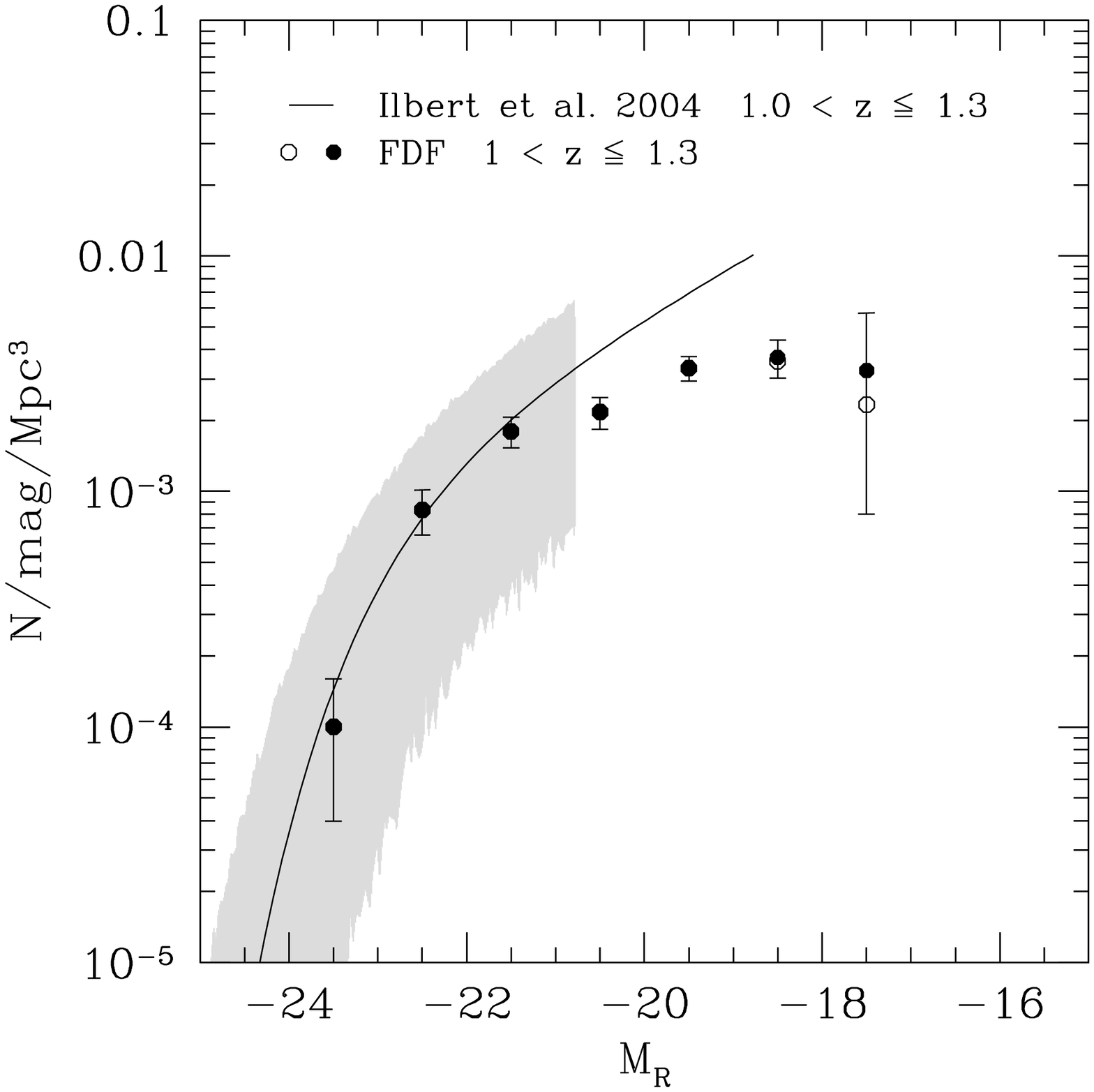}
\includegraphics[width=0.30\textwidth]{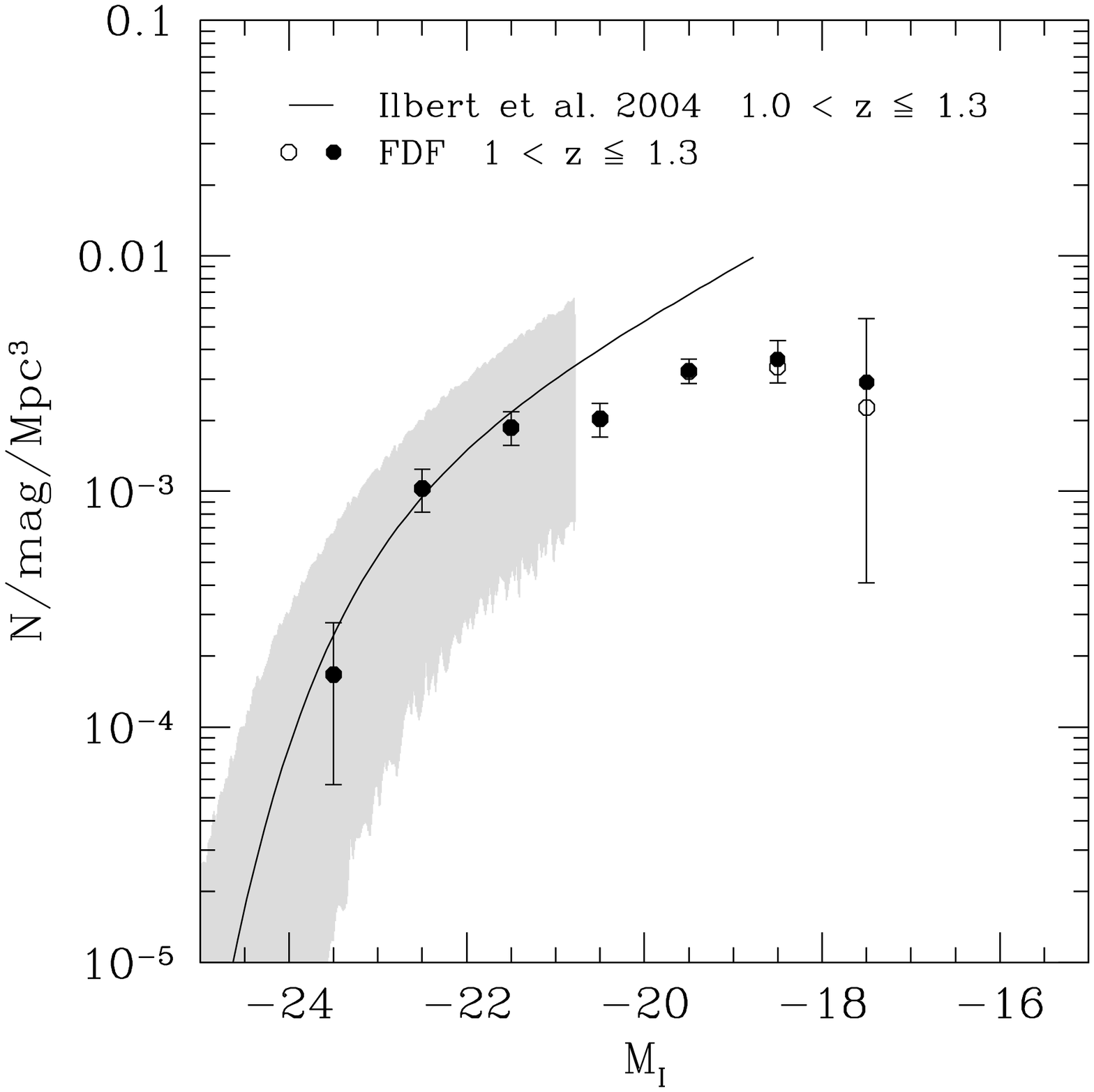}
\includegraphics[width=0.30\textwidth]{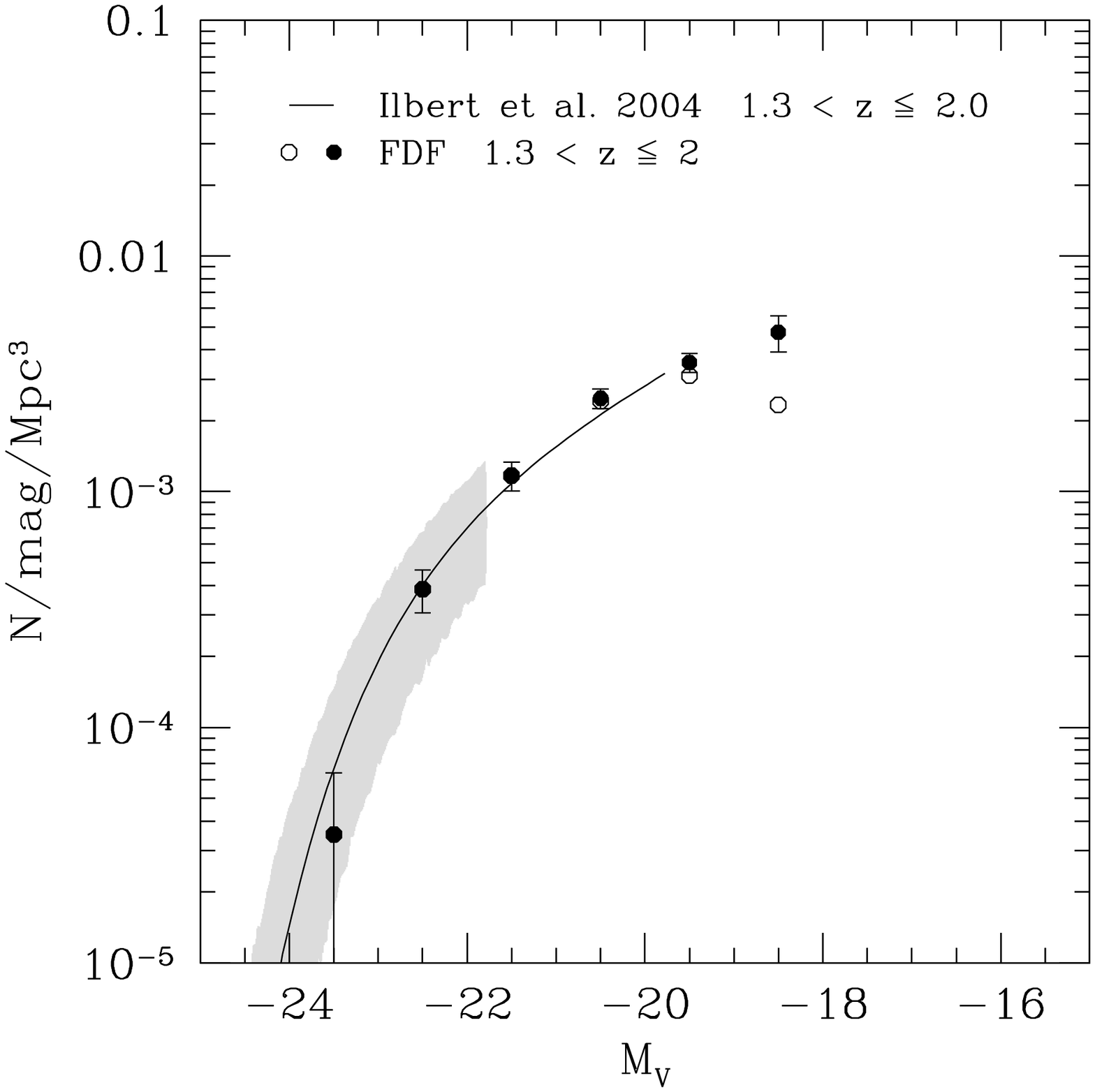}
\hfill
\includegraphics[width=0.30\textwidth]{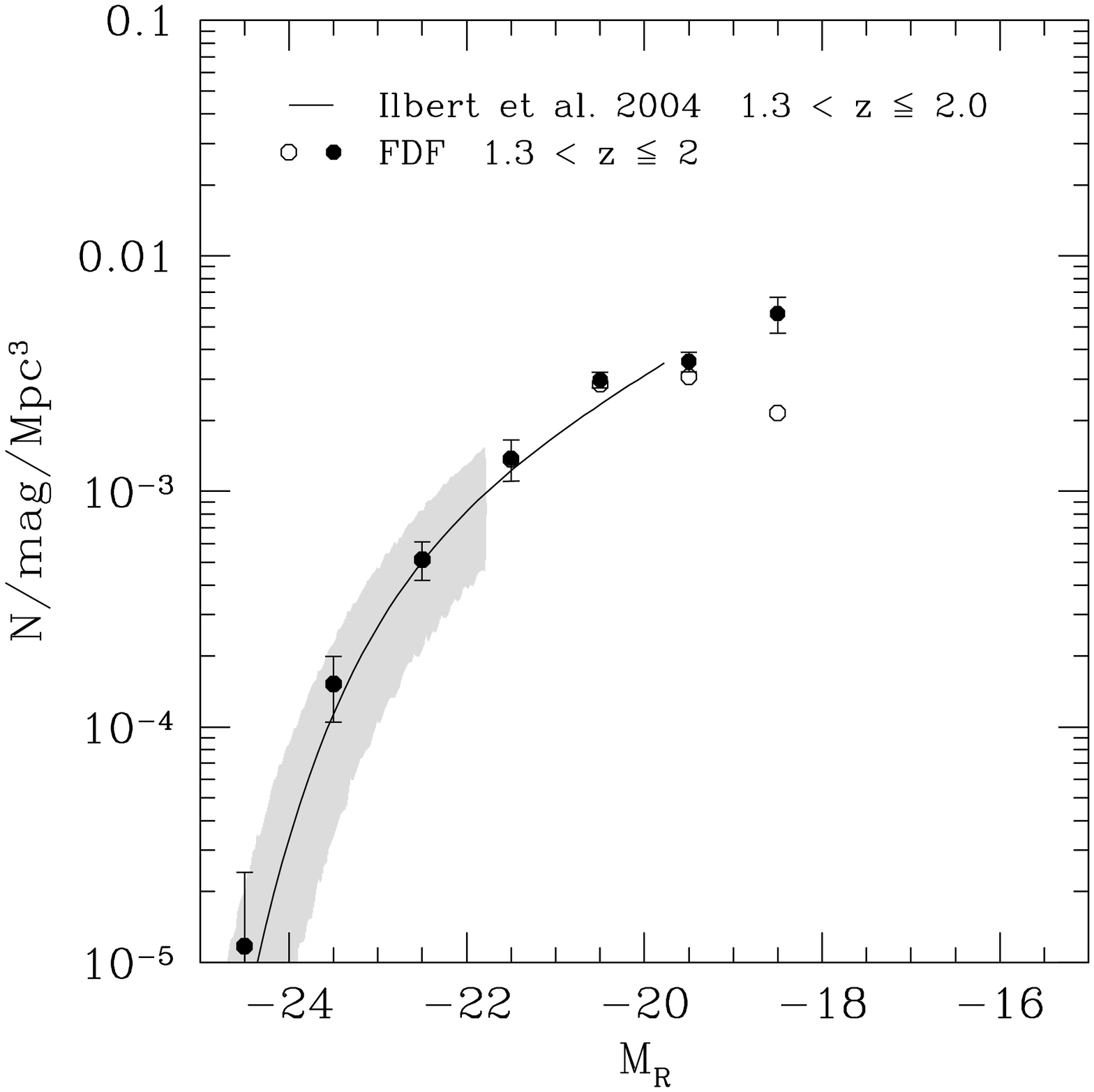}
\hfill
\includegraphics[width=0.30\textwidth]{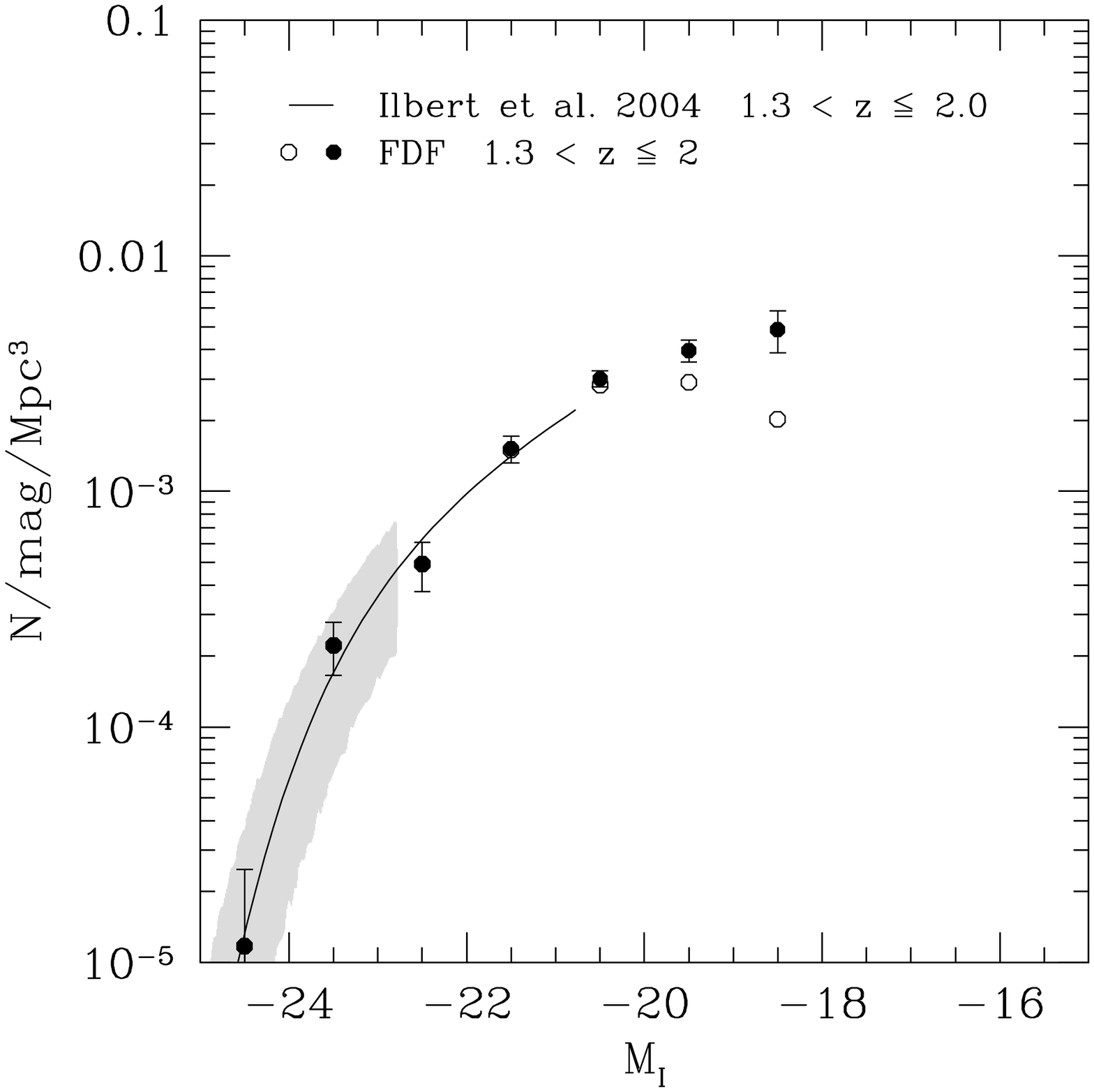}
\caption {\label{fig:lfred:lit_VRI_ilbert} Comparison of the V (left panels), R
  (middle panels), and I (right panels) band LF of the FDF with the
  Schechter function derived in \textit{\citet{ilbert:2}} (VVDS) at
\mbox{$0.20 <z\le 0.40$} (first row, light gray), 
\mbox{$0.40 <z\le 0.60$} (first row, dark  gray),
\mbox{$0.60 <z\le 0.80$} (second row, light gray),
\mbox{$0.80 <z\le 1.00$} (second row, dark  gray),
\mbox{$1.00 <z\le 1.30$} (third row), and 
\mbox{$1.30 <z\le 2.00$} (fourth row).
The shaded regions of all plots with $z \le 1$ are based on
$\Delta$M$^\ast$, $\Delta\phi^\ast$, and $\Delta\alpha$. Only in the
two high redshift bins (third and fourth row) the shaded region is
based only on $\Delta$M$^\ast$ and $\Delta\phi^\ast$. Please note that
we use the average error if the upper and lower values reported by
\citet{ilbert:2} disagree.}
\end{figure*}

\noindent\textit{\citet{ilbert:2}:}\\
\citet{ilbert:2} investigated the evolution of the galaxy LF from the
VIMOS-VLT Deep Survey (VVDS) in 5 restframe bands (U, B, V, R, I).
They used about 11000 objects with spectroscopic distance information
in the magnitude range $17.5 \le I \le 24.0$ to constrain the LF to
redshift $z \sim 2$.  In Fig.~\ref{fig:lfred:lit_VRI_ilbert} we
compare the V, R, and I band LF of the FDF with the Schechter function
derived in the VVDS survey for different redshift bins: \mbox{$0.20 <z
  \le 0.40$}, \mbox{$0.40 <z \le 0.60$}, \mbox{$0.60 <z \le 0.80$},
\mbox{$0.80 <z \le 1.00$}, \mbox{$1.00 <z \le 1.30$}, and \mbox{$1.30
  <z \le 2.00$}. Because of the limited sample size of the FDF at low
redshift we could not use the same local redshift binning as
\citet{ilbert:2}.  We therefore compare in
Fig.~\ref{fig:lfred:lit_VRI_ilbert} (first row) the VVDS Schechter
function at \mbox{$\langle z \rangle\sim 0.3$} (light gray) and
\mbox{$\langle z \rangle\sim 0.5$} (dark gray) with the FDF LF derived
at $0.2 <z\le 0.6$ as well as in Fig.~\ref{fig:lfred:lit_VRI_ilbert}
(second row) the Schechter function at \mbox{$\langle z \rangle\sim
  0.7$} (light gray) and \mbox{$\langle z \rangle\sim 0.9$} (dark
gray) with the FDF LF derived at $0.6 <z\le 1.0$.  There is a very
good agreement between the FDF data and the VVDS survey at all
redshifts under investigation if we compare only the magnitude range
in common to both surveys (shaded region).  \citet{ilbert:2} derived
the faint-end slope from shallower data if compared with the FDF which
have only a limited sensitivity for the latter. Nevertheless, in all
three bands the differences between the formal $\alpha$ derived in the
FDF ($\alpha_{V}=-1.25 \pm 0.03$ and $\alpha_{r' \& i'}=-1.33 \pm
0.03$ constant in redshift) and in the VVDS are compatible within
$1\sigma$ to $2\sigma$ up to redshift $z \sim 0.8$ (only one bin in
the I-band LF differs by slightly more than $2\sigma$). At higher
redshift we do not see the steep slope ($\sim -1.5$) as derived by
\citet{ilbert:2}.  The circumstance that in the FDF we are able to
follow the LF about \mbox{3 -- 4} magnitudes deeper may explain the
disagreement between the extrapolated faint-end slope of
\citet{ilbert:2} and the FDF result.\\

\begin{figure*}
\includegraphics[width=0.45\textwidth]{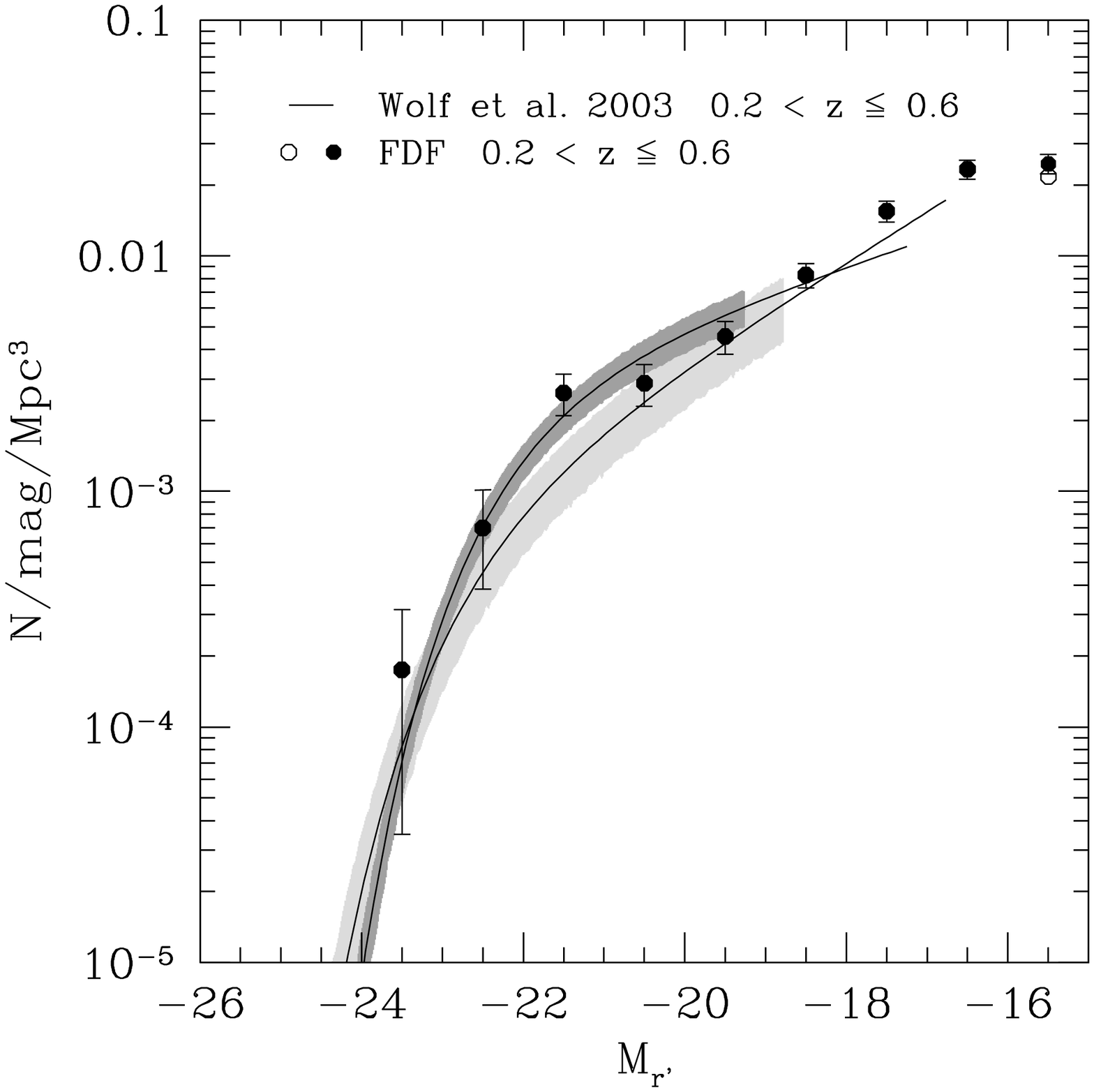}
\hfill
\includegraphics[width=0.45\textwidth]{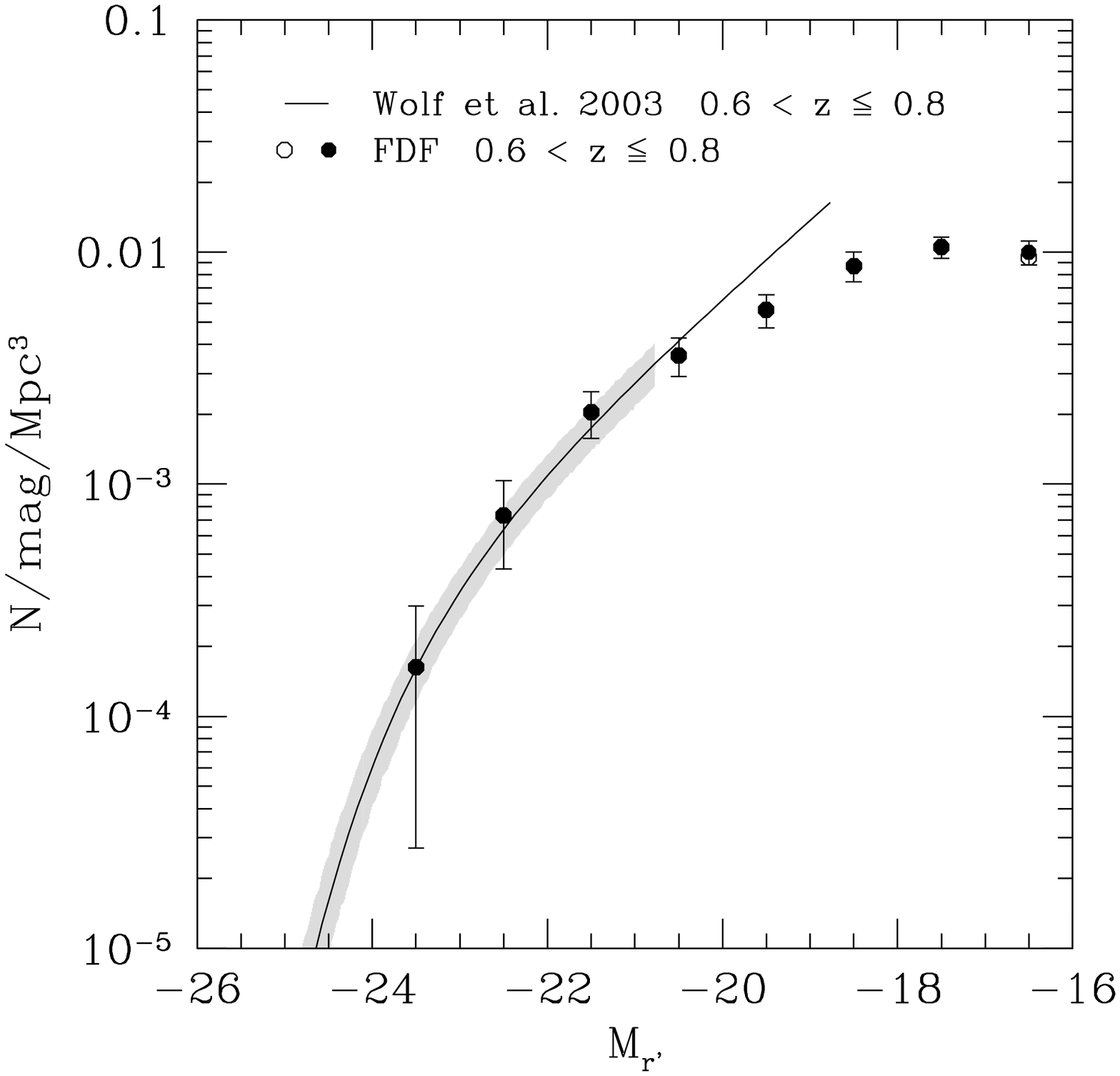}
\includegraphics[width=0.45\textwidth]{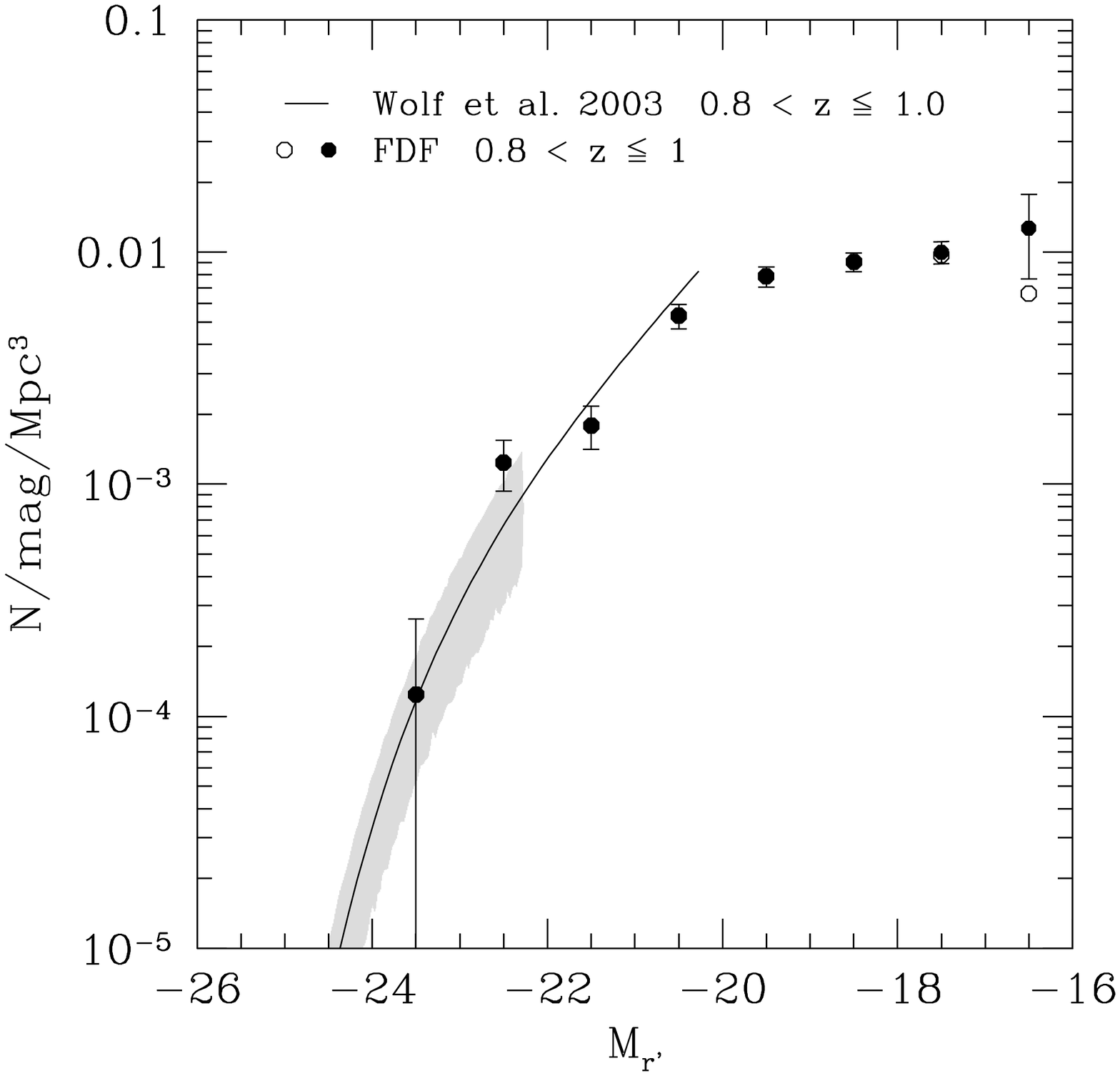}
\hfill
\includegraphics[width=0.45\textwidth]{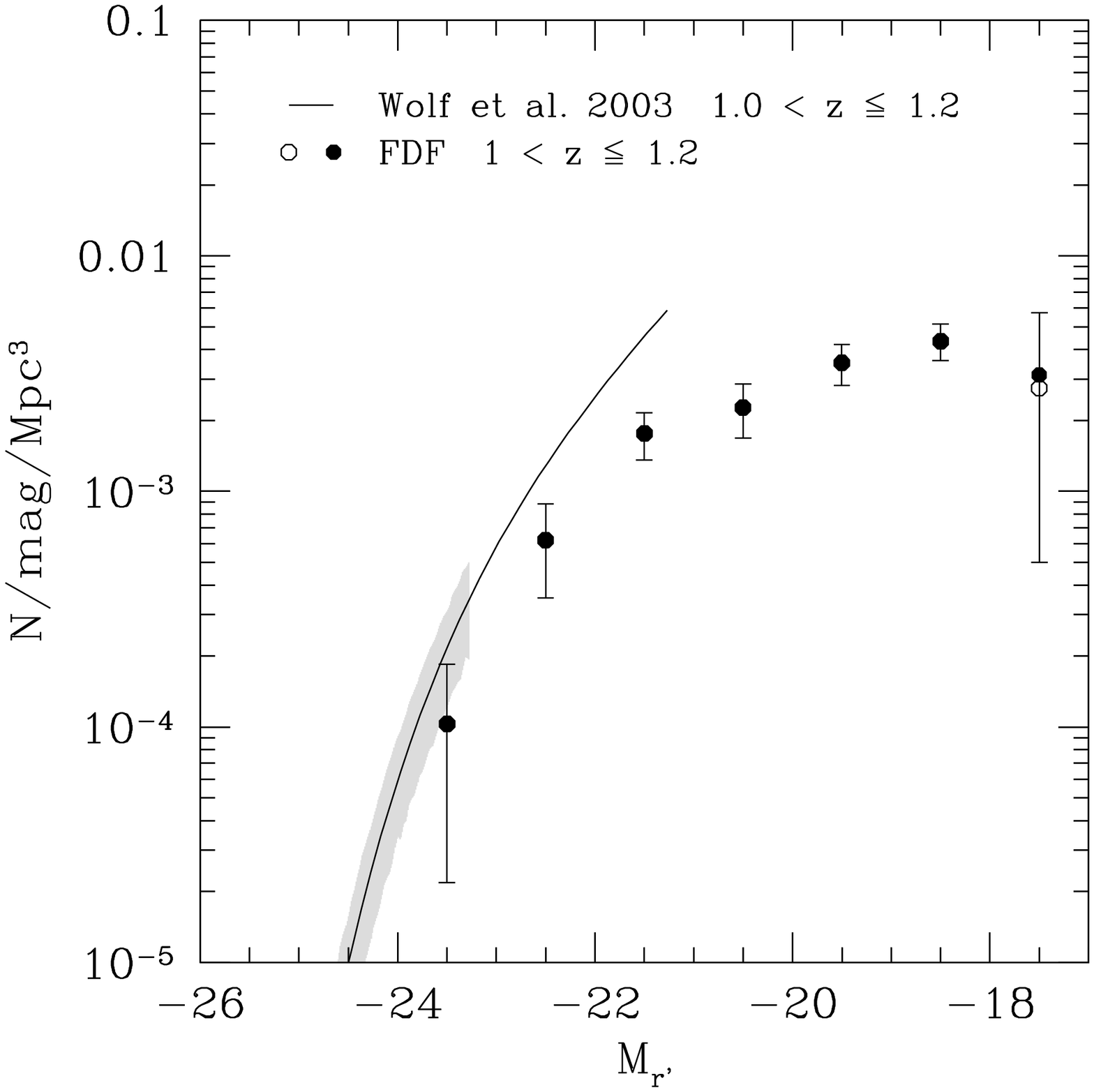}
\caption[Comparison of the FDF LF with \textit{\citet{combo17:1}}]
{\label{fig:lfred:lit_r_SDSS_combo}
Comparison of the LF in the r'-band of the FDF 
with the Schechter function derived in
\textit{\citet{combo17:1}}:
\mbox{$0.2 <z\le 0.4$} (upper left panel, light gray), 
\mbox{$0.4 <z\le 0.6$} (upper left panel, dark grey), 
\mbox{$0.6 <z\le 0.8$} (upper right panel), 
\mbox{$0.8 <z\le 1.0$} (lower left panel), 
\mbox{$1.0 <z\le 1.2$} (lower right panel). 
The shaded regions of nearly all plots are based on $\Delta$M$^\ast$,
$\Delta\phi^\ast$, and $\Delta\alpha$. Only in the highest redshift
bin (lower right panel) the shaded region is based only on
$\Delta$M$^\ast$ and $\Delta\phi^\ast$. The cut-off at low luminosity
indicates the limiting magnitude of the sample.  }
\end{figure*}

\noindent\textit{\citet{combo17:1}:}\\
In Fig.~\ref{fig:lfred:lit_r_SDSS_combo} we compare the r'-band LF of
the FDF with the R-band selected luminosity function derived in the
COMBO-17 survey \citep{combo17:1} for different redshift bins: 0.2 --
0.6, 0.6 -- 0.8, 0.8 -- 1.0, 1.0 -- 1.2.  Because of the limited
sample size of the FDF at low redshift we could not use the same local
redshift binning as \citet{combo17:1}.  We compare therefore in
Fig.~\ref{fig:lfred:lit_r_SDSS_combo} (upper left panel) the COMBO17
Schechter function at \mbox{$\langle z \rangle\sim 0.3$} (light gray)
and \mbox{$\langle z \rangle\sim 0.5$} (dark gray) with the FDF LF
derived at $0.2 <z\le 0.6$.  There is a very good agreement between
the FDF data and the COMBO-17 survey at all redshifts under
investigations if we compare only the magnitude range in common to
both surveys.  Although the number density of the FDF seems to be
slightly higher for the restframe UV LF (FDFLF~I), this is not the
case if we compare the LF in the R-band.  \citet{combo17:1} derived
the faint-end slope from relatively shallow data which have only a
limited sensitivity for the latter. This may explain the disagreement
between the extrapolated faint-end slope of
\citet{combo17:1} and the FDF result.\\

\noindent\textit{\citet{chen:1}:}\\
The galaxy sample analyzed by \citet{chen:1} contains $\sim 6700$
H-band selected galaxies (within 847 arcmin$^2$) in the HDFS region
with complementary optical U, B, V, R, and I colors, and $\sim 7400$
H-band selected galaxies (within 561 arcmin$^2$) in the Chandra Deep
Field South region with complementary optical V, R, I, and z' colors.
The galaxy sample is part of the Las Campanas Infrared Survey
\citep[LCIR][]{marzke:1, mccarthy:1} and based on photometric
redshifts.

\begin{figure*}
\includegraphics[width=0.33\textwidth]{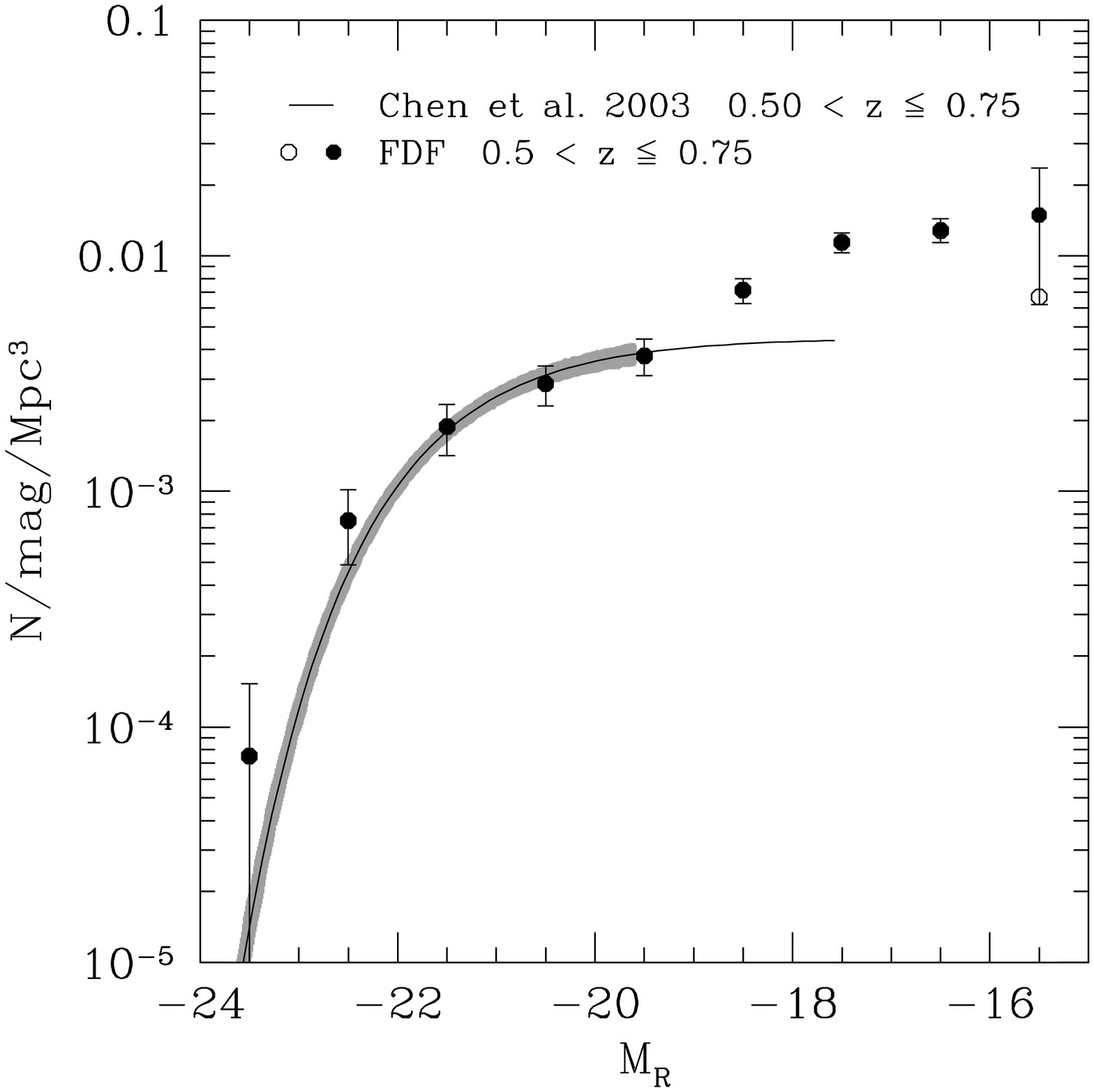}
\includegraphics[width=0.33\textwidth]{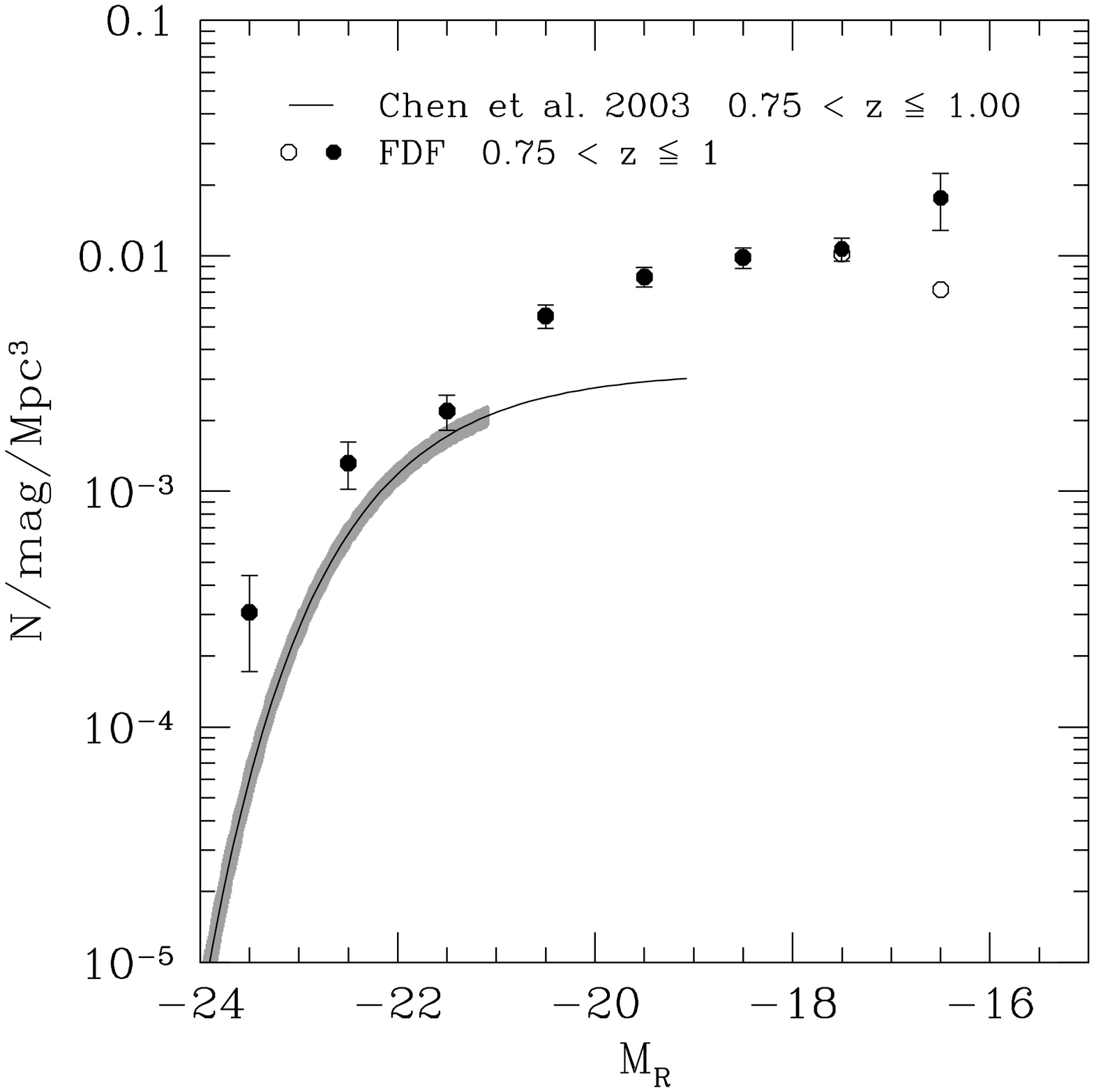}
\includegraphics[width=0.33\textwidth]{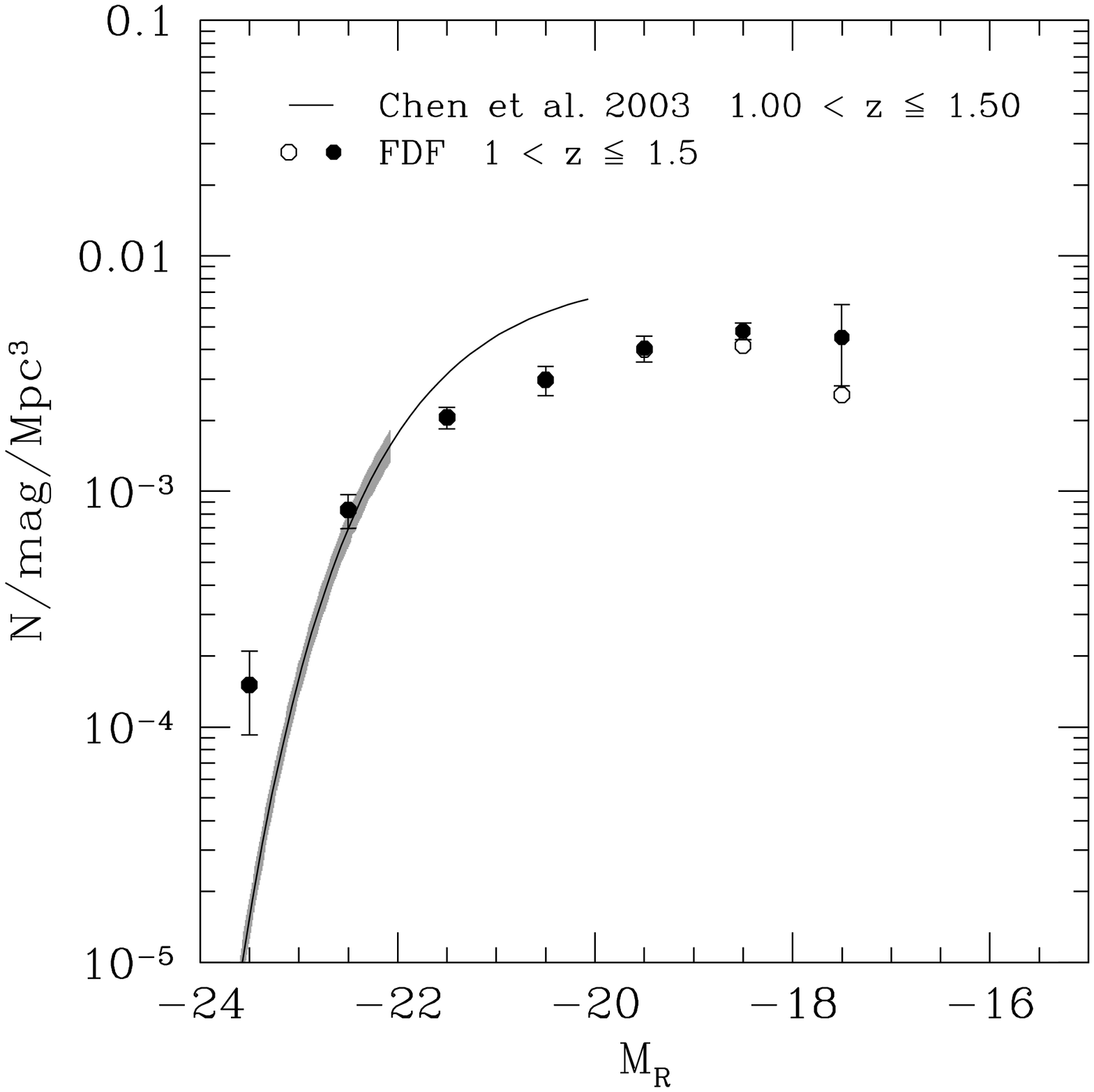}
\caption[Comparison of the FDF luminosity
  function with \textit{\citet{chen:1}}]
{\label{fig:lfred:chen}Comparison of the luminosity
  function in the R-band of the FDF with the Schechter function
  derived in \textit{\citet{chen:1}}:
\mbox{$0.50 <z\le 0.75$} (left panel), 
\mbox{$0.75 <z\le 1.00$} (middle panel), and  
\mbox{$1.00 <z\le 1.50$} (right panel). 
The shaded region is based on $\Delta$M$^\ast$, $\Delta\phi^\ast$ and
$\Delta\alpha$ for \mbox{$0.50 <z\le 0.75$} (left panel).  For
\mbox{$0.75 <z\le 1.00$} (middle panel) and \mbox{$1.00 <z\le 1.50$}
(right panel) the shaded region is based only on $\Delta$M$^\ast$ and
$\Delta\phi^\ast$. The cut-off at low luminosity indicates the
limiting magnitude of the sample.  }
\end{figure*}

Fig.~\ref{fig:lfred:chen} shows a comparison of the R-band luminosity
function derived by \citet{chen:1} with the LF in the FDF for three
different redshift bins: 0.50--0.75 (left panel), 0.75--1.00 (middle
panel), and 1.00--1.50 (right panel).  There is a good agreement
between the FDF LF and the Schechter function derived by
\citet{chen:1} in the lowest redshift bin (\mbox{$z=0.50$--$0.75$}) if
we compare only the magnitude range in common to both surveys.  At
intermediate redshift (\mbox{$z=0.75$--$1.00$}) the number density of
the bright end of the FDF LF is slightly higher than in \citet{chen:1}.
On the other hand, for the highest redshift bin
(\mbox{$z=1.00$--$1.50$}) the number density of the bright end derived
by \citet{chen:1} roughly agrees with the
results obtained in the FDF.\\

\begin{figure*}[tbp]
\includegraphics[width=0.33\textwidth]{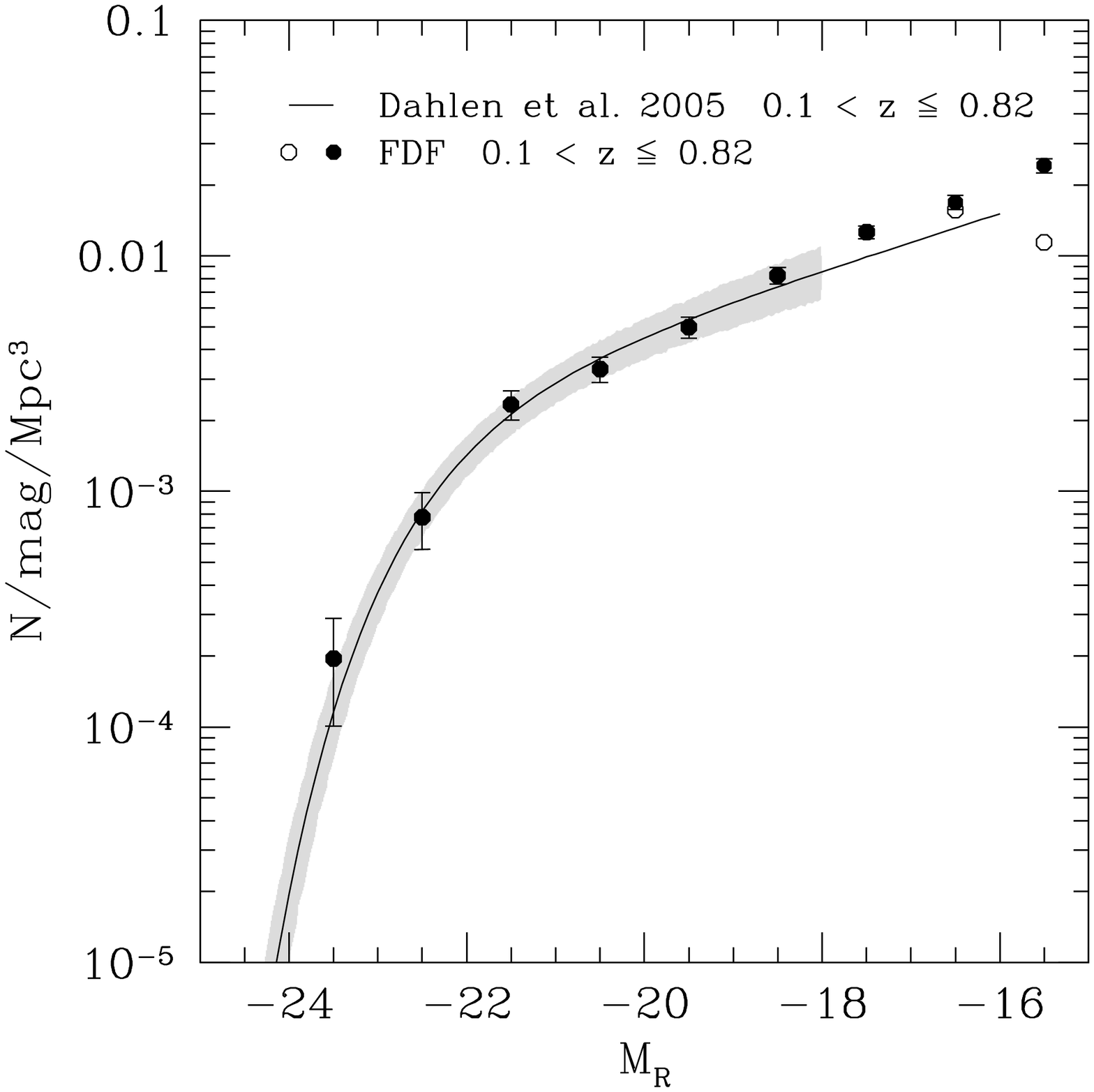}
\includegraphics[width=0.33\textwidth]{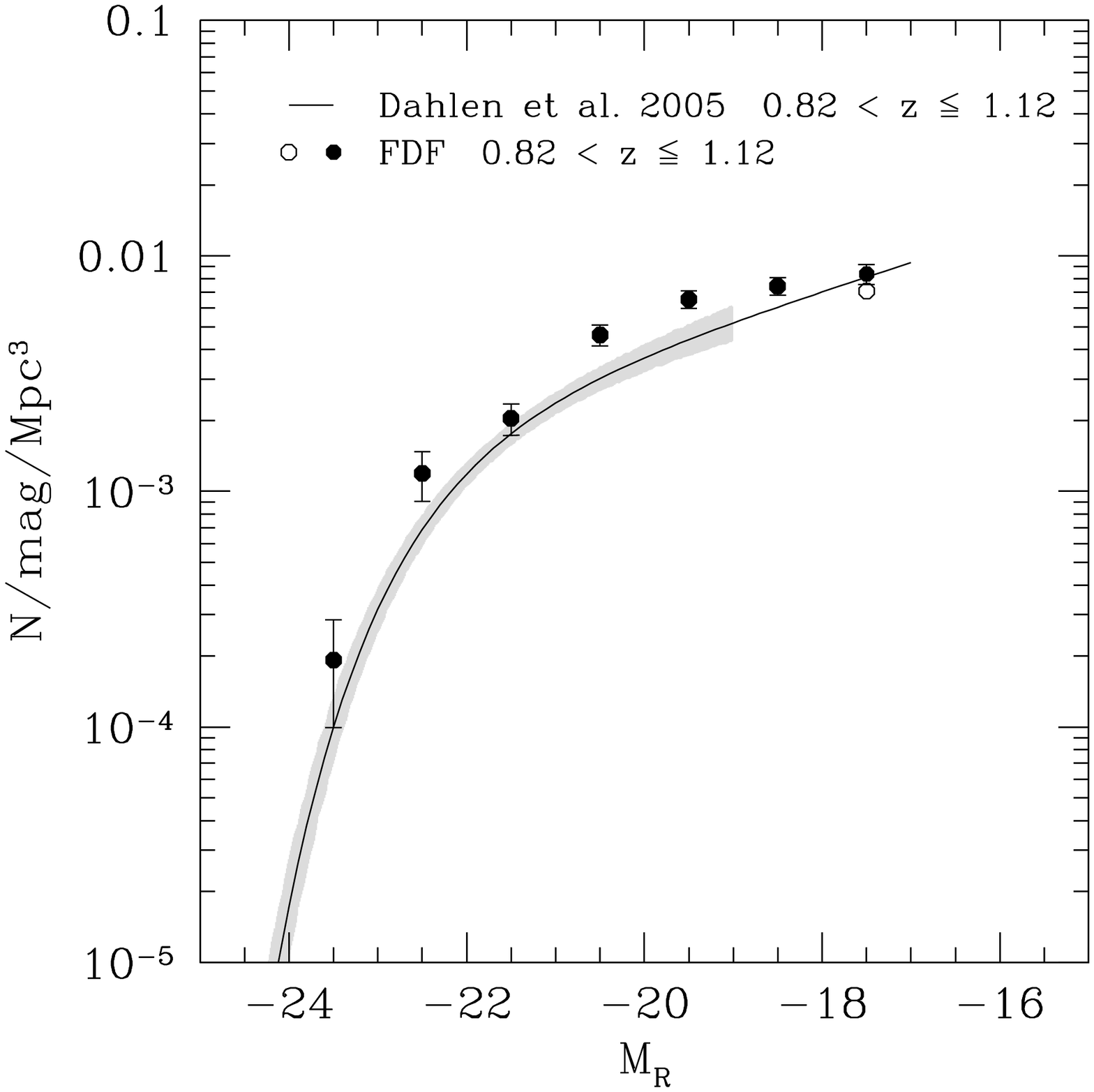}
\includegraphics[width=0.33\textwidth]{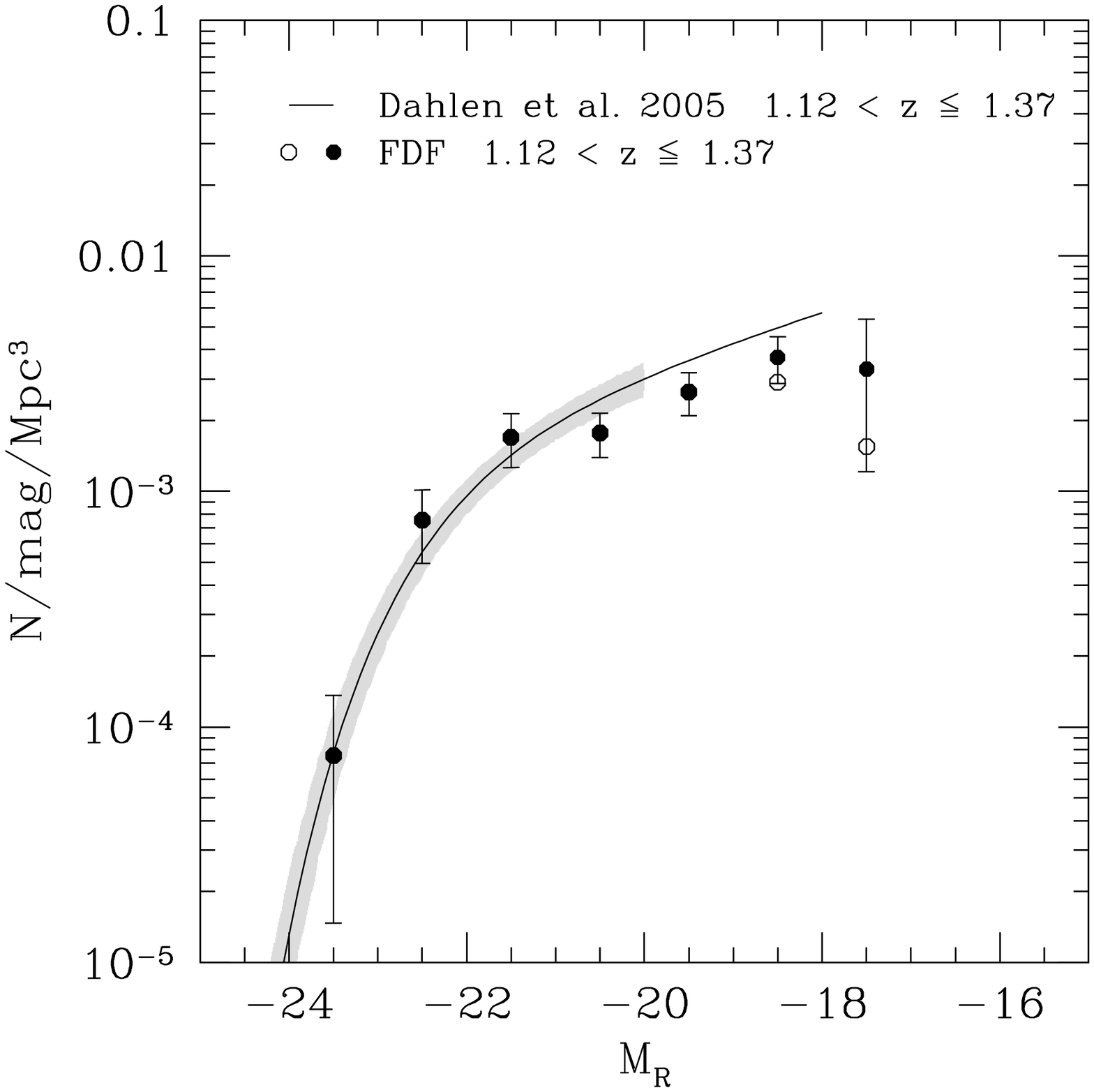}
\includegraphics[width=0.33\textwidth]{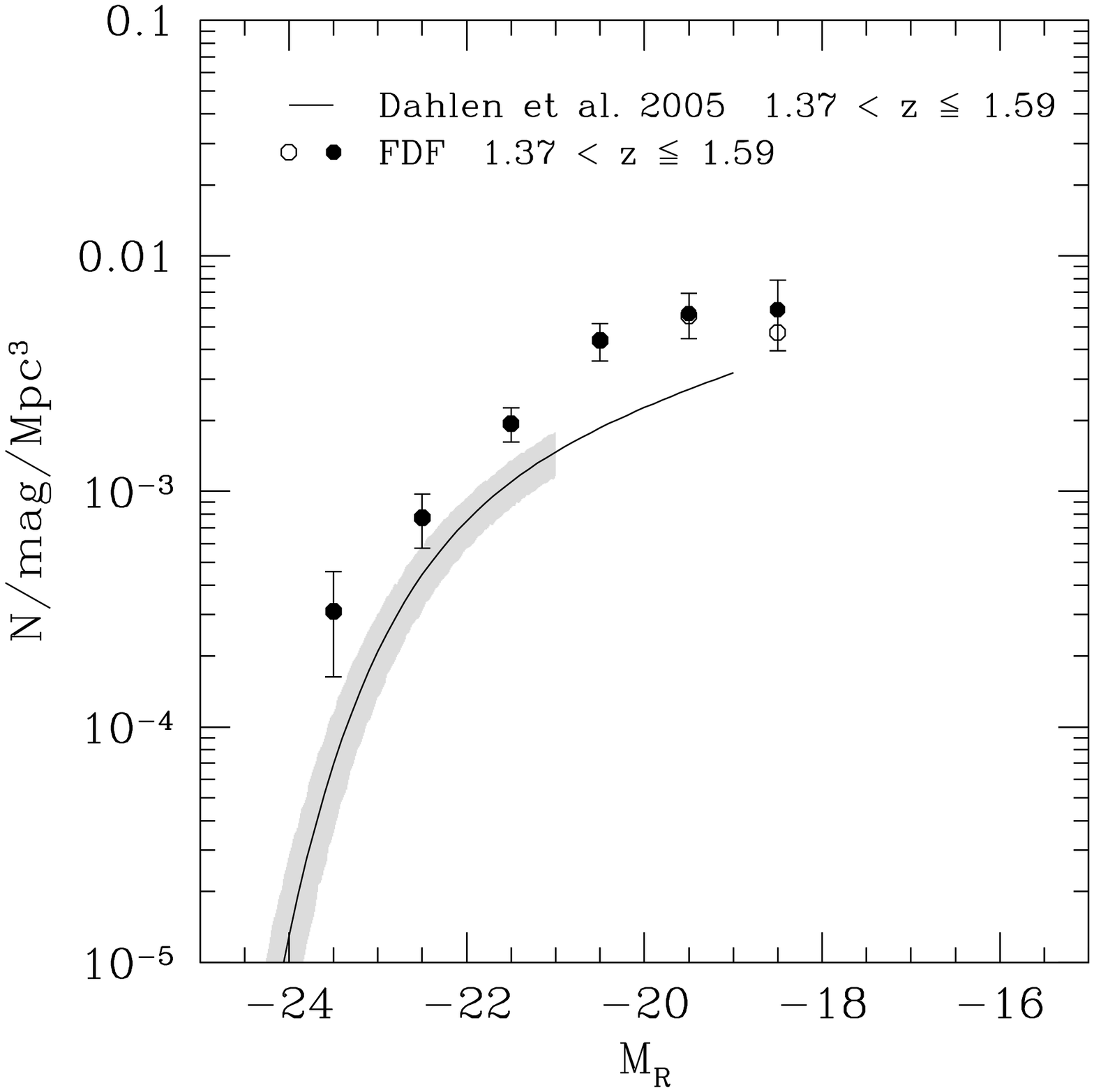}
\includegraphics[width=0.33\textwidth]{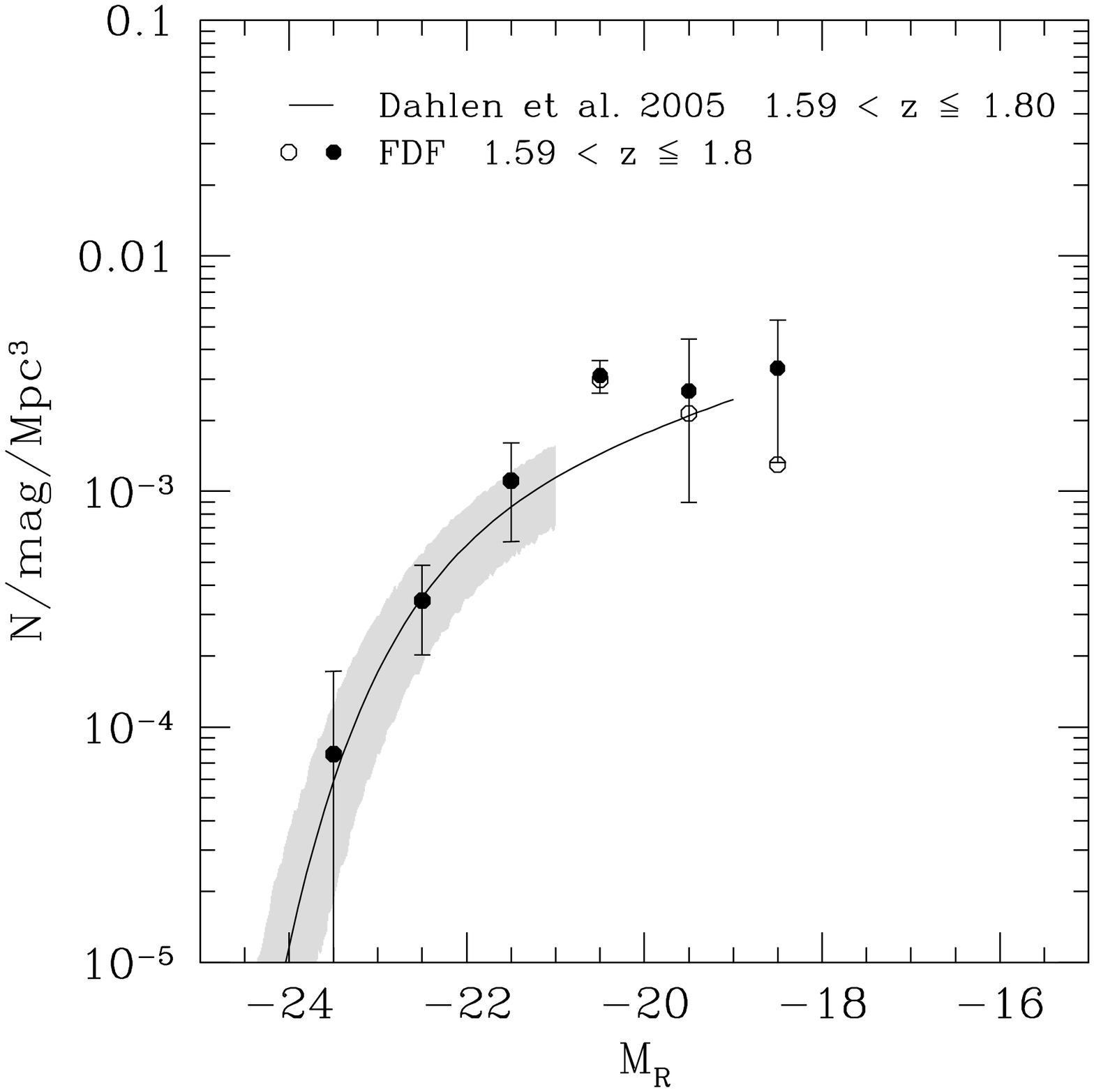}
\includegraphics[width=0.33\textwidth]{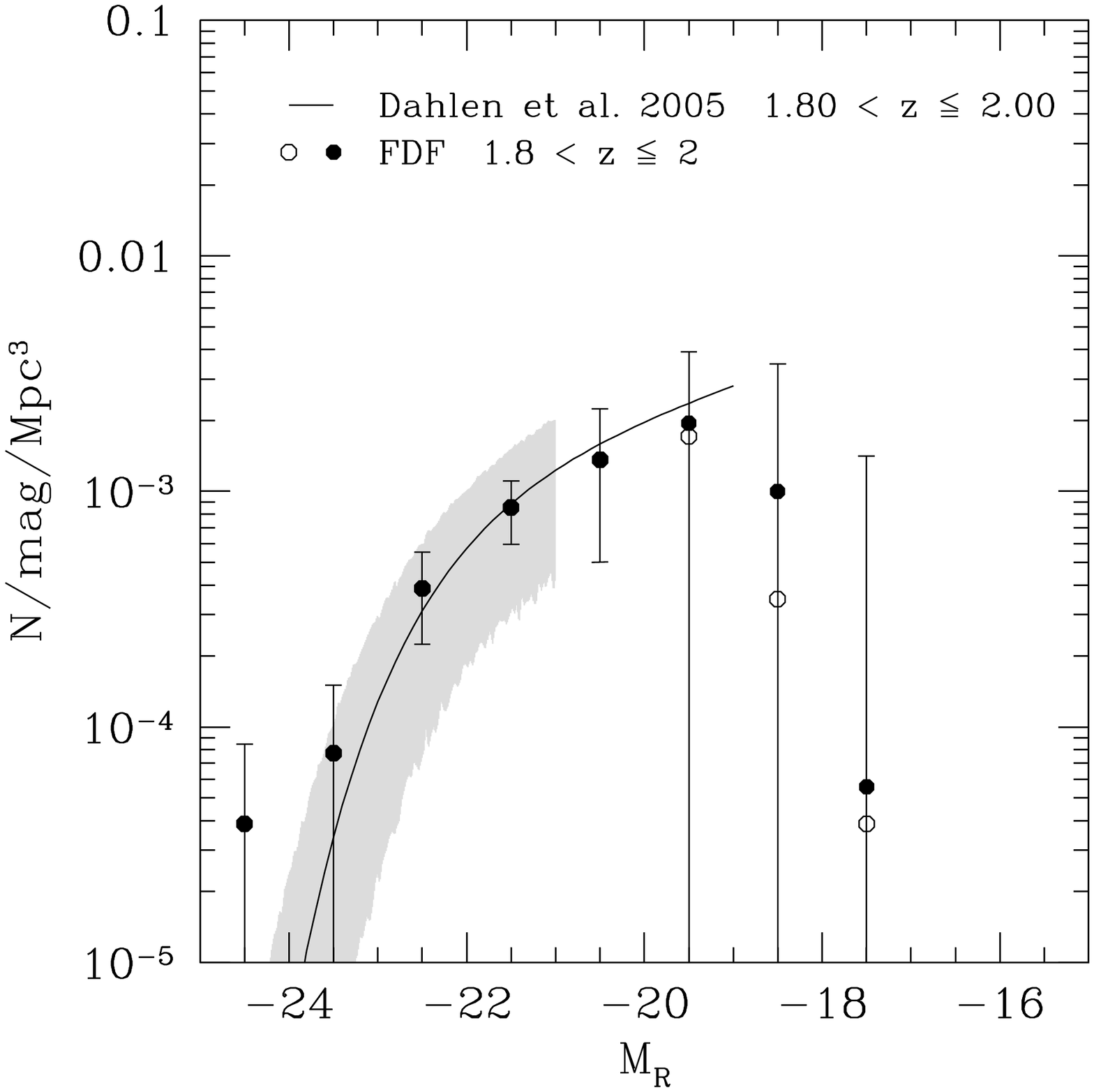}
\caption[Comparison of the R-band FDF LF with \citet{dahlen:1}] 
{\label{fig:lfred:lit_R_dahlen}
Comparison of the R-band LF of the FDF 
with the Schechter function derived in
\textit{\citet{dahlen:1}} at
\mbox{$0.10 <z\le 0.82$},
\mbox{$0.82 <z\le 1.12$},
\mbox{$1.12 <z\le 1.37$},
\mbox{$1.37 <z\le 1.59$},
\mbox{$1.59 <z\le 1.80$}, and 
\mbox{$1.80 <z\le 2.00$}
(from upper left to lower right panel).  The shaded region is based on
$\Delta$M$^\ast$, $\Delta\phi^\ast$ and $\Delta\alpha$ for all panels.
The cut-off of the shaded region at low luminosity indicates the
limiting magnitude of the sample.}
\end{figure*}

\noindent\textit{\citet{dahlen:1}:}\\
\citet{dahlen:1} used HST and ground-based $U$ through $K_s$
photometry in the GOODS-S Field to measure the evolution of the R-Band
luminosity function out to $z\sim2$.  They combine a wider area ($\sim
1100$ arcmin$^2$), optically selected ($R <24.5$) catalog with a
smaller area ($\sim 130$ arcmin$^2$) but deep NIR selected
($K_{s}<23.2$) catalog. Distances are based on photometric redshifts
with an accuracy of \mbox{$\Delta z / (z_{spec}+1) \sim 0.12$}
(\mbox{$\sim 0.06$} after excluding $\sim$ 3\% of outliers).  To
determine the restframe R-band galaxy luminosity function out to
$z\sim2$ they used the deep $K_{s}$ selected catalog.  A comparison
between the R-band LF of \citet{dahlen:1} and the FDF is shown in
Fig.~\ref{fig:lfred:lit_R_dahlen}. There is a very good agreement in
nearly all redshift bins. Only at \mbox{$0.82 <z\le 1.12$} and
\mbox{$1.37 <z\le 1.59$} the characteristic density in the FDF seems
to be slightly higher.\\


To summarize we can say, that the LFs derived in the FDF in general
show a very good agreement with other observational datasets from the
literature. At the bright end of the LF most of the datasets agree
within $1\sigma$.  Differences between the extrapolated Schechter
function of the literature and the measured faint-end in the FDF can
be attributed to the shallower limiting magnitudes of the other
surveys.

\section{Comparison with model predictions}
\label{sec:lfred:model}

\begin{figure*}[tbp]
\includegraphics[width=0.33\textwidth]{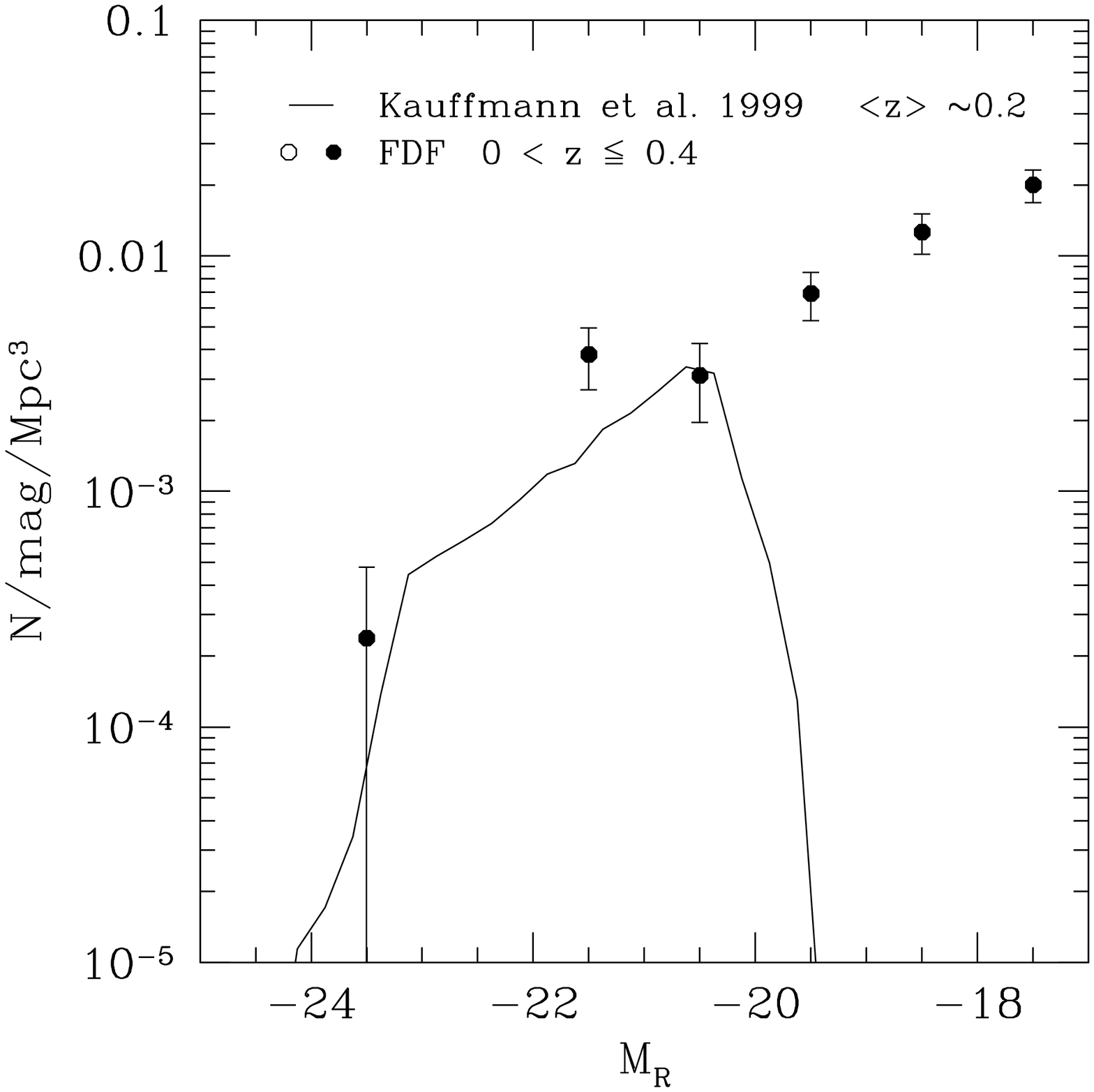}
\includegraphics[width=0.33\textwidth]{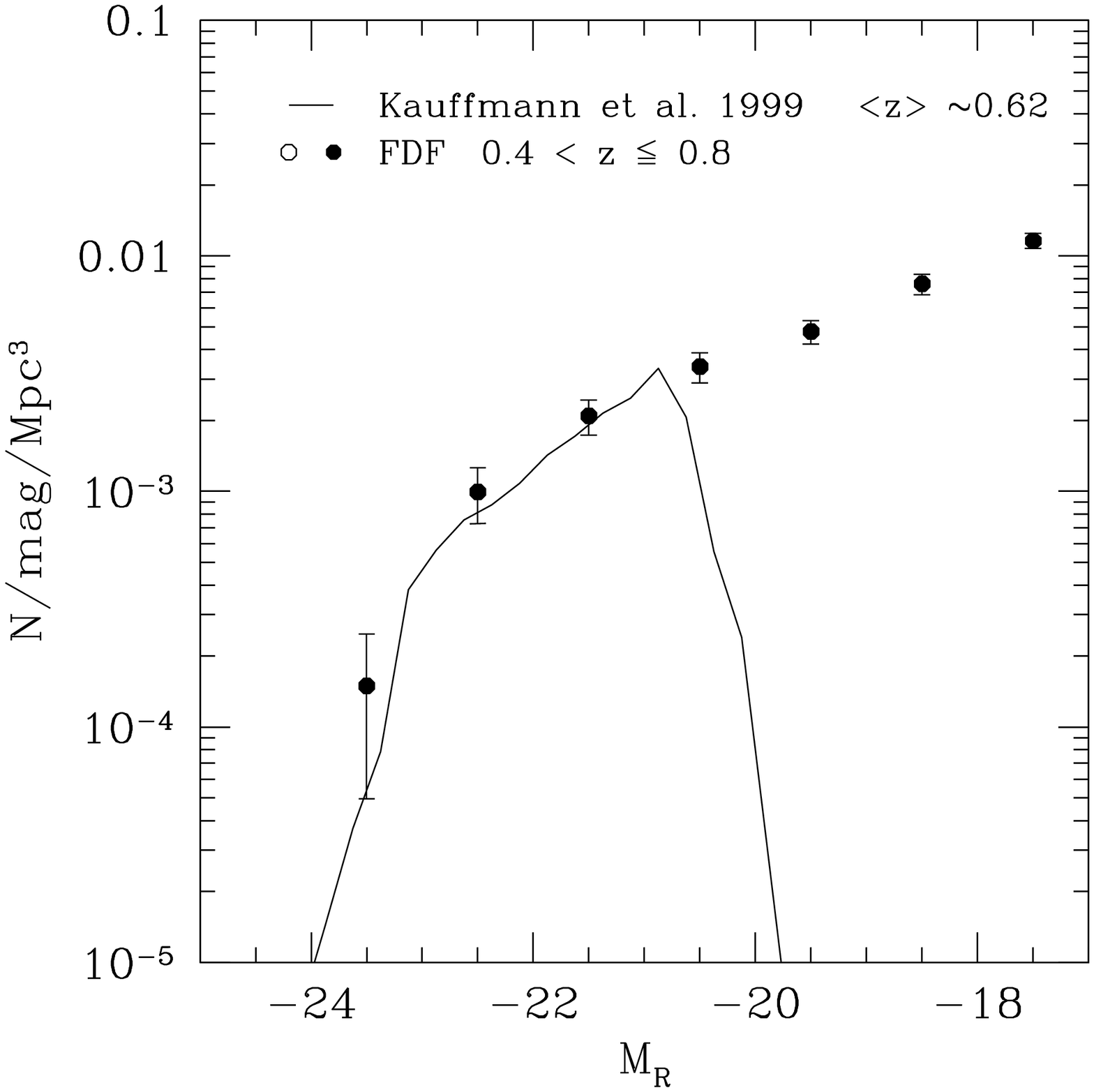}
\includegraphics[width=0.33\textwidth]{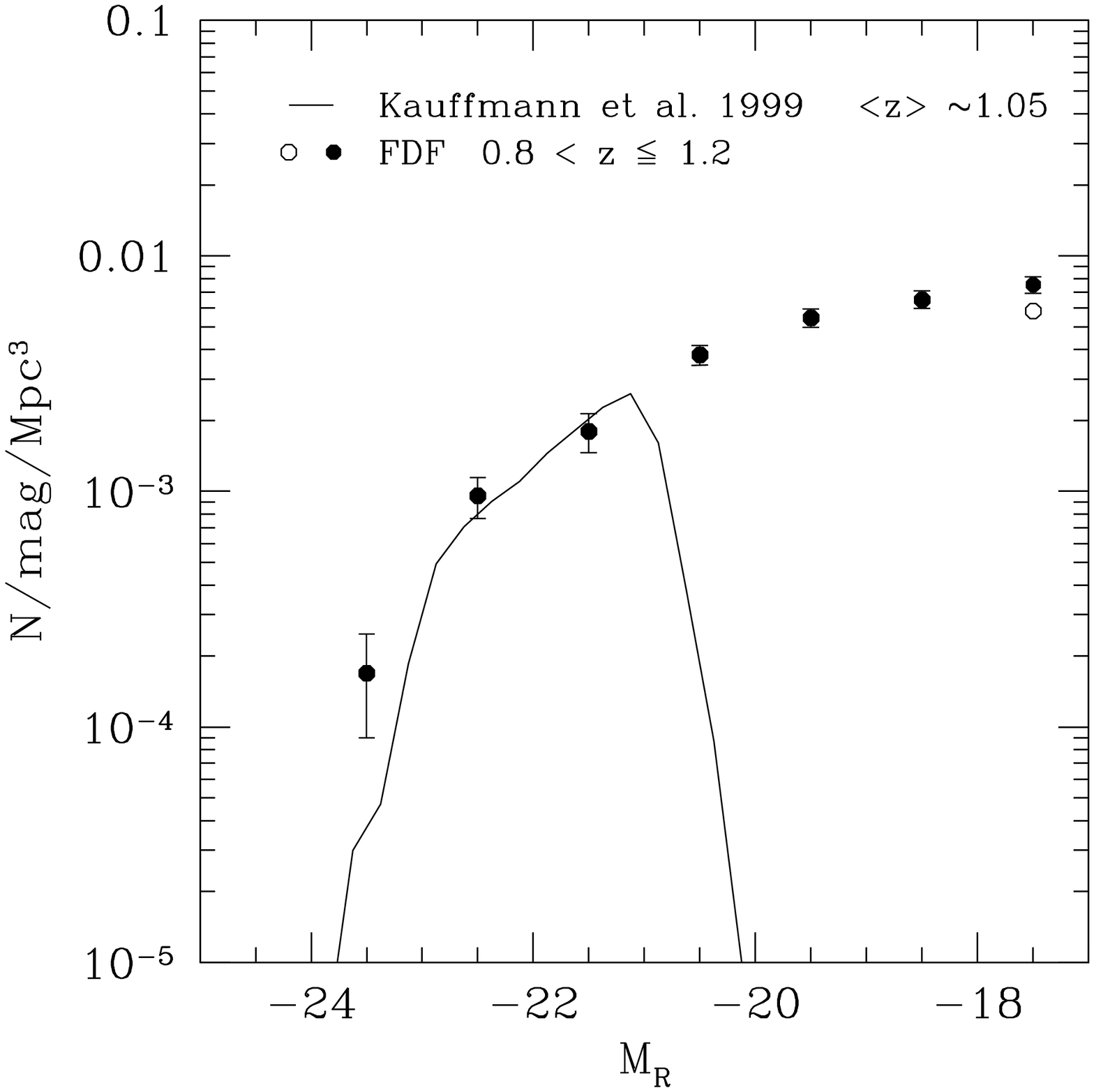}
\caption[Comparison of the R-band FDF LF with model
predictions of \citet{kauffmann:2}]
{\label{fig:lfred:lit_R_kauffmann}
Comparison of the R-band LF of the FDF with
predictions based on 
\textit{\citet{kauffmann:2}} (solid line):
\mbox{$\langle z \rangle\sim 0.20 $},
\mbox{$\langle z \rangle\sim 0.62 $}, and
\mbox{$\langle z \rangle\sim 1.05 $}, 
(from left to right panel).  The filled (open) symbols
show the LF corrected (uncorrected) for $V/V_{max}$.
The drops of the theoretical curves towards the faint-end is caused by
the limited mass resolution of the models, see \citet{kauffmann:2} for
details.}
\end{figure*}
\begin{figure*}[tbp]
\includegraphics[width=0.33\textwidth]{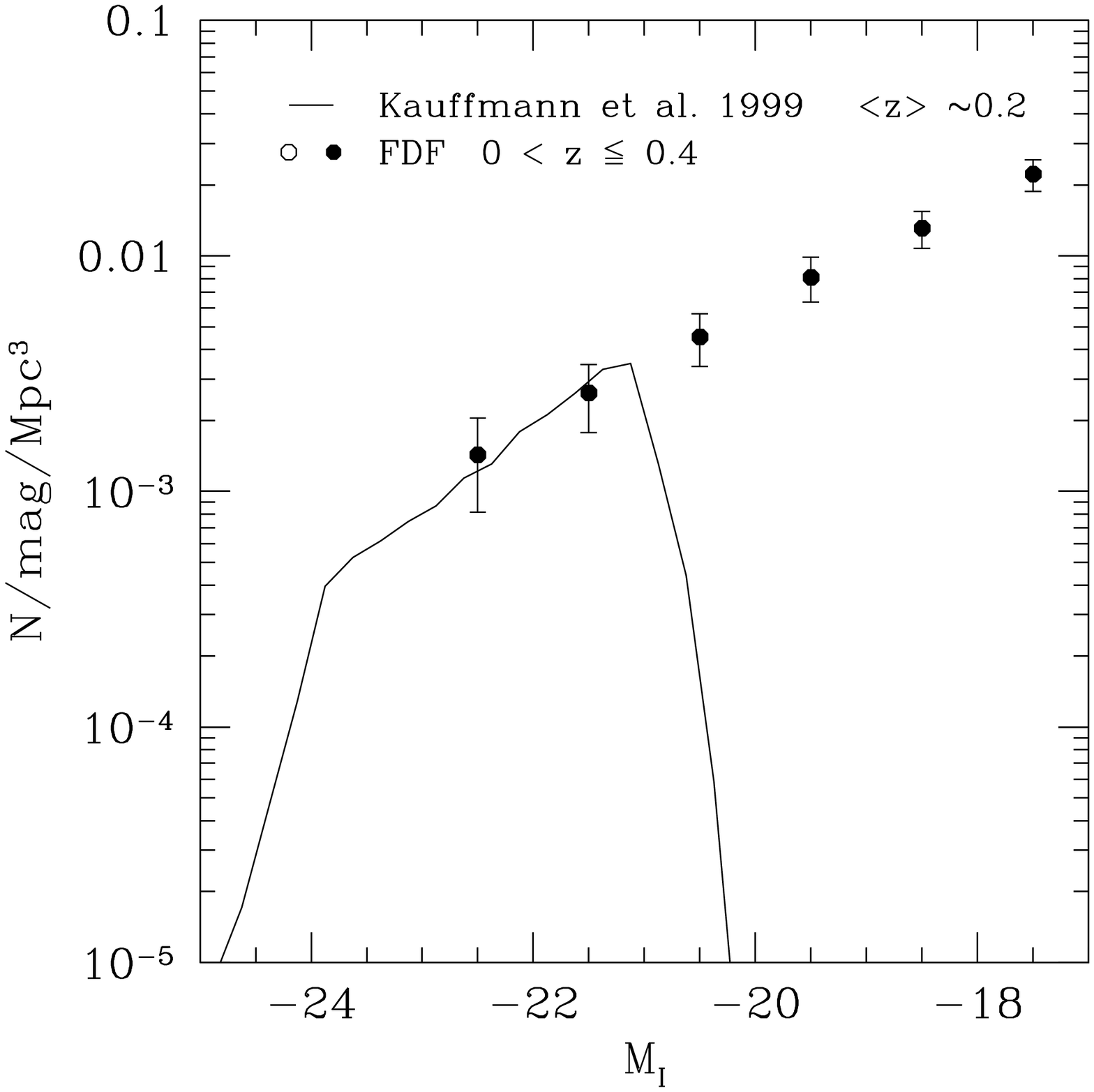}
\includegraphics[width=0.33\textwidth]{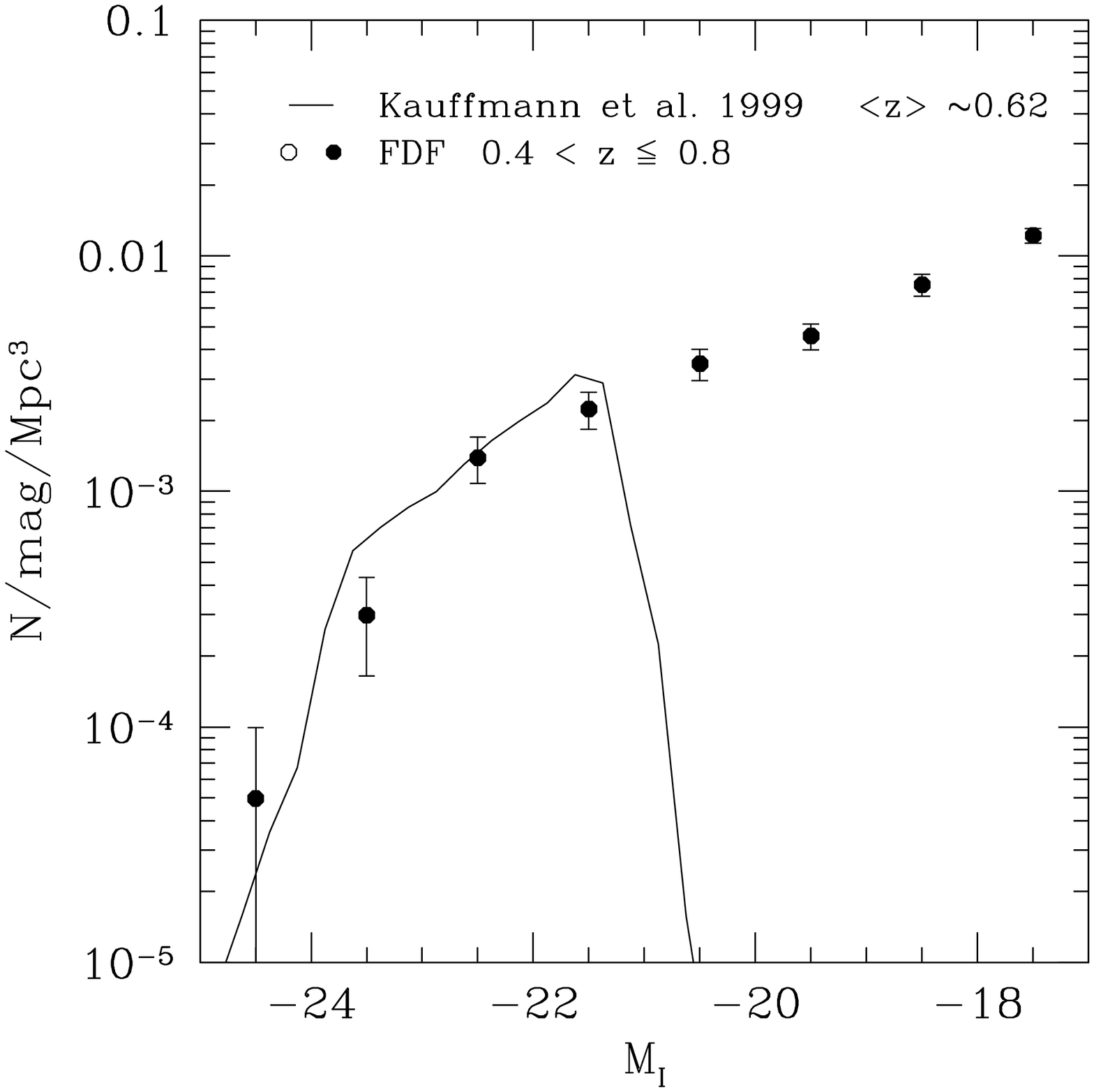}
\includegraphics[width=0.33\textwidth]{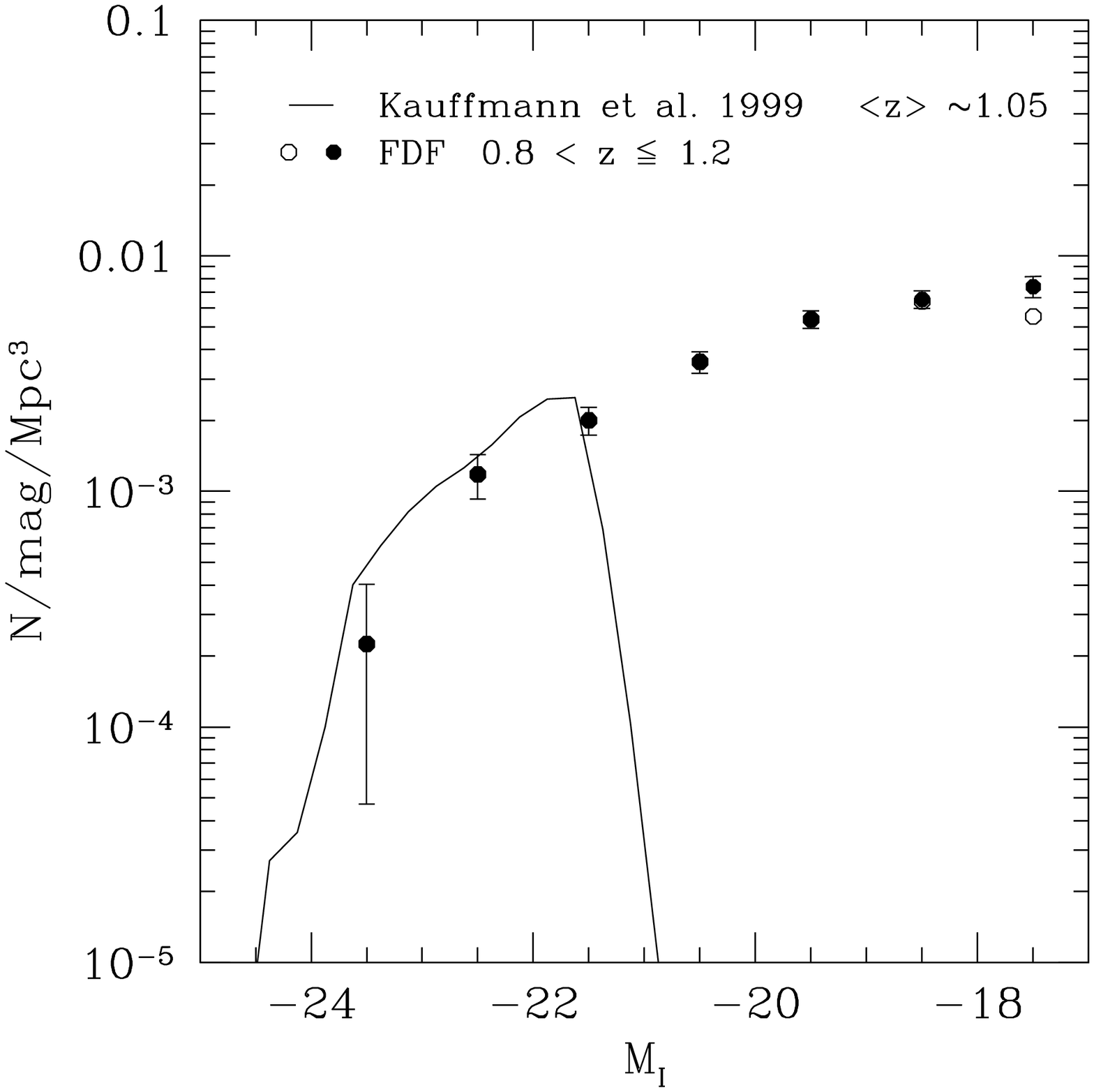}
\includegraphics[width=0.33\textwidth]{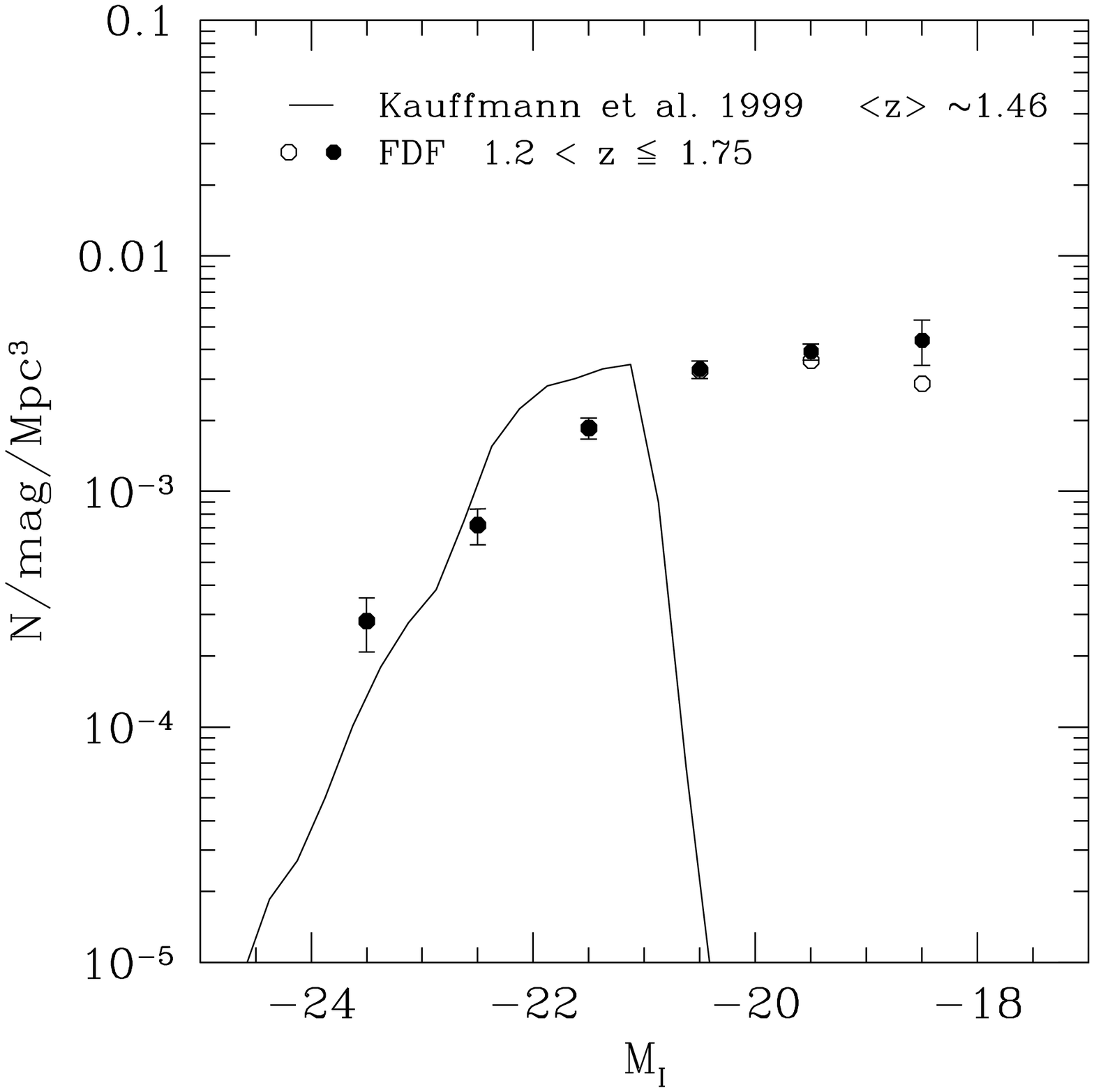}
\includegraphics[width=0.33\textwidth]{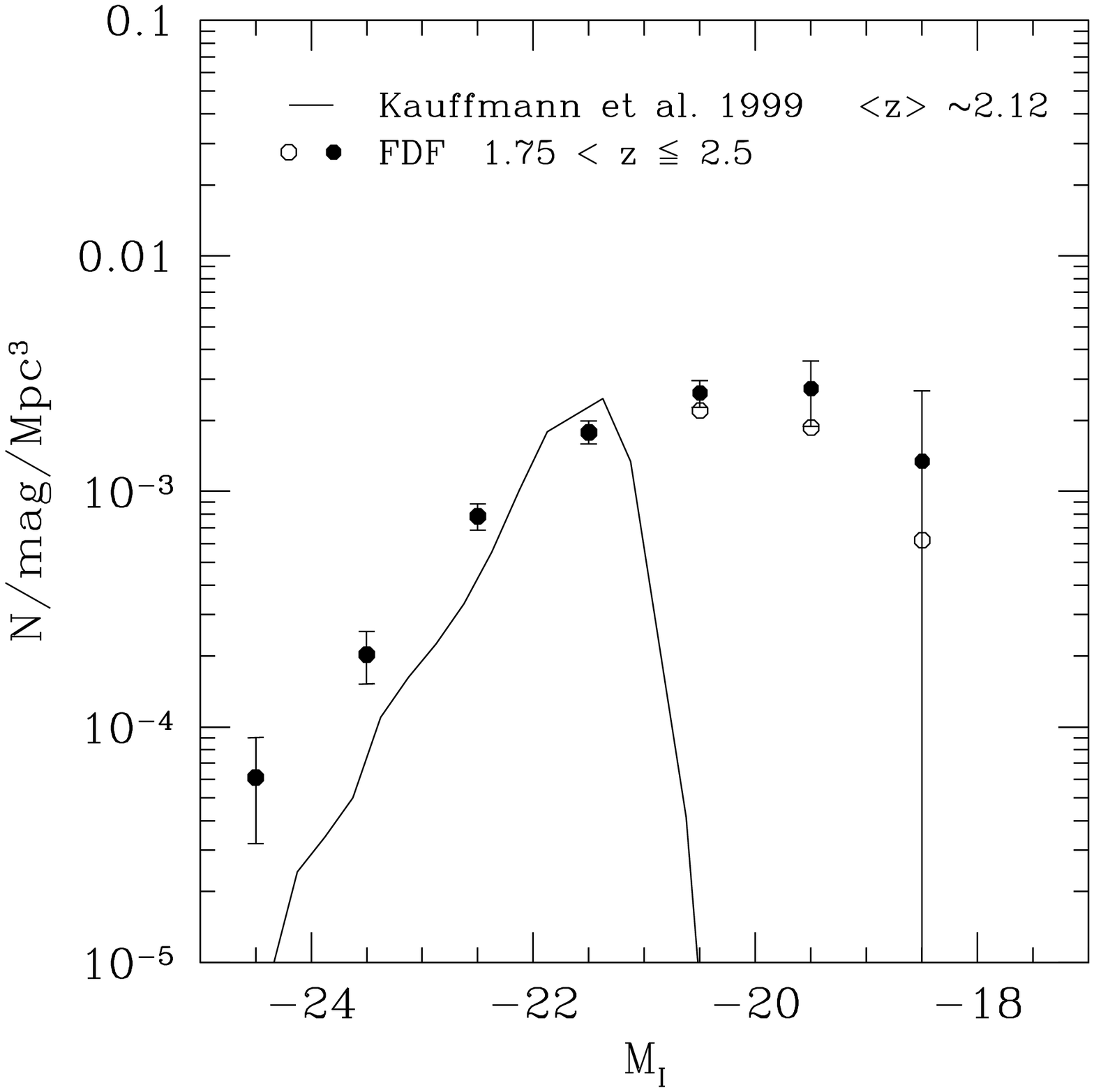}
\includegraphics[width=0.33\textwidth]{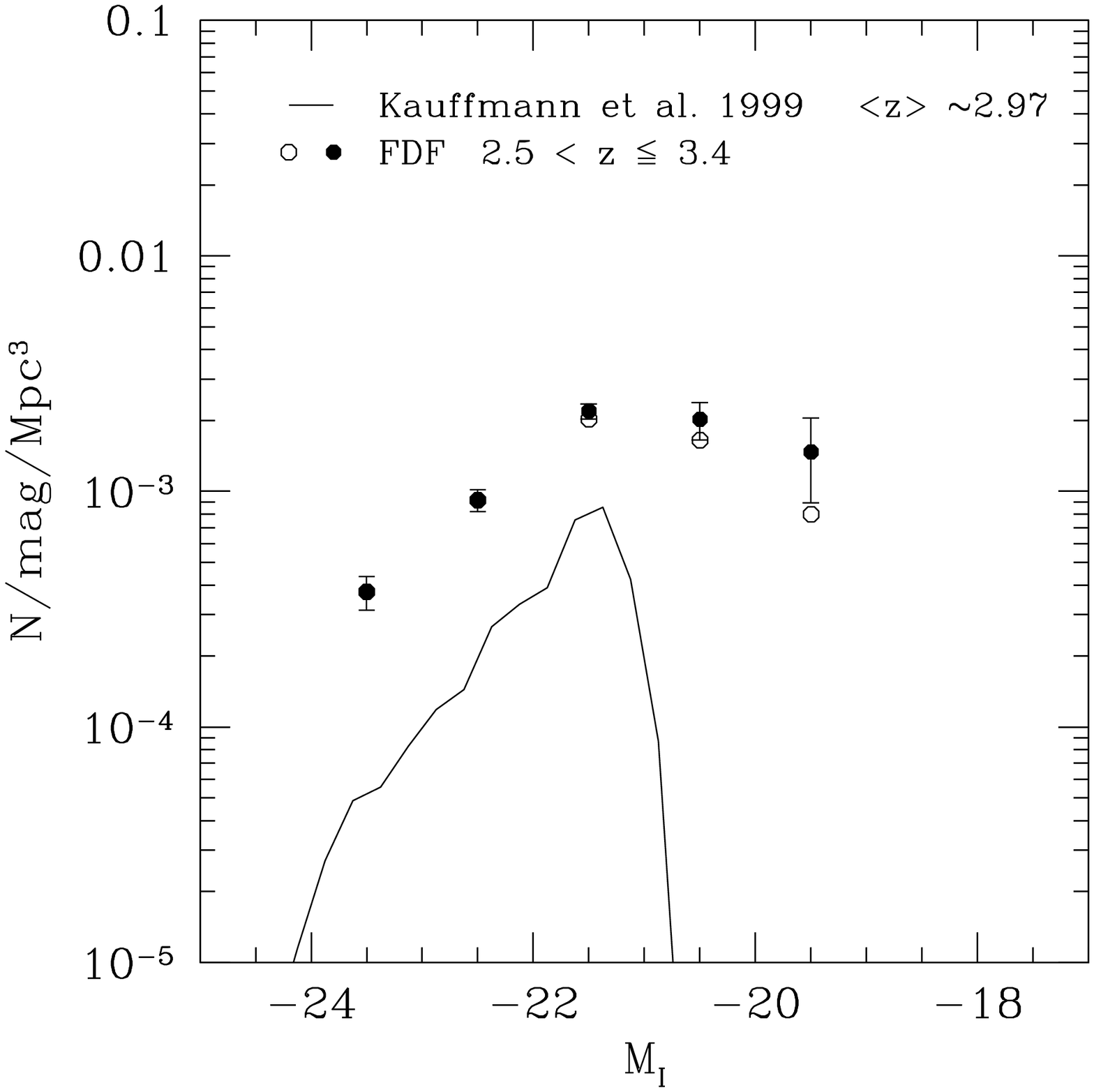}
\caption[Comparison of the I-band FDF LF with model
predictions of \citet{kauffmann:2}] 
{\label{fig:lfred:lit_I_kauffmann}
Comparison of the I-band LF of the FDF with
predictions based on 
\textit{\citet{kauffmann:2}} (solid line):
\mbox{$\langle z \rangle\sim 0.20 $},
\mbox{$\langle z \rangle\sim 0.62 $},
\mbox{$\langle z \rangle\sim 1.05 $},
\mbox{$\langle z \rangle\sim 1.46 $},
\mbox{$\langle z \rangle\sim 2.12 $}, and 
\mbox{$\langle z \rangle\sim 2.97 $}
(from upper left to lower right panel).  The filled (open) symbols
show the LF corrected (uncorrected) for $V/V_{max}$.
The drops of the theoretical curves towards the faint-end is caused by
the limited mass resolution of the models, see \citet{kauffmann:2} for
details.}
\end{figure*}

As discussed for example in \citet{benson:1} different physical
processes are involved in shaping the bright and the faint-end of the
galaxy LF.  Therefore, it is interesting to compare LFs predicted by
models with observational results to better constrain those processes.
In this section we compare the R-band and I-band LFs in different
redshift bins with model predictions of \citet{kauffmann:2}.\\
In Fig.~\ref{fig:lfred:lit_R_kauffmann} we show the R-band luminosity
function of the FDF together with the semi-analytic model predictions
by \citet{kauffmann:2}\footnotemark
\footnotetext{The models were taken from:\\
  http://www.mpa-garching.mpg.de/Virgo/data\_download.html} for
\mbox{$\langle z \rangle\sim 0.20 $}, \mbox{$\langle z \rangle\sim
  0.62 $}, \mbox{$\langle z \rangle\sim 1.05 $}, whereas in
Fig.~\ref{fig:lfred:lit_I_kauffmann} we show the I-band LF in the
redshift bins \mbox{$\langle z \rangle\sim 0.20 $}, \mbox{$\langle z
  \rangle\sim 0.62 $}, \mbox{$\langle z \rangle\sim 1.05 $},
\mbox{$\langle z \rangle\sim 1.46 $}, \mbox{$\langle z \rangle\sim
  2.12 $}, and \mbox{$\langle z \rangle\sim 2.97 $}.  For the R-band
no semi-analytic model predictions are available for redshifts larger
than \mbox{$\langle z \rangle\sim 1.05 $}.\\
There is a good agreement between model (solid lines) and measured LFs
in the R-band.  Also for the I-band there is a good agreement between
the models and the luminosity functions derived in the FDF up to
redshift \mbox{$\langle z \rangle\sim 1.46 $} (of course at $z\approx
0$ the model is tuned to reproduce the data).  At \mbox{$\langle z
  \rangle > 1.46 $} the discrepancy increases as the model does not
contain enough bright galaxies.  Unfortunately, the models only
predict luminosities for massive galaxies and because of lack of
resolution do not predict galaxy number densities for faint
galaxies.\\

\section{Summary and conclusions}
\label{sec:lfred:summary_conclusion}

In this paper we use a sample of about 5600 I-band selected galaxies
in the FORS Deep Field down to a limiting magnitude of I $ = 26.8$ mag
to analyze the evolution of the LFs in the r', i', and z' bands over
the redshift range \mbox{$0.5 < z < 3.5$}, thus extending the results
presented in FDFLF~I to longer wavelengths.  All the results are based
on the same catalog and the same state of the art photometric
redshifts (\mbox{$\Delta z / (z_{spec}+1) \le 0.03$} with only $\sim
1$\% outliers) as in FDFLF~I.
The error budget of the luminosity functions includes the photometric
redshift error of each single galaxy as well as the
Poissonian error.\\
Because of the depth of the FDF we can trace the LFs deeper than most
other surveys and therefore obtain good constraints on the faint-end
slope $\alpha$ of the LF. A detailed analysis of $\alpha$ leads to
similar conclusions as found in FDFLF~I for the blue regime: the
faint-end of the red LFs does not show a large redshift evolution over
the redshift range \mbox{$0.5 \lsim z \lsim 2.0$} and is compatible
within $1\sigma$ to $2\sigma$ with a constant slope in most of the
redshift bins.  Moreover, the slopes in r', i', and z' are very similar
with a best fitting slope of $\alpha=-1.33 \pm 0.03$ for the combined
bands and redshift intervals
considered here.\\
Interestingly, an analysis of the slope of the LFs as a function of
wavelength shows a prominent trend of $\alpha$ to steepen with
increasing wavelength: \mbox{$\alpha_{UV \& u'}=-1.07 \pm 0.04$}
$\rightarrow$ \mbox{$\alpha_{g' \& B}=-1.25 \pm 0.03$} $\rightarrow$
\mbox{$\alpha_{r' \& i' \& z'}=-1.33 \pm 0.03$}.  To better understand
this wavelength-dependence of the LF slope, we analyze the
contribution of different galaxy types to the overall LF by
subdividing our galaxy sample into 4 typical SED types with restframe
U-V colors between 2.3 -- 1.9, 2.0 -- 1.6, 1.6 -- 0.9, and 0.9 -- 0
for SED type 1, 2, 3, and 4, respectively. Therefore, in a rough
classification one can refer to SED types 1 and 2 (SED type 3 and 4)
as red (blue) galaxies.\\
Although in the UV regime the overall LF is completely dominated by
extremely star-forming galaxies, the overall LF in the red regime is
mainly dominated by early to late type galaxies at the bright end, but
extremely star-forming galaxies at the faint-end. The relative
contribution of the different SED type LF to the overall LF clearly
changes as a function of analyzed waveband resulting (at the depth of
the FDF) in a steeper slope for the overall LF in the red
regime if compared to the blue regime.\\
To quantify the contribution of the different SED types to the total
luminosity density, we derive and analyze the latter in the UV and in
the red bands as a function of redshift: The contribution of type 1
and 2 galaxies to the UV LD is negligible at all analyzed redshifts as
SED type 3 and 4 completely dominate. On the other hand, the relative
contribution to the overall luminosity density of type 1 and 2
galaxies is of the same order or even exceeds those of type 3 and 4 in
the red bands.

We investigate the evolution of M$^\ast$ and $\phi^\ast$ (for a fixed
slope $\alpha$) by means of the redshift parametrization introduced
in FDFLF~I.  
 Based on the FDF data (\textit{Case~1} and
  \textit{Case~2}), we find only a mild brightening of M$^\ast$ and
  decrease of $\phi^\ast$ with increasing redshifts in all three
  analyzed wavebands. If we follow the evolution of the characteristic
  luminosity from \mbox{$\langle z \rangle\sim 0.5$} to \mbox{$\langle
    z \rangle\sim 3$}, we find an increase of $\sim$~0.8 magnitudes in
  the r', and $\sim$~0.4 magnitudes in the i' and z' bands.
  Simultaneously the characteristic density decreases by about 40 \%
  in all analyzed wavebands.
We compare the LFs with previous observational datasets and discuss
discrepancies. As for the blue bands, we find good/very good agreement
with most of the datasets especially at the bright end.  Differences
in the faint-end slope in most cases can be attributed to the
shallower limiting magnitudes of the other surveys.

We also compare our results with predictions of semi-analytical models
at various redshifts.  The semi-analytical models predict LFs which
describe the data at low redshift very well, but as for the blue
bands, they show growing disagreement with increasing redshifts.
Unfortunately, the models only predict luminosities for massive
galaxies and therefore, a comparison between the predicted and
observed galaxy number densities for low luminosity galaxies ($L \lsim
L^\ast$) could not be done.

\begin{acknowledgements}
  
   We thank the anonymous referee for his/her helpful comments
    which improved the presentation of the paper.  
We acknowledge
  the support of the ESO Paranal staff during several observing runs.
  This work was supported by the \emph{Deut\-sche
    For\-schungs\-ge\-mein\-schaft, DFG}, SFB 375
  Astro\-teil\-chen\-phy\-sik, SFB 439 (Galaxies in the young
  Universe), the Volkswagen Foundation (I/76\,520) and the Deutsches
  Zentrum f\"ur Luft- und Raumfahrt (50\,OR\,0301).

\end{acknowledgements}

\bibliographystyle{aa}
\bibliography{literature}

\end{document}